\documentclass[aip, jcp, reprint]{revtex4-2}

\usepackage{graphicx}
\usepackage{amsmath}
\usepackage{array}[=2016-10-06]
\usepackage{amsthm}
\usepackage{amssymb}
\usepackage{xcolor}
\usepackage[hidelinks]{hyperref}
\usepackage[normalem]{ulem} 
\hypersetup{
  colorlinks=true,
  linkcolor=black,   
  citecolor=black,   
  filecolor=black,   
  urlcolor=black      
}

\usepackage{tikz}     
\usetikzlibrary{arrows.meta}
\usepackage{bibunits} 

\defaultbibliography{polycubes_references_abbrev,revtexControl}
\defaultbibliographystyle{aipnum4-2}

\definecolor{spA}{RGB}{166,206,227}  
\definecolor{spB}{RGB}{253,191,111}  
\definecolor{spC}{RGB}{178,223,138}  
\definecolor{spD}{RGB}{202,178,214}  
\definecolor{spE}{RGB}{251,154,153}  
\definecolor{spF}{RGB}{255,255,153}  

\newcommand{\pctile}[8]{%
  \fill[#3] (#1,#2) rectangle ++(1,1);
  \draw[line width=0.6pt] (#1,#2) rectangle ++(1,1);
  \node[font=\footnotesize] at (#1+0.5,#2+0.5) {$#4$};
  \node[font=\scriptsize] at (#1+0.5,#2+0.83) {$#5$};
  \node[font=\scriptsize] at (#1+0.83,#2+0.5) {$#6$};
  \node[font=\scriptsize] at (#1+0.5,#2+0.17) {$#7$};
  \node[font=\scriptsize] at (#1+0.17,#2+0.5) {$#8$};
}
\newcommand{\pcpanel}[3]{\node[anchor=west,font=\small\bfseries] at (#1,#2) {(#3)};}
\newcommand{\pcnote}[3]{\node[anchor=west,font=\footnotesize] at (#1,#2) {#3};}

\newtheorem{proposition}{Proposition}

\begin{document}
\begin{bibunit}
\nocite{REVTEX42Control,aip41Control}

\title{How symmetry and complexity shape three-dimensional programmable self-assembly}

\author{Prarthana Agrawal}
\affiliation{Rudolf Peierls Centre for Theoretical Physics, University of Oxford, United Kingdom OX1 3PU}

\author{Joakim Bohlin}
\affiliation{Department of Physics and Astronomy, Chalmers University of Technology, Gothenburg SE-412 96, Sweden}

\author{Ard A. Louis}
\email[]{ard.louis@physics.ox.ac.uk}
\affiliation{Rudolf Peierls Centre for Theoretical Physics, University of Oxford, United Kingdom OX1 3PU}


\begin{abstract}
Self-assembly enables complex structures to emerge from simple building blocks through local interaction rules.  Are some structures intrinsically easier to design, that is, do they admit more sets of assembly rules than others? And, can this be predicted from their geometry? We address these questions using a three-dimensional polycube model with programmable face-to-face binding, defining a structure's \textit{assembly rule probability} as the chance that a randomly chosen rule set produces it. For polycubes, this probability decreases exponentially with \textit{assembly complexity}, the minimum information required to specify a structure’s assembly rules. Intuitively, this follows because structures with low assembly complexity are generated not only by minimal rule sets but also by numerous more elaborate ones, and so arise disproportionately often. Symmetry can reduce assembly complexity by allowing the same building blocks and local interactions to be reused across different parts of a structure.  Indeed, we find that the highest probability structures come from a tiny subset of highly symmetric structures. However, symmetry alone does not determine assembly complexity. Structures with the same symmetry can differ substantially in the amount of assembly information they require, depending on how effectively their global symmetry can be exploited by local growth rules. Overall, our results link information, symmetry, and assembly rule probability and suggest that identifying which geometric features permit rule reuse is central to designing minimal programmable self-assembling systems. 
\end{abstract}


\maketitle 

\section{Introduction}
\label{sec:intro}

Assembling a complex structure from its components requires information that specifies how these individual components interact and organize~\cite{Lehn2002, Whitesides2002a, hagan_equilibrium_2021, Cademartiri2015, Kushner1969}. In a top-down approach, assembly information is stored in a centralized blueprint that directs the assembly process, much like a robot assembling parts on an assembly line. But information can also be encoded locally in the interactions between components, allowing ordered structures to emerge spontaneously without external control. This bottom-up process, known as self-assembly, underlies many biological assemblies such as viral capsids~\cite{Caspar1962physical, Perlmutter2015mechanisms, Hagan2014modeling}, lipid membranes~\cite{Israelachvili2010}, microtubules~\cite{Wade2009microtubule}, and protein complexes~\cite{ Cademartiri2015, Kushner1969}. 

These biological systems have inspired the design of artificially programmed interaction rules between components so that target structures can be assembled, even at scales where individual components cannot be manipulated directly~\cite{Doty2012}. A particularly well-developed example is DNA nanotechnology~\cite{Seeman1982}, where approaches such as DNA origami~\cite{ Rothemund2006folding}, DNA bricks~\cite{Ke2012dnabricks}, DNA wireframe structures~\cite{benson2015dna, veneziano2016designer}, and DNA cubes~\cite{shiu2025acoustically, Diep2025universal}  enable the construction of remarkably complex nanoscale structures~\cite{seeman2017dna}.  Another promising approach includes DNA-coated colloids~\cite{Jacobs2025dnacolloid}, which enable the design of patchy-particle systems into complex  structures~\cite{liu2024inverse}.

The information required to self-assemble a given finite structure can be locally encoded in two contrasting ways. In \emph{fully addressable} designs, every component and interaction is uniquely specified~\cite{frenkel_order_2015, Cademartiri2015, jacobs_self-assembly_2016,Bohlin2023}. Such designs provide precise control over assembly but are information-intensive, difficult to scale, costly to implement~\cite{Hubl2025accessing}, and need thermodynamic annealing protocols~\cite{jacobs_self-assembly_2016,fonseca2018multi} to assemble due to the high configurational entropy cost.  At the other extreme are \emph{minimal} designs, which generate complex structures by reusing the smallest set of components and interactions~\cite{Ahnert2010, Johnston2022, Bohlin2023}. Such designs require substantially less assembly information and reduce synthesis costs. They may also have larger thermodynamic windows for high-yield assembly~\cite{Bohlin2023}.  Nature frequently exploits this strategy. For example, in viral capsid assembly, large symmetric shells are constructed from repeated copies of a small number of protein subunits~\cite{Caspar1962physical, Hagan2014modeling}.  It is an open question whether these near-minimal solutions arise because they confer adaptive advantages~\cite{Crick1956virus, Caspar1962physical, wales1998symmetry}, or whether they occur preferentially simply because they are easier to find by evolutionary search~\cite{Johnston2022}.

Some structures can be assembled from relatively simple minimal designs, reusing many copies of only a few component types, whereas others require designs that are close to fully addressable assembly. These differences  allowed Cartwright and Mackay  to introduce \emph{assembly complexity}, defined as the minimum information required to specify the assembly of a structure~\cite{Cartwright2012}, as a general measure of structural complexity. Traditionally, the complexity of crystal structures has been characterized in terms of symmetry, with highly symmetric structures regarded as simpler than those with lower symmetry. However, many biological and other self-assembled structures are aperiodic, making symmetry an incomplete measure of structural complexity. Cartwright and Mackay argued that assembly complexity provides a more general description because it applies equally to periodic and aperiodic structures~\cite{Cartwright2012,Buchanan2012}. Related information-theoretic approaches have also been used to quantify the structural complexity of inorganic crystals~\cite{Krivovichev2014}. This work motivates the first question our paper will address: \textit{1) How well does assembly complexity correlate with symmetry?}

A second question follows from a simple intuition: structures whose assembly complexity is much lower than that of a fully addressable design should admit many more alternative  assembly rules than structures whose minimal complexity is already close to this maximum. To quantify this intuition, we define the \textit{assembly rule probability} of a structure as a measure of how frequently it is generated when assembly rules are sampled at random. This naturally raises the question: \textit{2) How does assembly rule probability scale with assembly complexity and with symmetry?}

For insight into this second question we first go down a more abstract route, and  note that  assembly complexity  has parallels with Kolmogorov complexity $K(x)$ from algorithmic information theory (AIT)~\cite{ LiVitanyi}, defined as  the length of the shortest algorithm that outputs the string $x$ on a universal Turing machine (UTM). Unlike Kolmogorov complexity, which is uncomputable~\cite{LiVitanyi}, assembly complexity is grounded in the computable physical process of assembly itself.  

A central concept in AIT is \textit{algorithmic probability}. For a string $x$, the algorithmic probability  $P_U(x)$  is defined as the probability that a (suitably chosen) UTM $U$ will produce $x$ when supplied with a randomly chosen input program.  Since algorithmic probabilities for different UTMs differ only by machine-dependent multiplicative constants~\cite{LiVitanyi}, the UTM dependence can be omitted, and the probability written simply as $P(x)$. Levin's coding theorem~\cite{Levin1974, LiVitanyi}  then predicts that $P(x) \approx 2^{-K(x)}$.  Intuitively, if an output string can be produced by a short input program, then it can be produced by exponentially many other longer programs as well.    While this powerful theorem can only be proven for UTMs,  Dingle \emph{et al.}~\cite{Dingle2018} showed that for many non-universal input--output maps, an analogue to algorithmic probability of AIT, the computable probability $P(x)$ of generating output $x$ upon random inputs, can be bounded from above as $P(x)\lesssim 2^{-a\tilde K(x)-b}$, where \(\tilde K(x)\) is an appropriate approximation to the true Kolmogorov complexity and \(a>0\), \(b\) are map-dependent constants.   This \textit{simplicity bias} implies that high $P(x)$ outputs have low $\tilde{K}(x)$, and high $\tilde{K}(x)$ outputs have low $P(x)$, although low $\tilde{K}(x)$ does not necessarily imply high $P(x)$.  However, a statistical lower bound  constrains the total probability weight of outputs that lie far below the upper bound, so that the majority of the probability weight should not be too far from the upper bound~\cite{Dingle2020}. 

These AIT-based arguments are abstract but also broadly applicable. The predicted simplicity bias has been observed in diverse systems including two-dimensional polyomino self-assembly~\cite{Johnston2022}, Richard Dawkins' biomorphs~\cite{Martin2024},  a model of tooth development~\cite{dingle2025bounding}, and RNA secondary structure~\cite{Dingle2018,Johnston2022}. For functional non-coding RNA, the calculated $P(x)$ closely recapitulates the frequencies of folded structures observed in nature~\cite{dingle2015structure,Dingle2022,von2025rna}. The consistency of this behavior across such different systems suggests that assembly rule probability should likewise decrease exponentially with assembly complexity in a broad class of self-assembling systems.
 
\begin{figure*}
\includegraphics[width=\textwidth]{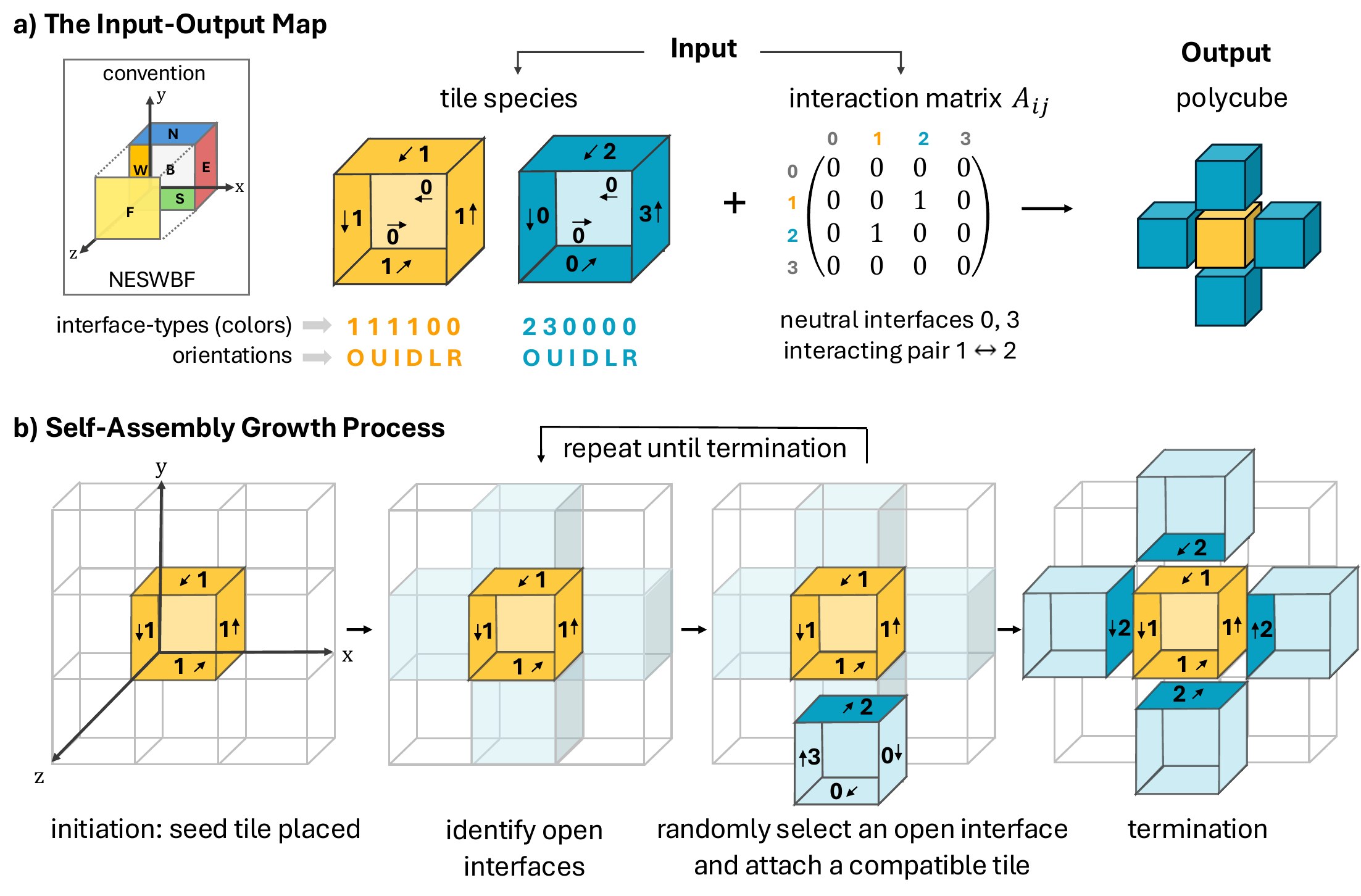}
\caption{\textbf{Polycube self-assembly model.} (a) \textit{Input–output map.}  
The assembly kit (\textit{input}) specifies a set of cube-shaped tile species and their interaction rules encoded in the matrix $A_{ij}$.  In this illustrative example, the kit contains $s=2$ tile species and $c=4$ interface types (colors). Interface types $1$ and $2$ form a complementary attractive pair, while interface types $0$ and $3$ are neutral. Each tile species has six faces, with each face specified by an interface type (integers $0$ to $c-1$) and an in-plane orientation (Out, Up, In, Down, Left, Right), listed in North-East-South-West-Back-Front (NESWBF) order. The leftmost panel shows the face-ordering convention and Cartesian axes used to define the orientations.  These local interaction rules deterministically produce the finite polycube (\textit{output}) shown on the right.
(b) \textit{Self-assembly growth process.}
Assembly begins with a \textit{seed} tile placed at the origin of the cubic lattice. This seed tile exposes open interfaces for attachment. At each subsequent step, one open interface is selected at random, an interface carrying the complementary type is selected from the assembly kit, and the corresponding tile is rotated into a compatible orientation and attached to the growing structure. Growth continues until no open interfaces remain, yielding the cross-shaped polycube shown. For visual clarity, neutral back and front faces are not labeled, and only bonding faces are labeled in the final frame.  In \textit{seeded assembly}, the first tile species in the assembly kit is used as the seed; in \textit{unseeded assembly}, the seed tile is chosen uniformly at random from the assembly kit.}
\label{Fig:polycube_model}
\end{figure*}

To address the two questions we posed above in a concrete setting, we study tile-based lattice models of self-assembly, which have the advantage of being tractable while still enabling a rich variety of structures to assemble.  Tile-based lattice models trace back to Wang tiles, introduced by Hao Wang in 1961 as square tiles with labeled edges that can be placed only when adjacent edges match~\cite{Wang1961}. Originally developed to study mathematical tilings, Wang tiles later became the foundation of algorithmic self-assembly~\cite{Patitz2012}. Building on Ned Seeman’s demonstrations of sequence-directed DNA self-assembly~\cite{Seeman1982}, Winfree recognized that DNA tiles could function as programmable tiles and developed the abstract Tile Assembly Model (aTAM)~\cite{winfree1998thesis, Rothemund2000program}. In the aTAM, structures grow from an initial seed by sequential tile addition governed by local glue-matching rules and binding-strength thresholds. Despite relying only on local interactions, the model can generate arbitrarily complex structures~\cite{Soloveichik2007,Rothemund2004} and even perform Turing-universal computation~\cite{winfree1998thesis, Doty2012}, establishing a formal connection between self-assembly and Kolmogorov complexity~\cite{Rothemund2000program, Adleman2001, Soloveichik2007}.

A two-dimensional model related to the aTAM is the polyomino model of self-assembly in which rotatable square tiles with fixed-strength, non-cooperative interactions assemble into connected polyominoes~\cite{Ahnert2010, Johnston2011, Greenbury2014}. Unlike the aTAM, the polyomino model is not Turing universal, but it provides a simple and physically motivated setting for studying programmable self-assembly. It has proved a useful model of protein quaternary structure assembly and provides a genotype–phenotype map in which genotypes encode assembly rules for square tiles (protein subunits), while phenotypes correspond to assembled polyominoes (protein complexes)~\cite{Ahnert2010, Johnston2011, Greenbury2014, Greenbury2016, manrubia2021genotypes}. Random sampling of these genotypes, as well as neutral evolution on the resulting genotype–phenotype map, produces, to first order at least, highly symmetric structures exponentially more often than asymmetric ones~\cite{Johnston2022}. Remarkably, the resulting symmetry distributions closely match those of protein complexes in the Protein Data Bank (PDB), suggesting that the prevalence of symmetry may arise not only through adaptive selection but also through simplicity bias: highly symmetric structures generally have lower assembly complexity and are therefore discovered more readily by evolutionary search.

Much of the work on programmable tile assembly has focused on two dimensions. Yet many biological and engineered assemblies, including protein complexes, are inherently three-dimensional. Here we study a three-dimensional analogue of the polyomino model, in which cubic tiles with programmable interactions assemble into connected polycubes~\cite{Bohlin2023, Diep2025universal, Evans2024}. This allows us to investigate how assembly complexity, assembly rule probability, and symmetry are related in three-dimensional self-assembly.

Our main finding is that assembly rule probability decreases approximately exponentially with assembly complexity, consistent with the simplicity bias previously observed in two-dimensional polyomino systems~\cite{Johnston2022}. Symmetry can reduce assembly complexity by allowing tile species and interface types to be reused, leading to a strong enhancement of certain highly symmetric structures under random sampling of assembly rules. However, symmetry alone does not determine assembly complexity or assembly rule probability. For example, a four-cube linear rod and a four-cube square have the same global symmetry, yet deterministic assembly of the rod requires substantially more information because its tile species and interface types cannot be reused as effectively. Assembly complexity therefore depends not only on global symmetry but also on how effectively local interaction rules can be reused during assembly.

The remainder of the paper is organized as follows. We first introduce the three-dimensional polycube model and define the assembly rule space, assembly rule probability, symmetry classification, and measures of assembly complexity. We then investigate how assembly rule probability depends on structure size, complexity, and symmetry. Finally, we examine why some symmetries reduce assembly complexity more effectively than others, and why structures with identical global symmetry can nevertheless differ substantially in their assembly rule probability.

\section{Model and Methods} \label{sec:model}

\subsection{Polycube model overview} \label{subsec:model_overview}

A polycube is a finite connected cluster of unit cubes joined face-to-face on a three-dimensional cubic lattice. The size of a polycube, denoted by $n$, is the number of constituent cubic tiles. In the polycube self-assembly model, structures emerge through local interactions between these tiles \cite{Bohlin2023}. The complete specification of an assembly system is given by an \emph{assembly kit}, consisting of \(s\) tile species, with \(c\) possible interface types assigned to their faces, and an interaction matrix defining which interface types bind.

Each cubic tile has six faces aligned with the Cartesian axes. Every face is assigned two attributes:

\begin{enumerate}
   \item \textbf{Interface type}: Each face carries an integer label
    \( i \in \{0, 1, \dots, c-1\} \) which defines its interface type. The label \( i = 0 \) represents a neutral interface that does not bind to anything, while non-zero labels correspond to active interface types that form interacting pairs with their designated complementary interfaces as specified by the interaction matrix. If \(c\) is even, the final label \(c-1\) is also considered neutral. Following the convention of Wang tiles~\cite{Wang1961}, these interface types are also referred to as \emph{colors}. 

    \item \textbf{Orientation}: Each face is additionally assigned an in-plane orientation \( r \in \) \{\texttt{O,U,I,D,L,R}\}, corresponding to \textit{Out}~$\swarrow$, \textit{Up}~$\uparrow$, \textit{In}~$\nearrow$, \textit{Down}~$\downarrow$, \textit{Left}~$\leftarrow$, and \textit{Right}~$\rightarrow$. The orientation is defined with respect to a right-handed coordinate system in which the observer looks down the \(+z\) axis towards the \(xy\)-plane (see Fig.~\ref{Fig:polycube_model}). This may be visualized as an arrow drawn on each face indicating its in-plane direction. Because each face lies in a distinct coordinate plane, only four of the six orientation states are geometrically admissible per face: North–South faces admit \{\texttt{L,R,I,O}\}, East–West faces admit \{\texttt{U,D,I,O}\}, and Front–Back faces admit \{\texttt{U,D,L,R}\}. Orientation determines how complementary faces must align during attachment.
\end{enumerate}
A tile species is completely specified by the interface types and orientations of its six faces listed in fixed NESWBF (North, East, South, West, Back, Front) order (see Fig.~\ref{Fig:polycube_model}):
\[
T_k = (s_i,s_r),
\]
where \(s_i\) and \(s_r\) denote the corresponding sequences of interface labels and orientations.

Tiles are free to rotate during the assembly process, and two tiles are regarded as the same species if they differ only by a rigid rotation. For example, consider the tile species
\[
T_1 = (\texttt{123456},\, \texttt{OUIDLR}).
\]
A rotation by \(90^\circ\) clockwise about the \(z\)-axis yields
\[
T_1^{\mathrm{rot}} = (\texttt{412356},\, \texttt{LORIUD}).
\]
Since this transformation preserves the relative arrangement of interface types and orientations, both representations correspond to the same tile species \(T_1\).

Interactions between interface types are specified by a symmetric binary matrix \( A_{ij} \) of dimension \( c \times c \),
\begin{equation}
     A_{ij} =
    \begin{cases}
    1, & \text{if interface types } i \text{ and } j \text{ attract},\\[4pt]
    0, & \text{if they do not bind.}
    \end{cases}   
\end{equation}

Each active (non-neutral) interface type is assigned a unique complementary partner such that odd integers bind exclusively to the subsequent even integers: \(1 \leftrightarrow 2, 3 \leftrightarrow 4,\) etc. Two faces bind only when their interface types are complementary and their orientations match.

\subsection{Self-assembly growth process} \label{subsec:growth_process}

Assembly proceeds through stochastic growth on the cubic lattice (Fig.~\ref{Fig:polycube_model}(b)). Each assembly run begins with the placement of an initial tile (seed) at the origin. In \emph{seeded} assembly, the first tile species in the assembly kit is used as the seed, whereas in \emph{unseeded} assembly the seed tile is chosen uniformly at random from the assembly kit.

\begin{figure}
\includegraphics[width=0.48\textwidth]{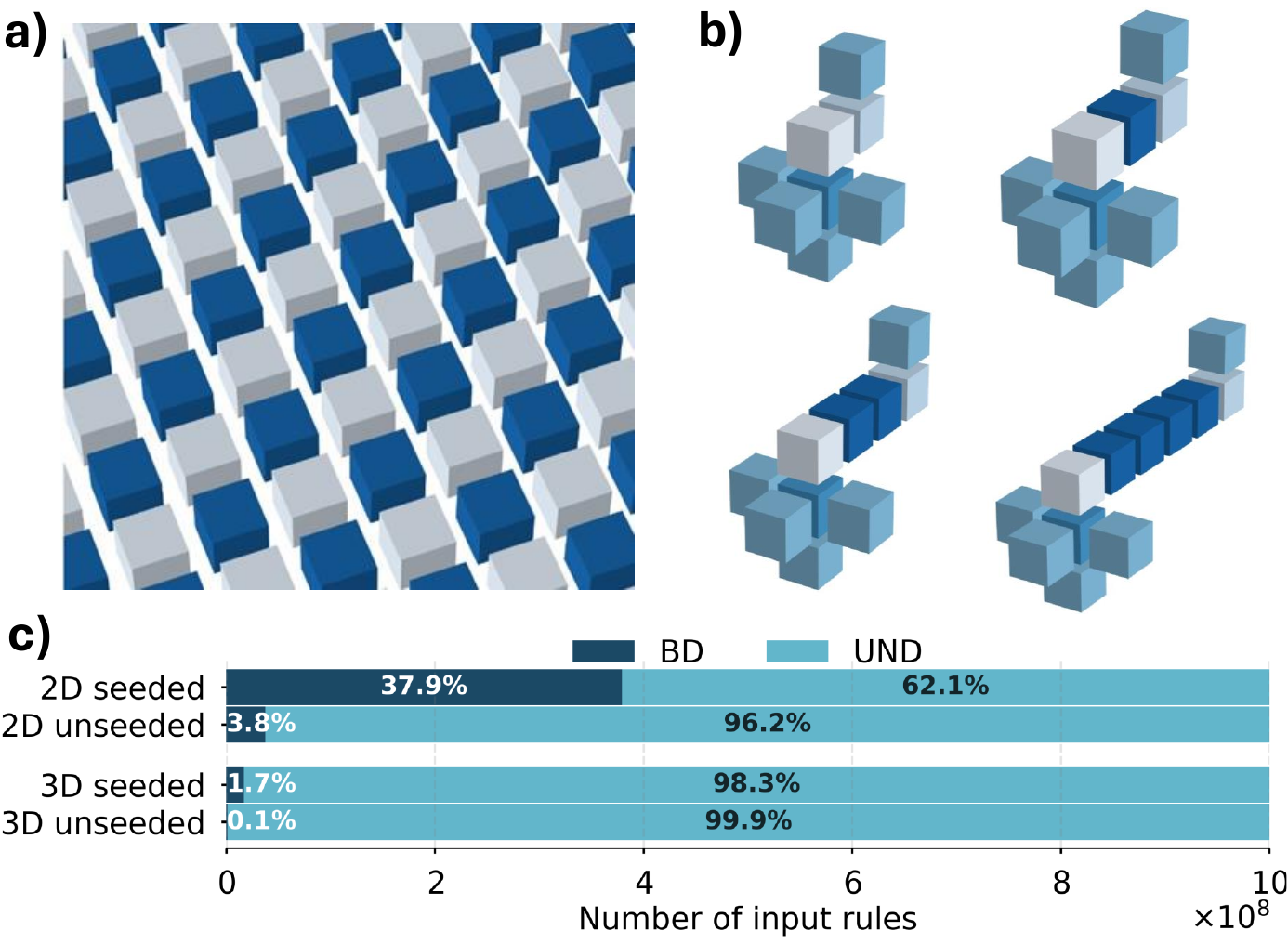}
\caption{\textbf{Classification of assembly outcomes.}
(a) \textit{Example of unbounded assembly.}
A rule set composed of tile species $(\texttt{131300},\ \texttt{OOOOLR})$ and $(\texttt{242400},\ \texttt{OOOOLR})$ from rule space $S^{3D}_{2s,5c}$ drives unrestricted growth, yielding an unbounded structure.
(b) \textit{Example of non-deterministic assembly.}
A rule set from $S^{3D}_{5s,7c}$ with tiles $(\texttt{100004},\ \texttt{IOOOLD})$, $(\texttt{002005},\ \texttt{OOOOLR})$, $(\texttt{000060},\ \texttt{OOOOUR})$, $(\texttt{055355},\ \texttt{OUIDLR})$, and $(\texttt{020100},\ \texttt{OUOULR})$ produces different structures across repeated stochastic runs, visible as variation in the number and placement of dark-blue tiles. Rule sets that produce bounded-deterministic assemblies are classified as BD, while those producing unbounded or non-deterministic assemblies, or both, are collectively classified as UND.
(c) \textit{Distribution of assembly behaviors.}
Fractions of BD and UND outcomes are shown for \(10^9\) sampled rule sets in \(S^{2D}_{5s,31c}\) and \(S^{3D}_{5s,31c}\) under seeded and unseeded assembly. Three-dimensional systems yield a smaller fraction of BD rule sets than two-dimensional ones, consistent with each tile having more faces through which growth can continue, which makes UND rule sets more likely. Unseeded assembly further suppresses the BD fraction due to variation in the choice of seed tile.}
\label{Fig:UND_assembly}
\end{figure}

Once the seed tile is placed, the growth of the polycube unfolds through the following steps:
\begin{enumerate}
    \item All exposed faces of placed tiles carrying non-neutral interface types that can bind to at least one interface present in the assembly kit are identified as \textit{open interfaces}.
    \item One open interface is selected uniformly at random.
    \item A face carrying the complementary interface type is selected from the assembly kit. When multiple such faces are available, on the same or different tile species, the selection is made uniformly at random.
    \item The tile species containing the selected face is rotated into a compatible orientation and attached to the growing assembly.
\end{enumerate}
Here, we assume infinite copies of these finitely many tile species throughout assembly. 
Steps $1$--$4$ are repeated until no open interfaces remain, in which case growth has terminated, or until the structure exceeds the imposed bounds on its maximum linear extent or total tile count, in which case the simulation is stopped and the assembly is classified as \emph{unbounded} (Fig.~\ref{Fig:UND_assembly}(a)).

If the assembly terminates in a bounded structure, we repeat the assembly process up to $k$ times for the same rule set, each time with different random choices of open interface (step 2) and of the face selected to bind to it (step 3). This tests whether the rule set uniquely determines the final structure despite variations in the order of attachment events. Rule sets for which all trajectories produce geometrically equivalent final structures, up to global cubic rotations and reflections, are classified as \emph{deterministic}; if different trajectories yield non-equivalent outcomes, the rule set is classified as \emph{non-deterministic} (see Sec.~\ref{SM_sec:inputspace} of the Supplementary Material (SM) for details).

Together, boundedness and determinism define four possible classes of rule sets: bounded and deterministic (BD), bounded and non-deterministic, unbounded and deterministic, and unbounded and non-deterministic. A bounded and deterministic rule set reproducibly assembles into the same finite structure across independent assembly trajectories. A bounded and non-deterministic rule set can, for example, produce different finite structures along different assembly trajectories, as in multifarious assemblies such as the werewolf in Ref.~\onlinecite{Bohlin2023}. An unbounded and deterministic rule set will generate the same indefinitely growing structure, such as a periodic crystal. An unbounded and non-deterministic rule set can grow without bound and also yield different structures along different trajectories.  We collectively classify these latter three classes as UND and restrict our subsequent analysis to BD rule sets. We refer to the finite structures they deterministically produce as \emph{BD assemblies}.

\subsection{Rule-space sampling}

The polycube model defines an input--output mapping from assembly rule sets \(I\) to assembled polycubes \(O\). For BD rule sets, each rule set produces a unique output structure, yielding the deterministic mapping \(f:I\rightarrow O\). Many different rule sets may produce the same output polycube, resulting in a many-to-one mapping.

For a fixed number of tile species \(s\) and interface types \(c\), the space of all rule-set encodings forms a discrete input space \(S_{s,c}\). We use superscripts \(2D\) and \(3D\) to distinguish two-dimensional polyomino and three-dimensional polycube rule spaces, respectively. For example, \(S^{3D}_{5s,31c}\) denotes the three-dimensional rule space with \(s=5\) tile species and \(c=31\) interface types, while \(S^{2D}_{5s,31c}\) denotes the corresponding two-dimensional rule space. 

In the two-dimensional polyomino model~\cite{Ahnert2010}, each of the four square tile edges is assigned one of \(c\) interface types. Since edge interfaces have no rotational degree of freedom, no orientation is required, giving
\[
|S^{2D}_{s,c}| = c^{4s}.
\]
In the three-dimensional polycube model, each of the six tile faces is assigned one of \(c\) interface types and one of four geometrically allowed orientations, giving
\[
|S^{3D}_{s,c}| = (4c)^{6s}.
\]

In both models, interface interactions follow the fixed pairing convention encoded in \(A_{ij}\). Even for modest parameters (e.g., \(s=2\), \(c=4\)), the three-dimensional rule space exceeds \(10^{14}\) rule sets, rendering exhaustive enumeration impractical. We therefore sample rule sets uniformly at random from \(S_{s,c}\) and simulate the assembly process for each. 

\subsection{Characterizing the output structures}
\label{subsec:output}

To characterize the output landscape generated by the polycube assembly model, each BD assembly is described by three main properties: its assembly rule probability, its geometric symmetry under cubic transformations, and its assembly complexity measured by the information required to specify its minimal assembly kit.

\begin{table}[t]
\centering
\begin{tabular}{|c|c|c|}
\hline
  \hfil \textbf{Class} & \hfil \textbf{Geometric Operation} & \hfil \textbf{\#Conjugates}\\
 \hline
 $\boldsymbol{I}$ & Identity & 1 \\
 $\boldsymbol{A}$ & $\pi/2$ rotation about face axis & 6 \\
 $\boldsymbol{B}$ & $\pi$ rotation about face axis & 3 \\
 $\boldsymbol{C}$ & $\pi$ rotation about edge axis & 6 \\
 $\boldsymbol{D}$ & $2\pi/3$ rotation about vertex axis & 8 \\
 $\boldsymbol{E}$ & Reflection in face plane & 3 \\
 $\boldsymbol{F}$ & Reflection in edge plane & 6 \\
 $\boldsymbol{H}$ & $\boldsymbol{D} \times \boldsymbol{K}$ & 8 \\
 $\boldsymbol{J}$ & $\boldsymbol{A} \times \boldsymbol{E}$ & 6 \\
 $\boldsymbol{K}$ & Inversion through cube center & 1 \\
 \hline
\end{tabular}
\caption{
\textbf{Cubic symmetry operations and their conjugacy classes.}
The cubic symmetry group \(O_h\) contains \(48\) symmetry operations, grouped into ten conjugacy classes~\cite{Lunnon1972}. The table lists each class, its geometric operation, and the number of conjugate operations belonging to that class. Classes \(\boldsymbol{I}\)--\(\boldsymbol{D}\) contain proper rotations, while classes \(\boldsymbol{E}\)--\(\boldsymbol{K}\) contain reflections, inversion, and improper rotations.}
\label{Table:cubic_operations}
\end{table}

\begin{table*}[htbp]
\centering
\small
\setlength{\tabcolsep}{5pt}
\begin{minipage}[t]{0.46\textwidth}
\centering
\begin{tabular}{ |c@{\hspace{0.7em}}c@{\hspace{0.7em}}c@{\hspace{0.7em}}c@{\hspace{1.1em}} >{\raggedright\arraybackslash}p{2.9cm}| }
\hline
\textbf{Class} & \textbf{Order} & \textbf{\#Conj.} & \textbf{Sch.} & \textbf{Class structure} \\
\hline
$I$  & 1  & 1 & $C_{1}$   & $\boldsymbol{I}$ \\
$A$  & 4  & 3 & $C_{4}$   & $\boldsymbol{I}+\boldsymbol{B}+2\boldsymbol{A}$ \\
$B$  & 2  & 3 & $C_{2}$   & $\boldsymbol{I}+\boldsymbol{B}$ \\
$C$  & 2  & 6 & $C_{2}$   & $\boldsymbol{I}+\boldsymbol{C}$ \\
$D$  & 3  & 4 & $C_{3}$   & $\boldsymbol{I}+2\boldsymbol{D}$ \\
$E$  & 2  & 3 & $C_{s}$   & $\boldsymbol{I}+\boldsymbol{E}$ \\
$F$  & 2  & 6 & $C_{s}$   & $\boldsymbol{I}+\boldsymbol{F}$ \\
$H$  & 6  & 4 & $S_{6}$   & $\boldsymbol{I}+\boldsymbol{K}+2\boldsymbol{D}+2\boldsymbol{H}$ \\
$J$  & 4  & 3 & $S_{4}$   & $\boldsymbol{I}+\boldsymbol{B}+2\boldsymbol{J}$ \\
$K$  & 2  & 1 & $C_{i}$   & $\boldsymbol{I}+\boldsymbol{K}$ \\
$BB$ & 4  & 1 & $D_{2}$   & $\boldsymbol{I}+3\boldsymbol{B}$ \\
$BC$ & 4  & 3 & $D_{2}$   & $\boldsymbol{I}+\boldsymbol{B}+2\boldsymbol{C}$ \\
$BE$ & 4  & 3 & $C_{2h}$  & $\boldsymbol{I}+\boldsymbol{B}+\boldsymbol{E}+\boldsymbol{K}$ \\
$BF$ & 4  & 3 & $C_{2v}$  & $\boldsymbol{I}+\boldsymbol{B}+2\boldsymbol{F}$ \\
$CE$ & 4  & 6 & $C_{2v}$  & $\boldsymbol{I}+\boldsymbol{C}+\boldsymbol{E}+\boldsymbol{F}$ \\
$CK$ & 4  & 6 & $C_{2h}$  & $\boldsymbol{I}+\boldsymbol{C}+\boldsymbol{F}+\boldsymbol{K}$ \\
$EE$ & 4  & 3 & $C_{2v}$  & $\boldsymbol{I}+\boldsymbol{B}+2\boldsymbol{E}$ \\
$CD$ & 6  & 4 & $D_{3}$   & $\boldsymbol{I}+2\boldsymbol{D}+3\boldsymbol{C}$ \\
$FF$ & 6  & 4 & $C_{3v}$  & $\boldsymbol{I}+2\boldsymbol{D}+3\boldsymbol{F}$ \\
\hline
\end{tabular}
\end{minipage}%
\hfill
\begin{minipage}[t]{0.52\textwidth}
\centering
\begin{tabular}{ |c@{\hspace{0.7em}}c@{\hspace{0.7em}}c@{\hspace{0.7em}}c@{\hspace{1.1em}} >{\raggedright\arraybackslash}p{4cm}| }
\hline
\textbf{Class} & \textbf{Order} & \textbf{\#Conj.} & \textbf{Sch.} & \textbf{Class structure} \\
\hline
$AB$  & 8  & 3 & $D_{4}$   & $\boldsymbol{I}+2\boldsymbol{A}+2\boldsymbol{C}+3\boldsymbol{B}$ \\
$AE$  & 8  & 3 & $C_{4h}$  & $\boldsymbol{I}+\boldsymbol{B}+\boldsymbol{E}+\boldsymbol{K}+2\boldsymbol{A}+2\boldsymbol{J}$ \\
$BFF$ & 8  & 3 & $D_{2d}$  & $\boldsymbol{I}+2\boldsymbol{F}+2\boldsymbol{J}+3\boldsymbol{B}$ \\
$CJ$  & 8  & 3 & $D_{2d}$  & $\boldsymbol{I}+\boldsymbol{B}+2\boldsymbol{C}+2\boldsymbol{E}+2\boldsymbol{J}$ \\
$EEE$ & 8  & 1 & $D_{2h}$  & $\boldsymbol{I}+\boldsymbol{K}+3\boldsymbol{B}+3\boldsymbol{E}$ \\
$EF$  & 8  & 3 & $C_{4v}$  & $\boldsymbol{I}+\boldsymbol{B}+2\boldsymbol{A}+2\boldsymbol{E}+2\boldsymbol{F}$ \\
$EFF$ & 8  & 3 & $D_{2h}$  & $\boldsymbol{I}+\boldsymbol{B}+\boldsymbol{E}+\boldsymbol{K}+2\boldsymbol{C}+2\boldsymbol{F}$ \\
$BD$  & 12 & 1 & $T$       & $\boldsymbol{I}+3\boldsymbol{B}+8\boldsymbol{D}$ \\
$CF$  & 12 & 4 & $D_{3d}$  & $\boldsymbol{I}+\boldsymbol{K}+2\boldsymbol{D}+2\boldsymbol{H}+3\boldsymbol{C}+3\boldsymbol{F}$ \\
$BBC$ & 16 & 3 & $D_{4h}$  & $\boldsymbol{I}+\boldsymbol{K}+2\boldsymbol{A}+2\boldsymbol{C}+2\boldsymbol{F}+2\boldsymbol{J}+3\boldsymbol{B}+3\boldsymbol{E}$ \\
$CCC$ & 24 & 1 & $T_{d}$   & $\boldsymbol{I}+3\boldsymbol{B}+6\boldsymbol{F}+6\boldsymbol{J}+8\boldsymbol{D}$ \\
$DEE$ & 24 & 1 & $T_{h}$   & $\boldsymbol{I}+\boldsymbol{K}+3\boldsymbol{B}+3\boldsymbol{E}+8\boldsymbol{D}+8\boldsymbol{H}$ \\
$R$   & 24 & 1 & $O$       & $\boldsymbol{I}+3\boldsymbol{B}+6\boldsymbol{A}+6\boldsymbol{C}+8\boldsymbol{D}$ \\
$G$   & 48 & 1 & $O_{h}$   & $\boldsymbol{I}+\boldsymbol{K}+3\boldsymbol{B}+3\boldsymbol{E}+6\boldsymbol{A}+6\boldsymbol{C}+6\boldsymbol{F}+6\boldsymbol{J}+8\boldsymbol{D}+8\boldsymbol{H}$ \\
\strut~ &  &  &  &  \\
\strut~ & & & & \\
\strut~ & & & & \\
\hline
\end{tabular}
\end{minipage}
\caption{
\textbf{Symmetry classes of free polycubes.}
The cubic symmetry group \(O_h\) contains \(98\) subgroups organized into \(33\) conjugacy classes~\cite{Lunnon1972}, each corresponding to a distinct symmetry class of free polycubes. The table lists the symmetry-class label (Class), subgroup order (Order), number of conjugate subgroups (\#Conj.), Schoenflies symbol (Sch.), and the subgroup class structure (Class structure).  The order gives the number of symmetry operations in the subgroup, while the class structure gives its composition in terms of the operation classes \(\boldsymbol{I}\)--\(\boldsymbol{K}\) defined in Table~\ref{Table:cubic_operations}. Ordinary labels \(I,A,B,\ldots\) denote polycube symmetry classes, whereas bold labels \(\boldsymbol{I},\boldsymbol{A},\boldsymbol{B},\ldots\) denote symmetry operation classes. The symmetry classes range from \(I\), containing only the identity operation, to \(G\), corresponding to the full cubic group with all \(48\) symmetry operations.}
\label{Table:lunnon_sym_classes}
\end{table*}

\subsubsection{Assembly rule probability}

Assembled polycubes are identified by their global geometry, independent of the specific arrangement of tile species. For example, the assemblies made with tile orders \(T_1\text{--}T_2\text{--}T_1\) and \(T_2\text{--}T_1\text{--}T_2\) are treated as the same structure because both correspond to the same three-tile linear polycube. Once tile identities are ignored, geometric structures may be identified under three standard equivalence conventions~\cite{Redelmeier1981, Golomb1994}:

\begin{itemize}
    \item \emph{Fixed polycubes}: two structures are considered distinct unless they differ only by translation.
    \item \emph{One-sided polycubes}: two structures are considered equivalent under translations and rotations, but mirror figures are treated as distinct.
    \item \emph{Free polycubes}: two structures are considered equivalent under translations, rotations, and reflections.
\end{itemize}

Throughout this work, outputs are classified as \emph{free polycubes}, identifying structures up to all 48 symmetry operations of the cube (Table~\ref{Table:cubic_operations}). The total number of free polycubes of size \(n\) is given by the On-Line Encyclopedia of Integer Sequences (OEIS) sequence A038119~\cite{OEIS2026}. For two-dimensional assemblies, we similarly classify outputs as \emph{free polyominoes}, with the total number of free polyominoes of size \(n\) given by OEIS sequence A000105~\cite{OEIS2026} (see SM Table~\ref{table:free_polycubes_polyominoes}).

Given this equivalence criterion, multiple BD rule sets may produce the same geometric structure or satisfy the same structural property. 
We define the \emph{assembly rule probability}, $P_{AR}(\cdot)$, as the fraction of BD rule sets that produce a specified outcome; UND rule sets are excluded. For a specific polyomino or polycube structure $p$,
\begin{equation}
P_{AR}(p)
=
\frac{\text{number of BD rule sets producing }p}
{\text{total number of BD rule sets}},
\end{equation}
whereas for assemblies of size $n$,
\begin{equation}
P_{AR}(n)
=
\sum_{p:\,|p|=n} P_{AR}(p).
\end{equation}
Thus, $P_{AR}(n)$ represents the fraction of all BD rule sets that produce an assembly of size $n$.

For comparisons within a fixed size class $n$, we write the size-normalized assembly rule probability as
\begin{equation}
P_{AR}(X \,|\,  n)
= \frac{1}{P_{{AR}}(n)}
  \sum_{\substack{p \in X \\ |p| = n}} P_{{AR}}(p),
\label{eq:size_normalized}
\end{equation}
where $X$ denotes a set of structures sharing a property, such as symmetry order $SO$, symmetry class $SC$, or assembly complexity $\tilde{K}$. Thus, $P_{AR}(X \,|\,  n)$ is the fraction of BD rule sets producing size-$n$ assemblies that also satisfy the specified property $X$.

In contrast to $P_{AR}$, which is defined over the space of BD rule sets, the \emph{morphospace probability} is defined over the space of geometrically distinct structures. For structures of size $n$, let ${M}_n$ denote the theoretical morphospace, i.e., the set of all geometrically distinct free polycubes or polyominoes of size $n$, with $|{M}_n|$ given by the corresponding OEIS enumeration. We define the size-normalized \emph{morphospace probability} as
\begin{equation}
P_M(X \,|\,  n)
=
{|\{p \in {M}_n : p \in X\}|}/{|{M}_n|},
\label{eq:morphospace_prob}
\end{equation}
which is the fraction of geometrically distinct free structures of size $n$ that satisfy the specified property or belong to category $X$.

The theoretical morphospace need not coincide with the set of structures observed by finite rule sampling. Some structures in ${M}_n$ may be inaccessible within the chosen rule space, while others may be accessible but not encountered in a finite sample. We therefore define the \emph{sampled morphospace}, $\hat{{M}}_n \subseteq {M}_n$, as the set of distinct size-$n$ structures observed during rule sampling. The corresponding \emph{sampled morphospace probability} is
\begin{equation}
P_{\hat{M}}(X \,|\,  n)
=
{|\{p \in \hat{{M}}_n : p \in X\}|}/
{|\hat{{M}}_n|},
\label{eq:morphospace_prob_sampled}
\end{equation}
which gives the fraction of observed size-$n$ structures satisfying the specified property $X$. If all theoretically possible size-$n$ structures are discovered, then $\hat{{M}}_n={M}_n$, and the sampled and theoretical morphospace probabilities coincide.

\subsubsection{Symmetry classification} 

The geometric symmetry of each assembled polycube is determined by its invariance under the full cubic symmetry group \(O_h\) (denoted \(O_3\) in Lunnon’s notation~\cite{Lunnon1972}). This group contains \(48\) symmetry operations, including proper rotations, reflections, inversion, and improper rotations. These operations can be grouped into ten conjugacy classes. Operations belonging to the same conjugacy class represent the same type of geometric transformation but act about symmetry-equivalent axes or planes. Table~\ref{Table:cubic_operations} lists these ten conjugacy classes together with their geometric interpretation and the number of conjugates (symmetry-equivalent operations in each class). For example, class label $\boldsymbol{B}$ refers to a $\pi$ rotation about a face axis and contains $3$ conjugate operations, one about each Cartesian axis.

Each polycube is invariant under a subgroup \(H \subseteq O_h\), where \(H\) consists of all cubic transformations listed in Table~\ref{Table:cubic_operations} that leave the polycube unchanged. The number of operations in \(H\) is the \emph{order} of the subgroup, which we refer to as its \emph{symmetry order}. Thus, a polycube with no non-trivial symmetry has symmetry order \(1\), corresponding only to the identity operation, whereas a polycube invariant under every operation in \(O_h\) has symmetry order \(48\).

The $98$ subgroups of \(O_h\) fall into $33$ conjugacy classes~\cite{Lunnon1972}. Two subgroups are conjugate within \(O_h\) if they contain the same geometric arrangement of symmetry operations but differ in the spatial directions along which those operations act.  These $33$ subgroup conjugacy classes enumerate all possible symmetry types of free polycubes~\cite{Lunnon1972}, and we refer to them as \emph{symmetry classes}. Table~\ref{Table:lunnon_sym_classes} lists each symmetry class together with its subgroup order, number of conjugate subgroups (\#Conj.), corresponding Schoenflies symbol (Sch.), and class structure in terms of the operation classes of \(O_h\) from Table~\ref{Table:cubic_operations}.

It is important to distinguish between conjugacy classes of individual symmetry operations, listed in Table~\ref{Table:cubic_operations}, and conjugacy classes of subgroups describing the overall symmetry of a polycube, listed in Table~\ref{Table:lunnon_sym_classes}. Following Lunnon’s notation~\cite{Lunnon1972}, we use the same labels ($A$, $B$, $C$, $\ldots$) for both levels of classification, but denote operation-based conjugacy classes in bold in Table~\ref{Table:cubic_operations} (e.g., \(\boldsymbol{A}\), \(\boldsymbol{B}\), \(\boldsymbol{C}\)) to distinguish them from the subgroup-based conjugacy classes in Table~\ref{Table:lunnon_sym_classes}. For example, in Table~\ref{Table:lunnon_sym_classes}, symmetry class \(A\) has order \(4\), three conjugate subgroups, and Schoenflies symbol \(C_4\). Its class structure is
\(\boldsymbol{I}+\boldsymbol{B}+2\boldsymbol{A}\).
These operations correspond to rotations by \(0\), \(\pi/2\), \(\pi\), and \(3\pi/2\) about a common fourfold face axis, i.e., a \(C_4\) axis in Schoenflies notation. The three conjugate subgroups arise by placing this fourfold axis along any one of the three mutually perpendicular face-axis directions of the cube, and therefore represent the same symmetry class of a free polycube.

Each assembled polycube is assigned to the symmetry class corresponding to its invariant subgroup \(H\). In the analysis, we further represent a symmetry class \(SC\) with symmetry order \(SO\) as \(SC_{SO}\). For example, \(A_4\) denotes symmetry class \(A\) with symmetry order \(4\), and \(G_{48}\) denotes symmetry class \(G\) with symmetry order \(48\). 

Similarly, the symmetries of two-dimensional polyominoes are described by the dihedral group of the square, $D_4$, whose subgroups fall into eight conjugacy classes. Grouping axis-parallel and diagonal mirror lines together, we classify polyominoes into six symmetry types: $C_1$ (no non-trivial symmetry), $C_2$ (twofold rotation), $C_4$ (fourfold rotation), $D_1$ (one mirror line), $D_2$ (two perpendicular mirror lines), and $D_4$ (full square symmetry).

\begin{figure*}[t]
\includegraphics[width=\textwidth]{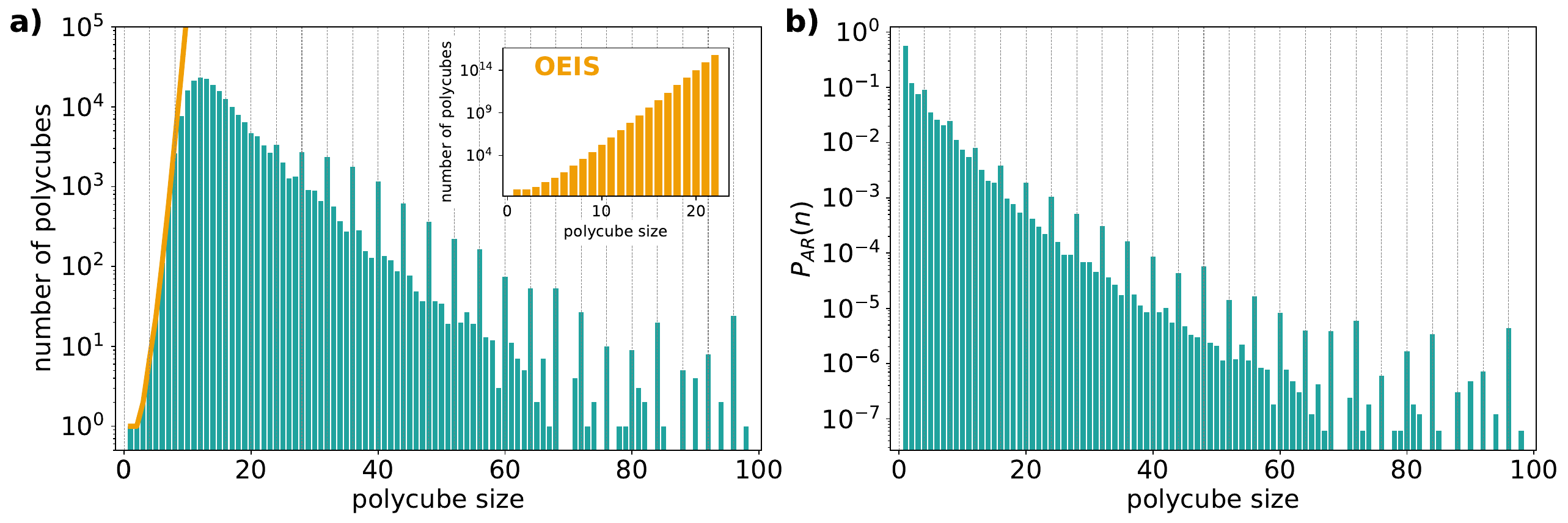}
\caption{
\textbf{Size distribution of polycubes generated from $10^9$ randomly sampled rule sets in the $S^{3D}_{5s,31c}$ rule space.}
(a) Number of distinct free polycubes generated by bounded-deterministic (BD) rule sets as a function of polycube size. The number of theoretically possible free polycubes is shown in yellow (OEIS sequence A038119)~\cite{OEIS2026}. All possible free polycubes up to size \(6\) are recovered from the sampled rule sets. Beyond this size, the fraction of the theoretical morphospace recovered decreases rapidly, while the number of possible polycubes continues to increase with size (inset).
(b) Assembly rule probability \(P_{AR}(n)\), giving the fraction of BD rule sets that produce polycubes of size \(n\). Assembly rule probability generally decreases with increasing polycube size, with pronounced local peaks occurring predominantly at sizes that are multiples of four, as indicated by the vertical gray lines.
}
\label{Fig:3D_5s31c_10^9_nmer_hist}
\end{figure*}

\subsubsection{Complexity measures}

We quantify assembly complexity using three complementary measures, each inspired by Kolmogorov complexity $K$. The minimal number of distinct tile species is given by \(\tilde{K}_s\). The minimal number of interface types (including the neutral type) is given by \(\tilde{K}_i\).  The third measure, \(\tilde{K}_{LZ}\), is obtained by applying a symmetrized version of the  Lempel--Ziv 1976 compression algorithm to the binary encoding of the input rule set~\cite{Lempel1976}, as done in Ref.~\onlinecite{Dingle2018}.  Prior to complexity calculation, assembly rules are simplified to remove redundant information. Details of the simplification procedure and complexity calculations are provided in SM Sec.~\ref{SM_sec:inputspace}.  

For structures that can be generated by multiple BD rule sets, each complexity measure is defined as the minimum value over all corresponding BD inputs. In practice, we cannot always exhaustively enumerate all possible rules, and so the minimum is taken over all rules that are found by sampling. That implies that our measures may be upper bounds.  However, simpler assembly descriptions typically occupy larger regions of rule space because there are more ways to add redundant rules. Therefore they are more likely to appear upon random sampling, and so we expect that for the most frequently assembled structures,  observed minima provide a good approximation to the true minima. As a check, we compared the sampled minimum complexities of all tetracubes with those obtained using the SAT-based inverse-design solver of Ref.~\onlinecite{Bohlin2023} and found that the two agree.

\subsection{Assembly simulation parameters}

Polycube assembly simulations were performed in the three-dimensional rule space \(S^{3D}_{5s,31c}\), consisting of \(s=5\) tile species and \(c=31\) interface types. Interface type $0$ was neutral, while all non-zero interfaces formed unique complementary pairs \((1 \leftrightarrow 2,\ 3 \leftrightarrow 4,\dots,29 \leftrightarrow 30)\) according to the interaction matrix \(A_{ij}\) defined earlier. The rule space \(S^{3D}_{5s,31c}\) contains more than \(10^{62}\) rule set encodings, from which \(10^9\) were sampled uniformly at random.

Similarly, polyomino assembly simulations were performed in the corresponding two-dimensional rule space \(S^{2D}_{5s,31c}\), using the same numbers of tile species and interface types and the same interface-pairing convention. This rule space contains more than \(10^{29}\) rule set encodings, from which \(10^9\) were likewise sampled uniformly at random.

Unless otherwise specified, simulations in both dimensions were performed in seeded mode, with the first tile species initiating the growth process. To classify assemblies as bounded or unbounded, the maximum linear extent was set to $16$ tiles and the maximum total assembly size to $100$ tiles. Rule sets producing bounded assemblies were then tested for determinism using up to $k=25$ independent stochastic trajectories. Further details and justification of these simulation parameters are provided in SM Sec.~\ref{sec:robustness}.

\begin{figure*}
\includegraphics[width=\textwidth]{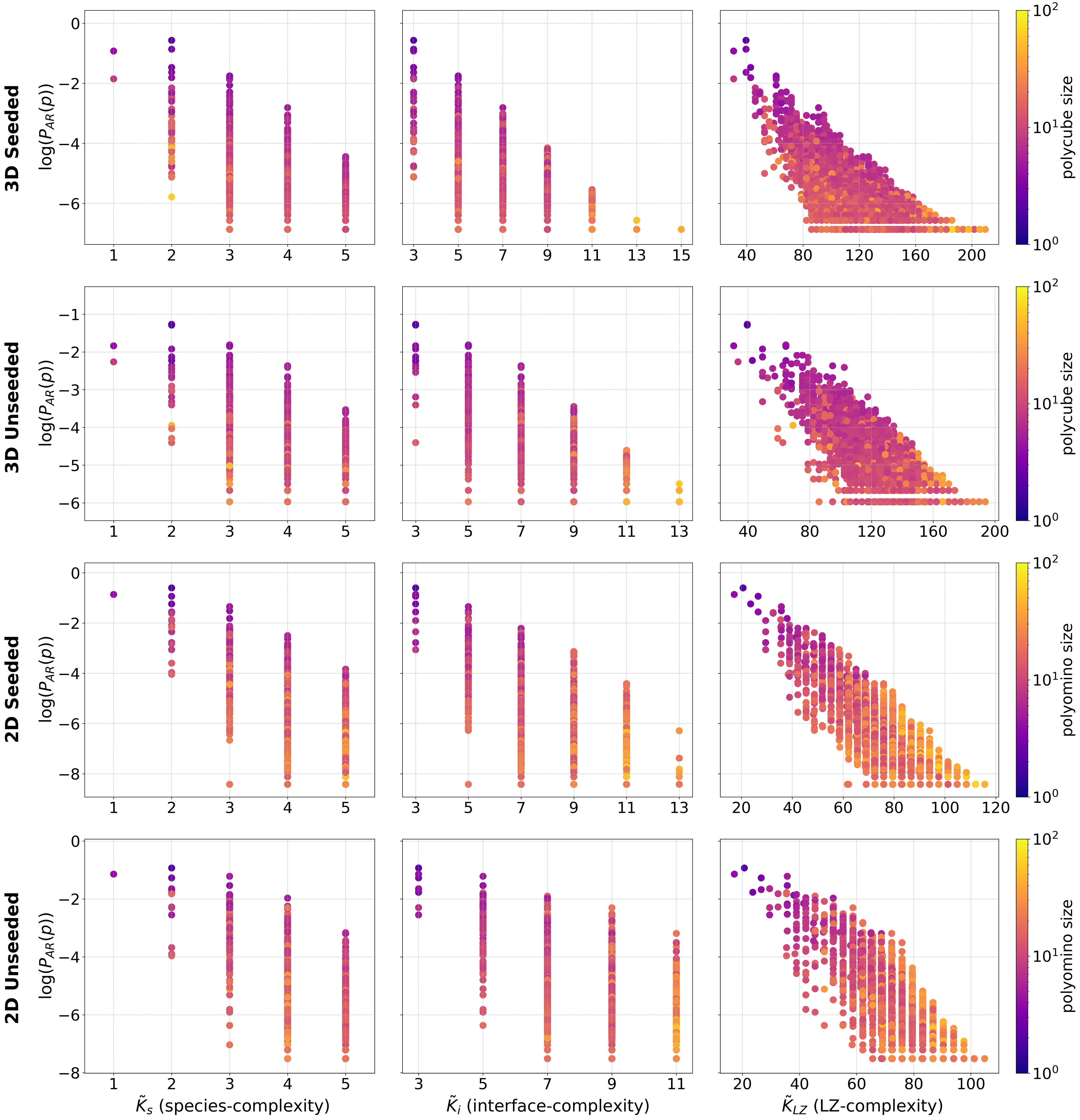}
\caption{
\textbf{Assembly rule probability versus assembly complexity.}
Each point represents a distinct polycube or polyomino \(p\) generated by BD rule sets among the \(10^9\) randomly sampled rule sets. Single-tile assemblies are omitted, and \(P_{AR}(p)\) is normalized over the remaining BD rule sets. The logarithm of the assembly rule probability, \(\log P_{AR}(p)\), is plotted against assembly complexity. From top to bottom, the rows show three-dimensional seeded, three-dimensional unseeded, two-dimensional seeded, and two-dimensional unseeded assembly in the corresponding \(S^{3D}_{5s,31c}\) and \(S^{2D}_{5s,31c}\) rule spaces. Columns show three measures of assembly complexity: species complexity \(\tilde{K}_s\), interface complexity \(\tilde{K}_i\), and Lempel--Ziv complexity \(\tilde{K}_{LZ}\). Color indicates polycube or polyomino size. Across both dimensions and assembly modes, assembly rule probability decreases with increasing assembly complexity, showing a pronounced bias towards structures with simpler assembly descriptions. Lower complexity and higher probability also correlate with smaller assemblies.}
\label{Fig:2D_3D_freq_vs_comp_for_all_complexities}
\end{figure*}

\section{Results and Discussion} 
\label{sec:results}

Sampling \(10^9\) rule sets from \(S^{3D}_{5s,31c}\) under seeded assembly yielded \(1.7\times10^7\) (1.7\%) BD rule sets, while the remaining \(98.3\%\) were unbounded, non-deterministic, or both (UND). Of the BD rule sets, \(56\%\) produced only a single cube, and the remaining \(7.3\times10^6\) (0.73\% of all sampled rule sets) assembled structures of two or more cubes. Together, the BD rule sets generated \(2\times10^5\) distinct free polycubes, reflecting the many-to-one mapping from rule sets to assembled structures.

For the corresponding two-dimensional system, sampling \(10^9\) rule sets from \(S^{2D}_{5s,31c}\) under seeded assembly yielded \(3.79\times10^8\) (37.9\%) BD rule sets, while \(62.1\%\) were UND. Of the BD rule sets, \(31\%\) produced only a single tile, and the remaining \(2.6\times10^8\) (26\% of all sampled rule sets) assembled polyominoes of two or more tiles. These BD rule sets generated \(2.4\times10^4\) distinct free polyominoes. As shown in Fig.~\ref{Fig:UND_assembly}(c), the fraction of BD rule sets is substantially lower in three dimensions than in two.

Having identified the BD rule sets and their corresponding assemblies, we next examine how assembly rule probability is distributed with respect to assembly size, assembly complexity, and symmetry.

\subsection{Size distribution of assembled structures}
\label{subsec:size_dist}

The number of mathematically possible free polycubes grows rapidly with polycube size \(n\) (OEIS~\cite{OEIS2026} sequence A038119, see SM Table~\ref{table:free_polycubes_polyominoes}). 
As shown in the inset of Fig.~\ref{Fig:3D_5s31c_10^9_nmer_hist}(a), the number of distinct free polycubes increases from just \(2\) at size \(3\) to \(112\) at size \(6\), reaching \(6.4\times10^{15}\) at size \(22\), the largest size for which exact enumeration is available. A conservative extrapolation suggests that by size $100$, the cumulative number of free polycubes exceeds $10^{80}$, comparable to the number of atoms in the observable universe. This illustrates that these spaces are hyper-astronomically large~\cite{Louis2016contingency}.

Figure~\ref{Fig:3D_5s31c_10^9_nmer_hist}(a) shows how much of this enormous morphospace is recovered by random sampling of BD assembly rules. Assemblies range from polycube size $1$ to $98$, approaching the imposed size cutoff of $100$. All $146$ possible free polycubes of sizes up to $6$ are recovered. Beyond this size, however, the fraction of the theoretical morphospace recovered decreases rapidly with increasing polycube size.

Figure~\ref{Fig:3D_5s31c_10^9_nmer_hist}(b) shows the assembly rule probability $P_{AR}(n)$, defined as the fraction of BD rule sets that generate polycubes of size $n$. Assembly rule probability generally decreases with increasing polycube size, indicating a strong bias towards smaller structures despite the much greater number of geometrically possible polycubes at larger sizes.

Figure~\ref{Fig:3D_5s31c_10^9_nmer_hist} also exhibits pronounced local peaks at polycube sizes that are predominantly multiples of four, both in the number of distinct polycubes recovered and in their assembly rule probability. These peaks are absent from the corresponding distribution of polycubes in the theoretical morphospace shown in the inset of Fig.~\ref{Fig:3D_5s31c_10^9_nmer_hist}(a). We return to the origin of these peaks later in our discussion of assembly complexity and symmetry. 

To examine the effect of sampling depth, SM Fig.~\ref{Fig:3D_5s31c_10^6_to_10^9_nmer_hist_rule_count_fraction} compares size distributions obtained from $10^6$ to $10^9$ sampled rule sets. Increasing the number of samples reveals additional distinct polycubes, particularly at larger sizes, while the overall distribution of assembly rule probability across sizes changes little. The overall size distributions also show similar patterns under seeded and unseeded assembly conditions and for two-dimensional polyomino assemblies (SM Fig.~\ref{Fig:2D_3D_5s31c_10^9_nmer_hist}).

Beyond finite sampling, the number of tile species and interface types available in our rule space $S^{3D}_{5s,31c}$ also limits which structures can be assembled, regardless of their existence in the theoretical morphospace. In SM Sec.~\ref{SM:tile_comp}, we derive the minimum assembly complexity for deterministic assembly of two simple structures. A linear rod of size $n$ requires $\lfloor n/2 \rfloor + 1$ tile species, whereas an L-shape of the same size requires $n-1$ species, close to the fully addressable limit of $n$. Consequently, a rule space with $s$ tile species cannot deterministically assemble L-shaped structures larger than $s + 1$, and therefore cannot recover the complete theoretical morphospace at larger sizes, even with exhaustive sampling.

\begin{figure*}
\includegraphics[width=\textwidth]{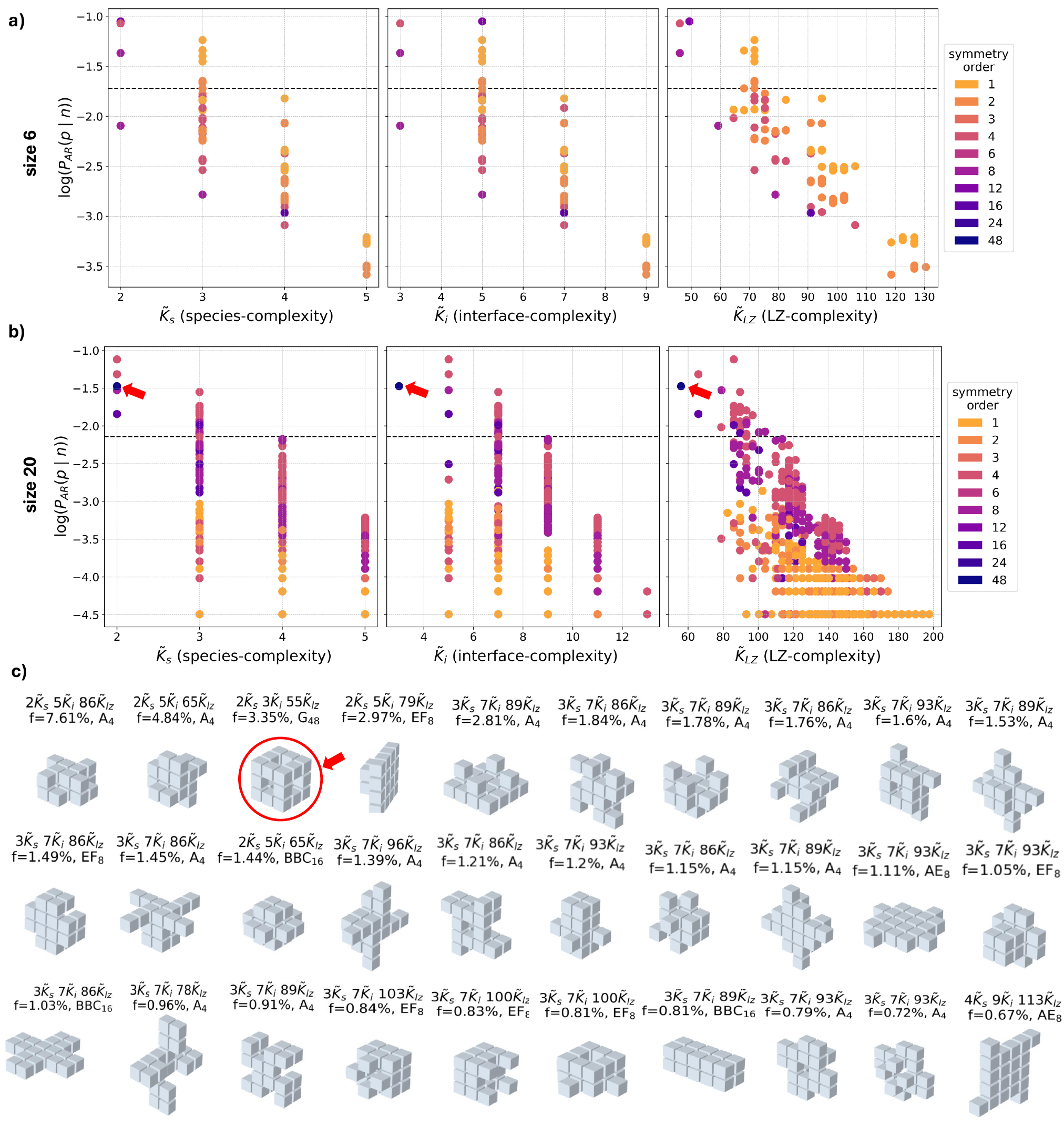}
\caption{
\textbf{Assembly rule probability decreases with complexity and is modulated by symmetry at fixed polycube size.}
Results are obtained from $10^9$ randomly sampled rule sets in the three-dimensional rule space $S^{3D}_{5s,31c}$ under seeded assembly.
({a}, {b}) Logarithm of the size-normalized assembly rule probability, $\log P_{AR}(p\,|\,  n)$, versus assembly complexity for distinct polycubes of sizes $6$ and $20$, respectively, generated by BD rule sets. Columns show species complexity $\tilde{K}_s$, interface complexity $\tilde{K}_i$, and Lempel--Ziv complexity $\tilde{K}_{LZ}$. Points are colored by symmetry order. Black dashed lines indicate the threshold above which polycubes collectively account for $50\%$ of the cumulative assembly rule probability at each size. At both sizes, assembly rule probability generally decreases with increasing complexity, and higher-symmetry structures tend to occur in the low-complexity, high-probability region. Red arrows in ({b}) indicate the size-$20$ polycube with the maximum symmetry order of $48$, circled in red in ({c}).
({c}) The thirty most frequently assembled polycubes of size $20$, ordered by decreasing assembly rule probability. Each structure is labeled by its three complexity values, assembly frequency $f$, and symmetry class, with the subscript indicating symmetry order. Here, $f$ is $P_{AR}(p\,|\,  n=20)$ expressed as a percentage. The red outline identifies the order-$48$ structure indicated in ({b}). An interactive visualization is available \href{https://akodiat.codeberg.page/polycubeViewer/}{online}. 
}
\label{Fig:3D_5s31c_10^9_20mer_4mer}
\end{figure*}

\subsection{Assembly rule probability decreases exponentially with assembly complexity}

Figure~\ref{Fig:2D_3D_freq_vs_comp_for_all_complexities} plots assembly rule probability against the three complexity measures introduced in Sec.~\ref{subsec:output}. Single-tile structures are excluded from these plots since they do not involve assembly. 
Across two and three dimensions, under both seeded and unseeded assembly, the observed patterns are consistent with the simplicity bias predicted by Dingle \emph{et al.}~\cite{Dingle2018}: structures with high assembly rule probability have low assembly complexity, while structures with high assembly complexity have low assembly rule probability. This bias towards simplicity can be quantified by an upper bound on assembly rule probability that decreases exponentially with assembly complexity, with most of the probability weight concentrated relatively close to the bound on a logarithmic scale (see SM Fig.~\ref{Fig:freq_vs_comp_with_fit}). These trends are observed for all three complexity measures, indicating that each of these measures captures something salient about the amount of information needed to assemble a structure.  

The color gradient in Fig.~\ref{Fig:2D_3D_freq_vs_comp_for_all_complexities} highlights that larger polycubes generally occupy the high-complexity, low-probability region of the plots. In other words, they typically require more assembly information to specify their deterministic growth and therefore admit fewer compatible rule sets than smaller polycubes. This difference in assembly complexity helps explain the pattern observed in Fig.~\ref{Fig:3D_5s31c_10^9_nmer_hist}, where larger structures typically arise less frequently under random sampling of assembly rules, even though there are exponentially more of them in the theoretical morphospace.

However, a closer examination of the color gradient reveals substantial variation in both assembly rule probability and assembly complexity among polycubes and polyominoes of similar size. We next examine how assembly rule probability depends on complexity when assembly size is held fixed.

\begin{figure*}
    \includegraphics[width=\textwidth]{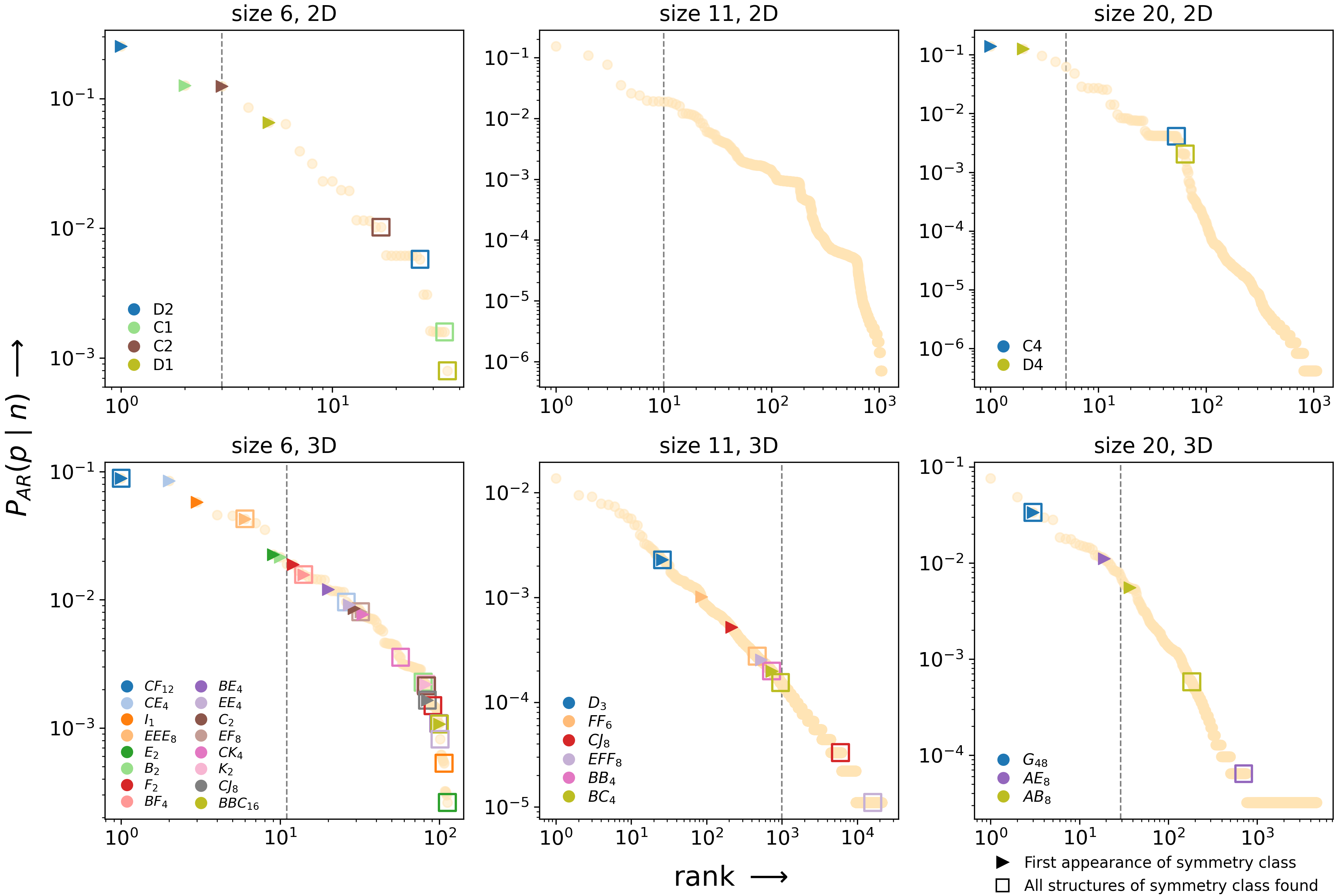}
    \caption{
    \textbf{Assembly rule probability versus rank at fixed structure size.}
    Each point represents a distinct polycube or polyomino \(p\) generated by BD rule sets. For each size \(n\), structures are ordered from highest to lowest size-normalized  assembly rule probability \(P_{AR}(p\,|\,  n)\). The top row shows two-dimensional polyominoes generated from \(S^{2D}_{5s,31c}\), and the bottom row shows three-dimensional polycubes generated from \(S^{3D}_{5s,31c}\), for sizes \(n=6\), \(11\), and \(20\). The gray dashed vertical line marks the rank at which the cumulative assembly rule probability reaches 50\%: the structures to its left together account for half of all BD rule sets producing structures of size \(n\). Filled triangles indicate the rank at which each symmetry class first appears, while open squares indicate the rank at which all structures belonging to that symmetry class have been encountered. Symmetry class markers are shown only for classes for which all theoretically possible structures at the given size are recovered by sampling; no such classes are found for polyominoes of size $11$.}
    \label{fig:freq_vs_rank}
\end{figure*}

\begin{figure*}
\includegraphics[width=\textwidth]{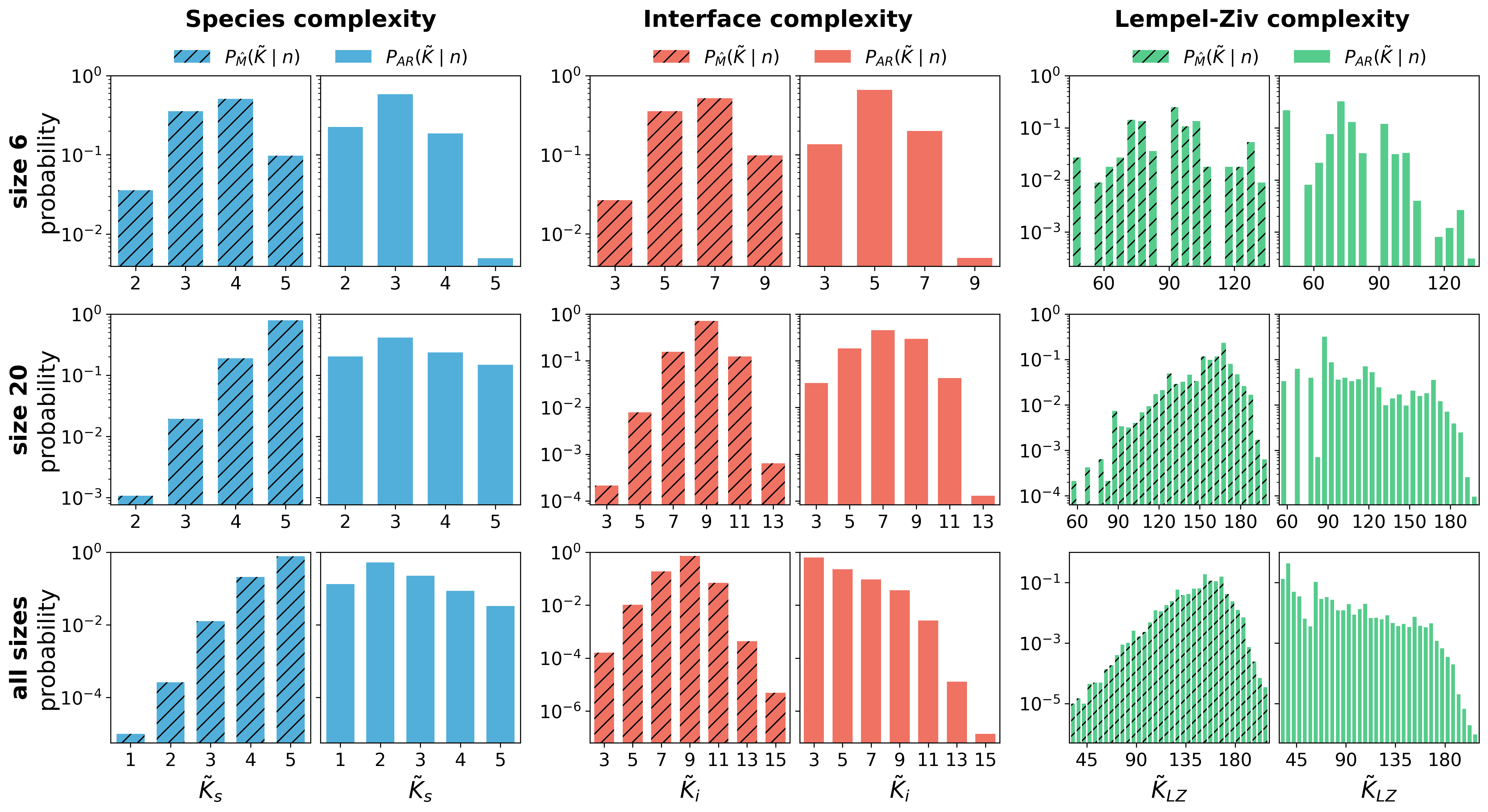}
\caption{
\textbf{Assembly complexity distributions at fixed polycube sizes.}
Assembly complexity distributions are shown for polycubes of sizes $6$ (top row) and $20$ (middle row), and for polycubes of all sizes combined (bottom row), generated by BD rule sets from $10^9$ randomly sampled rule sets in $S^{3D}_{5s,31c}$. Column groups show species complexity $\tilde{K}_s$, interface complexity $\tilde{K}_i$, and Lempel--Ziv complexity $\tilde{K}_{LZ}$. Within each group, hatched bars show the sampled morphospace probability $P_{\hat{M}}(\tilde{K}\,|\,  n)$, with each distinct polycube counted once, while solid bars show the assembly rule probability $P_{AR}(\tilde{K}\,|\,  n)$, obtained by counting the BD rule sets generating polycubes in each complexity bin. All $112$ possible free polycubes of size $6$ are recovered, so $\hat{M}_{n=6}=M_{n=6}$, whereas only a subset of the theoretical morphospace is recovered at size $20$. In the bottom row, the distributions are computed over sampled polycubes of all sizes, except single-cube assemblies. Values of $\tilde{K}_{LZ}$ are grouped into bins of width $5$. Across all plots, the assembly rule probability is shifted towards lower complexity relative to the sampled morphospace, showing that polycubes with simpler assembly descriptions are generated disproportionately often. 
}
\label{Fig:P(K)dist}
\end{figure*}

\subsection{Assembly rule probability at fixed assembly size}

Figure~\ref{Fig:3D_5s31c_10^9_20mer_4mer}(a) and (b) show the size-normalized assembly rule probability against assembly complexity for polycubes of sizes $6$ and $20$, respectively. At both sizes, assembly rule probability decreases with increasing complexity, with an approximately exponential upper bound, demonstrating that simplicity bias persists at fixed size. The corresponding empirical upper-bound fit for size $20$ is shown in SM Fig.~\ref{Fig:freq_vs_comp_20mer_with_fit}. 

The color gradient in Figs.~\ref{Fig:3D_5s31c_10^9_20mer_4mer}(a) and (b), indicating symmetry order, shows that the low-complexity, high-probability region is predominantly occupied by highly symmetric polycubes. Figure~\ref{Fig:3D_5s31c_10^9_20mer_4mer}(c) shows the $30$ most frequently assembled polycubes of size $20$. All $30$ have symmetry order $4$ or higher, indicating that highly symmetric polycubes are strongly favored under random sampling of assembly rules. A particularly striking example is the polycube with full cubic symmetry (order $48$), indicated by the red arrows in Fig.~\ref{Fig:3D_5s31c_10^9_20mer_4mer}(b) and circled in (c). Among the approximately $10^{14}$ possible free polycubes of size $20$, this is the \emph{only} structure possessing all $48$ symmetries of the cube. Yet it is generated by approximately $1$ in $30$ BD rule sets producing size-$20$ polycubes. It is therefore more than twelve orders of magnitude more likely to occur under random sampling of assembly rules than under uniform sampling of free polycubes of size $20$.

This strong preference for symmetric structures is not limited to a single polycube. At size $20$, the $29$ most probable polycubes together account for $50\%$ of the assembly rule probability, as indicated by the black dashed lines in Fig.~\ref{Fig:3D_5s31c_10^9_20mer_4mer}(b). All $29$ possess non-trivial symmetry and belong to just five symmetry classes: $A_4$, $AE_8$, $BBC_{16}$, $EF_8$, and $G_{48}$. Together, these classes represent only about $6\times10^{-12}$ of the theoretical morphospace (see SM Table~\ref{Table:OEIS}). Thus, half of the assembly rule probability is concentrated on structures drawn from an extraordinarily small fraction of all possible size-$20$ polycubes. This demonstrates the strong enrichment of symmetric structures under random rule sampling.

A complementary way to illustrate the uneven distribution of assembly rule probability among structures of the same size is through rank–probability plots, in which structures are ordered by decreasing probability. Figure~\ref{fig:freq_vs_rank} shows $P_{AR}(p\,|\,  n)$ against rank for polycubes and polyominoes of sizes $6$, $11$, and $20$. The corresponding distributions for all sizes up to $20$ are shown in SM Figs.~\ref{Fig:rank_superplot_3D} and~\ref{Fig:rank_superplot_2D}. At each size, assembly rule probability decreases rapidly with rank, spanning several orders of magnitude. When structures of all sizes are combined, the rank--probability distribution for polycubes approximately follows a power law (SM Fig.~\ref{Fig:rank_superplot_3D}), whereas that for polyominoes appears not to (SM Fig.~\ref{Fig:rank_superplot_2D}). For individual sizes, the distributions exhibit more complex behavior in both systems. These results further demonstrate that even among structures of the same size, the assembly map explores morphospace highly unevenly, with a strong bias towards certain structures over others.

The gray dashed lines in Fig.~\ref{fig:freq_vs_rank} mark the rank at which the cumulative assembly rule probability reaches $50\%$; further examples are shown in SM Figs.~\ref{Fig:rank_superplot_3D} and \ref{Fig:rank_superplot_2D}. In general, it is remarkable how tiny a fraction of all structures accounts for half of the BD rule sets. 
It is also interesting to note, as can be seen in SM Figs.~\ref{Fig:rank_superplot_3D} and \ref{Fig:rank_superplot_2D}, that sizes that are multiples of four need the fewest structures to account for 50\% of the BD rule sets.  This effect is more pronounced for polycubes than for polyominoes.  It is likely caused by the fact that symmetries are more efficient at reducing the assembly complexity for structures whose sizes are multiples of four, which are most commensurate with the square or cubic geometry of polyominoes and polycubes.   Note that this effect also explains why size $11$ has  more structures in the top $50\%$ set than size 20  does in Fig.~\ref{fig:freq_vs_rank}.  We return to assembly rule probabilities for  sizes that are multiples of four in Sec.~\ref{subsec:why_symmetries}.   

For classes whose structures are all recovered by our sampling, we mark the rank at which each symmetry class first appears (triangles) and the rank at which all of its possible structures have been found (squares) in Fig.~\ref{fig:freq_vs_rank}. At size $6$, all symmetry classes are fully recovered in both dimensions.
At larger sizes, complete recovery is observed mainly for highly symmetric classes, despite their rarity in the theoretical morphospace. 

Together, Figs.~\ref{Fig:3D_5s31c_10^9_20mer_4mer} and~\ref{fig:freq_vs_rank} show that polycubes with high assembly probability typically have a relatively low complexity, and a relatively high symmetry. Symmetry provides a mechanism for reducing assembly complexity by allowing the same assembly instructions to be reused across symmetry-related regions of a structure. Symmetric structures that require less assembly information may therefore be generated by more rule sets and occur more frequently under random rule sampling.   

However, the broad variation in assembly complexity and probability among structures with the same symmetry order in Fig.~\ref{Fig:3D_5s31c_10^9_20mer_4mer}, as well as differences between symmetry classes of the same order in Fig.~\ref{fig:freq_vs_rank}, suggests a more complex relationship between symmetry, complexity, and assembly rule probability. Before examining this relationship in more detail, we first consider how assembly rule probability is distributed across either complexity or symmetry.

\begin{figure*}
\centering
\includegraphics[width=\textwidth]{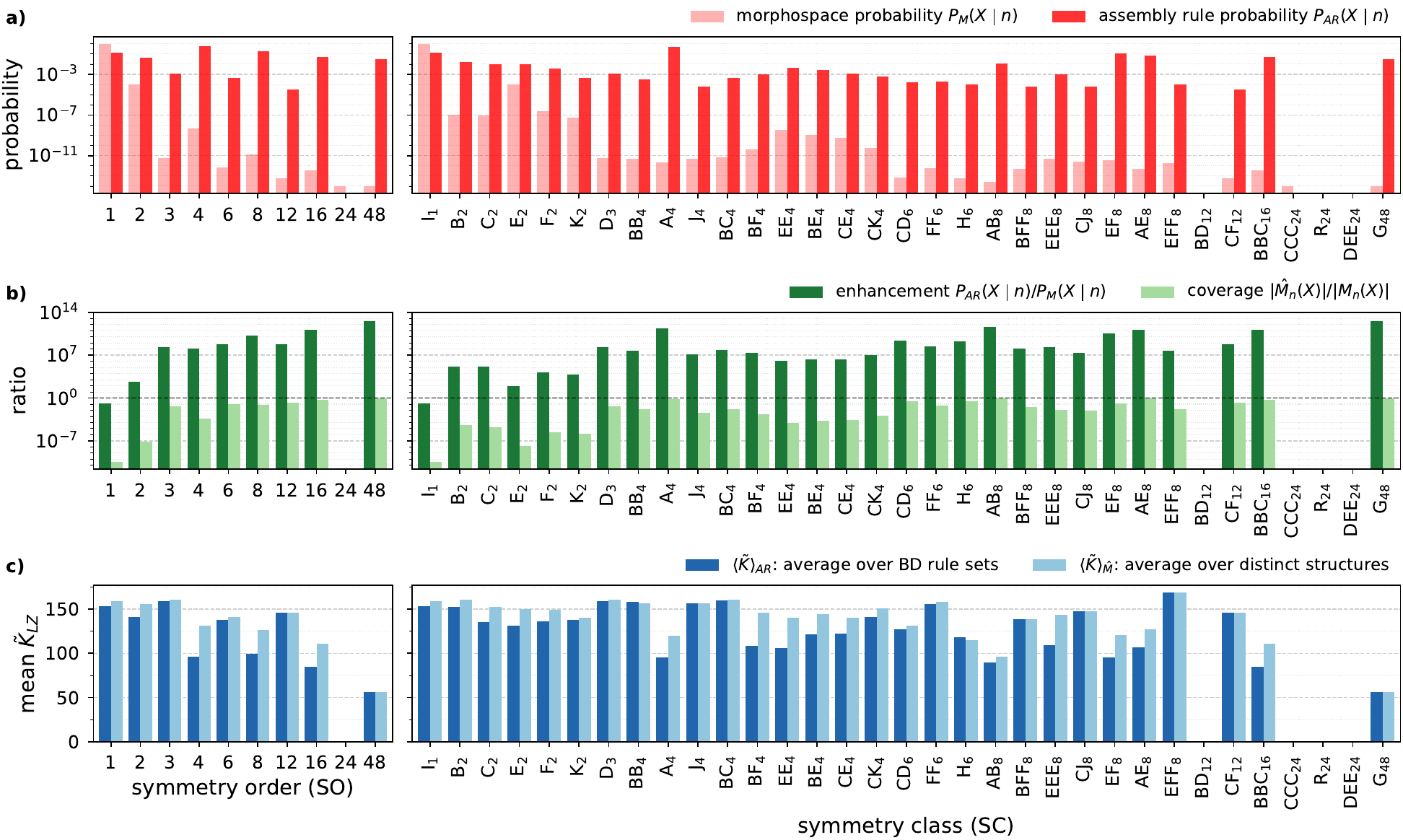}\\[6pt]
\caption{\textbf{Symmetry distributions of size-$20$ polycubes.}
Results are based on $10^9$ randomly sampled rule sets from $S^{3D}_{5s,31c}$, for polycubes of size $n = 20$.
In each panel, polycubes are grouped by symmetry order ($X = SO$, left) and by the symmetry classes ($X = SC$, right), with the symmetry order indicated by the subscript of each class label.
(a) \textit{Bias towards highly symmetric polycubes.}
Morphospace probability $P_M(X \,|\,  n)$ (light red) and assembly rule probability $P_{AR}(X \,|\,  n)$ (dark red).
Highly symmetric polycubes occur far more often under random sampling of assembly rules than their abundance in the theoretical morphospace would suggest, and symmetry classes of the same order can differ substantially in assembly rule probability.
(b) \textit{Enhancement and morphospace coverage.}
Enhancement factor $P_{AR}(X \,|\,  n)/P_M(X \,|\,  n)$ (dark green), measuring how strongly each symmetry order or class is over-represented relative to its abundance in the morphospace, and coverage $|\hat{M}_n(X)|/|M_n(X)|$ (light green), the fraction of all size-$20$ polycubes in $X$ that were recovered by sampling. The dashed line marks a ratio of $1$.
Highly symmetric classes are enhanced by many orders of magnitude, with some nearly fully recovered by sampling, whereas only a vanishing fraction of the low-symmetry morphospace is recovered.
(c) \textit{Mean assembly complexity across symmetry orders and classes.}
Mean Lempel--Ziv complexity averaged over BD rule sets, $\langle \tilde{K}_{LZ} \rangle_{AR} = \sum_{p} P_{AR}(p \,|\,  n, X)\,\tilde{K}_{LZ}(p)$ (dark blue), and averaged over the distinct polycubes of the sampled morphospace, $\langle \tilde{K}_{LZ} \rangle_{\hat{M}}$, with each structure counted once (light blue).
The most strongly enhanced classes in (b), such as $A_4$, $AB_8$, $EF_8$, $AE_8$, $BBC_{16}$ and $G_{48}$, have among the lowest mean complexities, and for most classes $\langle \tilde{K}_{LZ} \rangle_{AR}$ lies below $\langle \tilde{K}_{LZ} \rangle_{\hat{M}}$, showing that within a symmetry class the assembly rule probability is concentrated on its simplest structures.
}
\label{Fig:3D_prob_order_and_class_5s31c_10^9_20mer}
\end{figure*}

\subsection{Distribution of assembly rule probability over complexity}

So far, we have focused on the assembly rule probabilities of individual structures, finding that highly probable structures often have low assembly complexity and high symmetry. 
However, the number of possible structures grows exponentially with size (SM Table~\ref{table:free_polycubes_polyominoes}), and larger structures typically have higher assembly complexity, so we expect there to be many more complex structures than simple ones. 
Even at fixed size, low-symmetry structures greatly outnumber highly symmetric ones (SM Table~\ref{Table:OEIS}). Since lower symmetry is generally associated with higher assembly complexity, this suggests that complex structures may also be more numerous at a given size.
Therefore, although individual complex structures tend to have lower assembly rule probabilities, their much greater number may compensate for this. This raises the following question: Which effect dominates the total assembly rule probability $P_{AR}(\tilde{K})$ at a given complexity: the higher individual probabilities of simpler structures or the greater abundance of complex structures?

\begin{figure*}
\centering
\includegraphics[width=\linewidth]{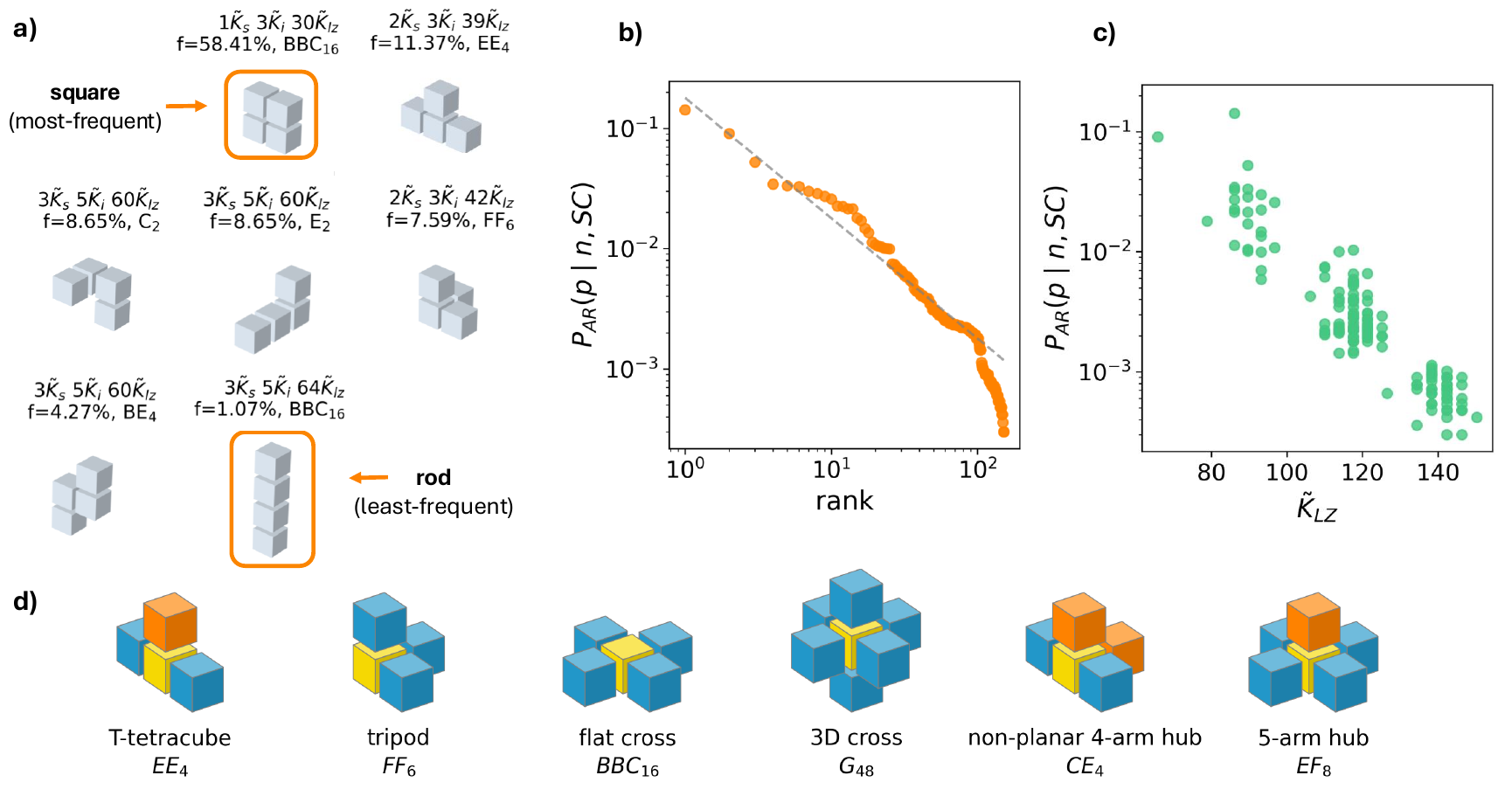}
\caption{\textbf{Global symmetry and assembly complexity.}
(a) \textit{The square and rod tetracubes share the same symmetry but differ in complexity.}
All seven free polycubes of size $4$ generated by bounded-deterministic (BD) rule sets from $10^9$ random samples in the three-dimensional rule space $S^{3D}_{5s,31c}$. Each structure is labeled by its species complexity $\tilde{K}_s$, interface complexity $\tilde{K}_i$, Lempel--Ziv complexity $\tilde{K}_{LZ}$, assembly frequency $f$, and symmetry class, with the subscript indicating symmetry order. Here, $f$ is the assembly rule probability $P_{AR}(p\,|\,  n=4)$ expressed as a percentage. The square and rod tetracubes (outlined in orange) share the same symmetry class, $BBC_{16}$, yet the square requires one tile species and is the most frequently assembled, while the rod requires three and is the least frequently assembled.
(b) \textit{Assembly rule probability versus rank for size-$20$ polycubes with $A_4$ symmetry class.} 
Of the $213$ free polycubes of size $20$ belonging to this symmetry class, $151$ are observed among the sampled BD rule sets. Structures are ranked in decreasing order of the symmetry class- and size-normalized assembly rule probability $P_{AR}(p\,|\,  n=20,SC=A_4)$, which spans nearly three orders of magnitude. The dashed line shows Zipf's law with exponent $1$.
(c) \textit{Assembly rule probability versus LZ complexity for size-$20$ polycubes with $A_4$ symmetry class.}
For the same $151$ observed structures as in (b), $P_{AR}(p\,|\,  n=20,SC=A_4)$ is plotted against the LZ complexity $\tilde{K}_{LZ}$.
(d) \textit{Local equivalence allows tile-species reuse beyond global symmetry.}
Cubes of the same color are related by global symmetry, with symmetry classes indicated below each structure. In the tripod, flat cross, and 3D cross, all arms are symmetry-equivalent (blue), whereas in the T-tetracube and the non-planar 4- and 5-armed hubs, they form two symmetry-distinct groups (blue and orange), suggesting three tile species from global symmetry alone. Nevertheless, all structures can be assembled using only two tile species ($\tilde{K}_s=2$), since every arm forms the same local connection to the central cube (yellow).
}
\label{Fig:4mers}
\end{figure*}

Arguments from AIT~\cite{LiVitanyi} suggest that the number of binary strings with Kolmogorov complexity $K$ grows as $2^K$, which would approximately counteract the $2^{-K(p)}$ individual probability  for a string $p$ expected from the AIT coding theorem~\cite{Levin1974}, resulting in a nearly flat probability distribution over complexity, $ P(K)\approx \mathcal{O}(1)$. An open question is how well this argument works for computable input-output maps~\cite{Erturk2026} with suitable approximations for complexity $\tilde{K}$. 
Interestingly, for deep neural networks on Boolean functions, which can be represented as strings, and where high expressivity means that all functions can be found, 
it has been shown that the distribution $P(\tilde{K})$ obtained upon random sampling of the weights (the analogue of sampling over assembly rules here) is nearly flat, at least on a log scale~\cite{mingard2025deep}.

By contrast, for RNA secondary structures, which can be represented as ternary strings, many low- and high-complexity strings do not correspond to valid folded secondary structures. 
Consequently, the distribution of complexities in the RNA morphospace does not follow the $2^{\tilde{K}}$ expectation for simple strings. This is one reason why, for RNA of length $L$, the probability distribution $P(\tilde{K}\,|\,  L)$ has an approximately Gaussian form, peaking at an intermediate complexity rather than remaining flat~\cite{von2025rna}.

What $P_{AR}(\tilde{K})$ to expect for polycubes and polyominoes is difficult to predict a priori, because the complexity distribution $P_M(\tilde{K})$ of the full morphospace is difficult to estimate analytically. However, we expect both very simple and very complex rule sets to be less likely to produce BD outcomes. Thus, $P_{AR}(\tilde{K})$ may peak at intermediate complexity, as observed for RNA secondary structures, although more complex behavior is also possible.

The left panel of each pair in Fig.~\ref{Fig:P(K)dist} (hatched bars) shows the complexity distributions of the sampled morphospace $P_{\hat{M}}(\tilde{K}\,|\, n)$ for polycubes of sizes $6$ and $20$, as well as $P_{\hat{M}}(\tilde{K})$ over all sizes. Corresponding distributions for additional polycube sizes are shown in SM Fig.~\ref{fig:P(K)_superplot}. At size $6$, all $112$ possible free polycubes are recovered, so the sampled morphospace coincides with the theoretical morphospace. At size $20$, however, only $4678$ distinct polycubes are recovered, compared with approximately $10^{14}$ in the theoretical morphospace. Although not all these structures are accessible in the rule space $S^{3D}_{5s,31c}$, many more should be accessible than are found in our sampling.  Since high-complexity structures generally have lower assembly rule probabilities, this means that the sampled morphospace distributions $P_{\hat{M}}(\tilde{K} \,|\,  20)$  and $P_{\hat{M}}(\tilde{K})$ will most severely underestimate the higher complexity parts of the full distribution. Nevertheless, even the sampled distributions show that there are typically many more complex structures than simpler ones.

The right panel of each pair in Fig.~\ref{Fig:P(K)dist} (solid bars) shows the corresponding assembly rule probability distributions $P_{AR}(\tilde{K}\,|\,  n)$ and $P_{AR}(\tilde{K})$. Unlike the morphospace distributions, these can be estimated without exhaustively sampling all possible structures. The required sampling depth is largely determined by the least probable complexity values, which generally receive sufficient samples in our simulations for reasonably accurate estimates.

Compared with the sampled morphospace, the assembly rule probability distributions are shifted towards lower complexity, reflecting the strong simplicity bias towards structures requiring less assembly information. The overall structure of the  $P_{AR}(\tilde{K}\,|\,  n)$ illustrates the general principle that  while simplicity bias implies that the most probable individual structures tend to be simple, this does not necessarily mean that random sampling of rules will predominantly produce simple structures, since  there are many more low probability complex  structures.  

\subsection{Distribution of assembly rule probability over symmetry orders and classes}

Just as for assembly complexity, we can examine how assembly rule probability is distributed across different symmetries. Figure~\ref{Fig:3D_prob_order_and_class_5s31c_10^9_20mer}(a) compares the morphospace probability $P_M(X\,|\,  n)$ with the assembly rule probability $P_{AR}(X\,|\,  n)$ across symmetry orders ($X=SO$) and symmetry classes ($X=SC$) for polycubes of size $n=20$. Corresponding distributions for other sizes are shown in SM Fig.~\ref{Fig:3D_prob_order_and_class_5s31c_10^9_15to20mer}. The theoretical morphospace is overwhelmingly dominated by low-symmetry structures (see SM Table~\ref{Table:OEIS}). For example, among the approximately $10^{14}$ free polycubes of size $20$, nearly all ($99.99\%$) belong to the asymmetric class $I_1$; only  $\approx 10^{10}$ have higher symmetry.  There is only one with  full cubic symmetry (class $G_{48}$). But, rather strikingly, 
under random sampling of BD rule sets its probability is within a factor of about four of that of all $I_1$ structures.  This enormous enhancement  of class $G_{48}$ is linked to the high symmetry and low assembly complexity of this structure. 

We also note that the probabilities of obtaining structures from all symmetry classes are quite similar (on a log scale), see also   other polycube sizes  in SM Fig.~\ref{Fig:3D_prob_order_and_class_5s31c_10^9_15to20mer}. The reason for this  pattern is that the rarer higher symmetry structures are more enhanced than the more common lower symmetry structures.  This composite effect of two separate trends  is analogous to what we observed for the $P_{AR}(\tilde{K})$ distributions.

Figure~\ref{Fig:3D_prob_order_and_class_5s31c_10^9_20mer}(b) shows the probability enhancement factor $P_{AR}(X\,|\,  n)/P_M(X\,|\,  n)$ (dark green), which measures how much more likely a randomly sampled BD rule set is to produce a structure belonging to a given symmetry order or class than would be expected from uniformly sampling the theoretical morphospace of that size. Only the lowest symmetry $I_1$ structures are underrepresented, whereas all other symmetries found at this size are enhanced, with the enhancement factor generally increasing with symmetry order, reaching over twelve orders of magnitude for the fully symmetric class $G_{48}$. However, symmetry classes of the same order can differ in their enhancement by several orders of magnitude. For example, among order-$4$ classes, the enhancement ranges from approximately $10^{6}$ for $EE_4$ to $10^{11}$ for $A_4$, showing that the total number of symmetry operations, or symmetry order, alone cannot explain the large differences in assembly rule probability between symmetry classes.

Figure~\ref{Fig:3D_prob_order_and_class_5s31c_10^9_20mer}(b) also shows the fraction of theoretically possible structures in each symmetry order or class recovered by our sampling (light green). This coverage is extremely low for asymmetric structures, with only $3\times10^{-11}$ of class $I_1$ recovered, but much higher for strongly enhanced symmetry classes. For example, we recover $71\%$ of class $A_4$, $56\%$ of $BBC_{16}$, and all structures belonging to $AB_8$, $AE_8$ and $G_{48}$. This shows that random sampling of assembly rules strongly favors symmetric structures, recovering a substantial fraction of some highly symmetric classes despite these classes representing only a minuscule fraction of the theoretical morphospace.

The strong enhancement of certain symmetry classes can be understood through their lower assembly complexity. Figure~\ref{Fig:3D_prob_order_and_class_5s31c_10^9_20mer}(c) shows the mean LZ complexity for each symmetry order and class, averaged either over BD rule sets $\langle \tilde{K}_{LZ} \rangle_{AR}$ (dark blue) or over distinct sampled structures $\langle \tilde{K}_{LZ} \rangle_{\hat{M}}$ (light blue). The most strongly enhanced classes, like $A_4$, $AB_8$, $EF_8$, $AE_8$, $BBC_{16}$ and $G_{48}$ have lower mean complexity. 
SM Fig.~\ref{Fig:3D_prob_enhance_complexity_minmax_top_20mer} extends this analysis to show the range of LZ complexities and the complexity of the most frequently assembled structure in each symmetry order and class. Corresponding results for unseeded assembly are shown in SM Fig.~\ref{Fig:3D_prob_enhance_complexity_minmax_top_20mer_unseeded}.

However, the relationship between symmetry and assembly complexity is not a strong or necessarily straightforward one when averaged over all structures with a particular symmetry. As discussed above, symmetry classes with the same order can differ substantially in both mean complexity and probability enhancement. Moreover, as we shall see below, even within a single symmetry class, structures can vary significantly in complexity and assembly rule probability. For nearly every symmetry class, $\langle \tilde{K}_{LZ} \rangle_{AR}$ is lower than $\langle \tilde{K}_{LZ} \rangle_{\hat{M}}$, indicating that simpler structures within the class account for a greater share of its assembly rule probability. Understanding these patterns requires examining in more detail  how global symmetry relates to  assembly complexity across  structures.

\subsection{Why some symmetries reduce assembly complexity more effectively}
\label{subsec:why_symmetries}

The extent to which symmetry can reduce the amount of information needed to specify the assembly rules depends on whether they permit the same local interactions to be reused during deterministic assembly. That may not always directly correlate with the number of possible symmetry operations. 

For example, rotations and reflections can differ in how effectively they allow local assembly rules to be reused. In our model, tile species can be rotated but not reflected during assembly. Proper rotations can therefore allow repeated motifs to be generated using rotated copies of the same tile species and interactions, whereas improper operations, such as reflections and rotoreflections, generally do not permit the same reuse without additional constraints on the interface patterns. Consider symmetry classes $A_4$ and $J_4$ (Table~\ref{Table:lunnon_sym_classes}). Both contain four symmetry operations, but $A_4$ is generated by a fourfold proper rotation, whereas $J_4$ is generated by a fourfold rotoreflection. This difference helps explain why structures belonging to $A_4$ generally have lower assembly complexity and much higher assembly rule probability than those belonging to $J_4$ (Fig.~\ref{Fig:3D_prob_order_and_class_5s31c_10^9_20mer}). Among the order-$4$ symmetry classes, $A_4$ has the fewest structures in the theoretical morphospace of size-$20$ polycubes, yet the highest assembly rule probability. Of the $30$ most frequently assembled size-$20$ polycubes shown in Fig.~\ref{Fig:3D_5s31c_10^9_20mer_4mer}(c), $18$ belong to this class.

However, proper rotational symmetry does not guarantee efficient reuse of assembly rules. If a symmetry operation exchanges two cubes across their shared bond, assigning the same tile species to both positions would place identical interface types on the two contacting faces. Since our model allows binding only between distinct complementary interface types, such a bond cannot form. These constraints can limit the reduction in assembly complexity even when the structure possesses several proper rotational symmetries. This is relevant to the symmetry classes $BB_4$ and $BC_4$, which contain only proper rotations but show much smaller probability enhancements than $A_4$. Similarly, for planar polycubes, reflection through the plane containing all cube centers leaves every cube in place. Although this reflection contributes to the global symmetry order, it provides no additional opportunities for tile-species reuse.

We now return to the peaks in assembly rule probability at polycube sizes that are multiples of four, observed in Fig.~\ref{Fig:3D_5s31c_10^9_nmer_hist}.  They arise  in particular from the strong preference for structures possessing fourfold rotational symmetry, particularly class $A_4$.  More generally, the relatively enhanced preference for higher symmetry BD structures is also reflected in an elevated mean symmetry order (SM Fig.~\ref{Fig:mean_sym_order_vs_size}) for these sizes.  More specifically, when a fourfold rotation axis passes between cubes rather than through them, the cubes occur in groups of four related by rotation. The same tile species and interface types can be reused across these groups, reducing their assembly complexity.   Structures that can take advantage of this have sizes that are multiples of four, and this explains the peaks in  Fig.~\ref{Fig:3D_5s31c_10^9_nmer_hist}.  

\subsection{Why structures with the same symmetry class can vary significantly in their assembly complexity and probability}

The preceding discussion illustrates how  different symmetry operations can vary in how efficiently they reduce assembly complexity. However, even structures belonging to the same symmetry class can require substantially different amounts of assembly information. 

A striking example is provided by the seven free tetracubes shown in Fig.~\ref{Fig:4mers}(a). The square and the linear rod (outlined in orange) both belong to symmetry class $BBC_{16}$ and therefore possess the same global symmetry. Nevertheless, deterministic assembly of the rod requires at least three tile species and five interface types, whereas the square requires only one species and three interface types. Consistent with this difference, the square has the highest assembly rule probability among tetracubes, while the rod has the lowest.

The difference arises from variations in how their symmetry operations act on the individual cubes. In the rod, the fourfold rotation axis passes through every cube, leaving their positions unchanged. By contrast, the fourfold axis of the square passes through its empty center, mapping each cube to the other three positions. The square can therefore reuse the same tile species at all four positions, whereas the rod cannot exploit its fourfold symmetry in this way. Furthermore, the rod's central bond prevents its two central cubes from sharing a species under the complementary binding rules discussed previously (see SM Sec.~\ref{SM:tile_comp} for an analytic calculation of the assembly complexity of rods).

Another example of variation within the same symmetry class can be seen among size-$20$ polycubes belonging to class $A_4$. Of the $213$ theoretically possible structures in this class, our sampling recovers $151$. Although all possess the same fourfold rotational symmetry, their assembly rule probabilities span almost three orders of magnitude (Fig.~\ref{Fig:4mers}(b)). This variation is accompanied by quite substantial differences in their assembly complexity (Fig.~\ref{Fig:4mers}(c)). 
This illustrates how structures with identical global symmetry operations can require different amounts of assembly information.

Conversely, local rule reuse does not always require global symmetry. Figure~\ref{Fig:4mers}(d) illustrates this using polycubes consisting of a central cube with single-cube arms. The T-tetracube, tripod, flat cross, and three-dimensional cross have symmetry orders $4$, $6$, $16$, and $48$, respectively, yet all can be assembled using just two tile species: one for the central cube and one for the arms. For the T-tetracube, global symmetry does not relate the stem to the side arms, but all three can share a species because they form the same local connection to the central cube. Similarly, the non-planar four-armed and five-armed hubs require only two species, even though their global symmetries would suggest three (the three colors in Fig.~\ref{Fig:4mers}(d)). These examples show that local rule reuse can reduce assembly complexity beyond what global symmetry would predict.

We have given a few examples  showing that local rule reuse can depend on other structural properties beyond symmetry, such as compactness and the constraints of deterministic growth.  Predicting more generally which features besides symmetry influence assembly complexity and assembly rule probability, and how their relative importance varies with structure size, remains an important direction for future work.

\section{Conclusion}
\label{sec:conclusion}

We studied how assembly complexity and symmetry shape the distribution of structures generated by a three-dimensional polycube model of programmable self-assembly. By randomly sampling assembly rules and analyzing the resulting bounded-deterministic structures, we find that polycubes with high assembly probability have low assembly complexity,  consistent with the exponential enhancement predicted by simplicity bias for input–output maps~\cite{Dingle2018,Dingle2020}. This extends earlier findings for two-dimensional polyominoes~\cite{Johnston2022} to a substantially larger and geometrically richer three-dimensional assembly space. 

A striking finding is that  the set of structures with the highest assembly rule probabilities is hugely enriched in high symmetry structures.
For example, the size-20 polycube with full cubic symmetry is generated more than twelve orders of magnitude more frequently than would be expected from its abundance in morphospace.  For these structures, symmetry provides an important mechanism for reducing assembly complexity, as symmetric structures can  reuse tile species and interfaces, allowing them to be specified by shorter assembly rules. As a result, they can occupy disproportionately large regions of rule space. 

Nevertheless, assembly complexity is under-determined by symmetry. 
Structures with the same symmetry order, or even belonging to the same symmetry class, can require very different amounts of assembly information. This is because not all symmetry operations allow tile species and interfaces to be reused equally effectively under the local binding rules. Moreover, repeated local patterns can reduce assembly complexity even when the corresponding positions are not related by global symmetry. 

Our results suggest that the application of Cartwright and Mackay's~\cite{Cartwright2012} idea that assembly complexity provides a more general measure of structural complexity than symmetry alone is not so straightforward for finite-sized assemblies.  Their main application was to extended structures, which would be unbounded deterministic structures in our language of assembly, and it would be interesting to study these in  more detail for polycubes and polyominoes.  One possible application would be to the protein crystals that create surface layers coating bacteria and archaea~\cite{Doye2024}. 

Future work should also investigate which structural features beyond global symmetry allow tile species and interfaces to be reused. These may include local structural motifs, compactness, and the number of possible assembly pathways, which could be studied using graph-theoretic approaches~\cite{agrawal2026successive}. Another important direction is to extend our analysis beyond the cubic lattice to other particle geometries and continuous space, where different interaction constraints may influence the relationship between symmetry and assembly complexity.

Understanding the links between geometry, assembly complexity, and assembly rule probability may help guide the design of programmable nanostructures, colloidal assemblies, and other engineered self-assembling systems where minimizing design complexity is both scientifically and technologically desirable. Recent studies of programmable shells have similarly shown that the arrangement of symmetry elements can influence the trade-off between design complexity and assembly fidelity~\cite{Tyukodi2025magic}. More broadly, relaxing strict deterministic assembly constraints may offer further opportunities to reduce the number of distinct building blocks, as explored in semiaddressable self-assembly~\cite{Hubl2025accessing}. These findings may also have implications for evolution, where biases towards structures requiring simpler assembly rules could contribute to non-adaptive explanations for the prevalence of symmetry and other structural features in protein complexes~\cite{Johnston2022}.

\begin{acknowledgments}
We thank John Mason for helpful communications about and contributions to the OEIS enumeration of free polycubes by symmetry class, and Abdurrahman Hadi Erturk and Andrew Turberfield for helpful discussions.
\end{acknowledgments}

\section*{Data Availability}
The code and data supporting this study are openly available through Zenodo at \url{https://doi.org/10.5281/zenodo.22045750}.
Figs.~\ref{Fig:2D_3D_freq_vs_comp_for_all_complexities} and~\ref{Fig:3D_5s31c_10^9_20mer_4mer} can be interactively visualized at \url{https://akodiat.codeberg.page/polycubeViewer/}.

\putbib
\end{bibunit}

\clearpage
\begin{bibunit}
\nocite{REVTEX42Control,aip41Control}

\renewcommand{\theHfigure}{S\arabic{figure}}
\renewcommand{\theHtable}{S\arabic{table}}
\renewcommand{\theHequation}{S\arabic{equation}}

\renewcommand{\theequation}{S\arabic{equation}}
\onecolumngrid

\begin{center}

{\Large\bfseries Supplementary Material}\\[0.6em]




\end{center}

\vspace{1em}


\renewcommand{\thefigure}{S\arabic{figure}}
\setcounter{figure}{0}

\renewcommand{\thetable}{S\arabic{table}}
\setcounter{table}{0}

\renewcommand{\thesection}{\Alph{section}}
\setcounter{section}{0}

\renewcommand{\thesubsection}{\thesection.\arabic{subsection}}

\setcounter{equation}{0}

\section{Determinism testing and complexity calculations}
\label{SM_sec:inputspace}

\subsection{Determinism testing}
\label{SM_subsec:ND}

A rule set is classified as {deterministic} if all tested trajectories terminate in the same free polycube, i.e., if the resulting structures are equivalent under the operations of $O_h$. Final structures are compared as free polycubes, i.e. including reflections, although tiles only rotate. This is justified because mirror-image structures require the same assembly information: any rule set assembling a structure has a mirror-image counterpart, with the same numbers of tile species and interface types, that assembles its mirror image.

To test determinism, we assemble the rule set along $k=25$  growth trajectories, generated by different random choices of open and complementary interfaces. The possible growth trajectories of a rule set can be represented by a choice tree, in which branching corresponds to alternative choices of open and complementary interfaces during assembly. For seeded assembly, we construct this tree and explore all possible trajectories when their number does not exceed $k$. Otherwise, the test is limited to $k$ sampled trajectories. For unseeded assembly, the choice of seed species introduces an additional branching level. We therefore do not construct the choice tree and instead perform $k$ independent assembly runs, selecting the seed species randomly in each run.

When only a subset of trajectories is sampled, non-determinism can be detected but cannot be excluded. Consequently, some rule sets classified as bounded-deterministic (BD) may admit unsampled trajectories leading to different structures. The sensitivity of the classification to the simulation parameters is examined in Sec.~\ref{sec:robustness}.

\subsection{Simplification of assembly rules and complexity calculations}
\label{SM_subsec:complexity}

Randomly sampled rule sets can contain information that does not affect the assembly process, such as non-neutral (active) interface types whose complementary partner is absent from the assembly kit. Retaining such redundant information would inflate the measured complexity without changing the assembled structure. We therefore calculate complexities
from a reduced and simplified rule set, constructed separately for each of the
trajectories used in the determinism test.

Different trajectories of the same BD rule set can use different subsets of the tile species and interface types present in the assembly kit, even when they produce the same final structure. For example, an interface type may have its complementary partner present in the kit but never participate in a bond because the corresponding attachment site becomes sterically blocked before that partner can attach. For each trajectory, we therefore retain only the tile species incorporated into the assembly, listed in order of first
incorporation starting with the seed, and the interface types that form at least one bond; all other faces of the retained tiles are set to the neutral type. The reduced rule set is then simplified as follows:

\begin{enumerate}
    \item Label $0$ denotes the neutral interface. When the number of
    interface types $c$ is even, the final interface type $c-1$ is also neutral under our interaction convention and is therefore relabeled as $0$. For the odd values of $c$ used in the simulations reported here, this step has no effect.
    \item Every active interface type whose complementary partner is absent from the assembly kit is set to the neutral type $0$, since it cannot participate in a bond. The orientation of every neutral face, including faces that were neutral before simplification, is set to a fixed reference orientation, \texttt{O} for the N, E, S, and W faces and \texttt{U} for the B and F faces, so that neutral faces carry no orientational information.
    \item  The remaining active interface types are relabeled consecutively, reading the tiles in order of first incorporation and the faces of each tile in NESWBF order. When an interface type is first encountered, it is assigned the next available odd label and its complementary partner the following even label, so that the complementary-pair structure of $A_{ij}$ is preserved.
    
    For example, with orientations omitted for clarity, suppose the incorporated tiles of a rule set in $S^{3D}_{2s,5c}$ carry the interface sequences $\{343202,002434\}$, where $1\leftrightarrow2$ and $3\leftrightarrow4$. Types $3$ and $4$ form bonds, whereas type $2$ cannot, because its partner $1$ is absent. Setting type $2$ to $0$ gives $\{343000,000434\}$, which becomes $\{121000,000212\}$ after consecutive labeling.
\end{enumerate}

From the simplified rule set of each trajectory, $\tilde{K}_s$ is the number of tile species and $\tilde{K}_i$ the number of interface types, including the neutral type. Since active interfaces are retained only in bonded complementary pairs, $\tilde{K}_i=1+2m$, where $m$ is the number of complementary pairs used.

To calculate $\tilde{K}_{LZ}$, tile species whose faces are all neutral are first removed, since they carry no information relevant to assembly. The simplified rule set is then converted to a binary string by listing the tile species in order of first incorporation and the six faces of each tile in NESWBF order. Each interface label is represented using $\lceil\log_2 c\rceil$ bits, where $c$ is the number of interface types in the rule space, and each orientation is represented using three bits:
\texttt{O}$\to$000, \texttt{U}$\to$001, \texttt{I}$\to$010,
\texttt{D}$\to$011, \texttt{L}$\to$100, and \texttt{R}$\to$101.
For example, in a polycube rule space with $c=5$ interface types, each face requires $3+3=6$ bits and each tile $36$ bits. For 2D polyominoes, orientations are omitted and only interface types are converted to binary.
Following Dingle \emph{et al.}~\cite{Dingle2018}, for a binary string $x$ of
length $n$,
\begin{equation}
\tilde{K}_{LZ}(x) =
\begin{cases}
\log_2 n, & x = 0^n \text{ or } 1^n,\\[2pt]
\log_2 n \,\dfrac{N_w(x_1\ldots x_n) + N_w(x_n\ldots x_1)}{2}, & \text{otherwise},
\end{cases}
\end{equation}
where $N_w$ is the number of words in the Lempel--Ziv 1976 parsing~\cite{Lempel1976}.
Evaluating the complexity for both the forward and reversed strings removes the dependence on reading direction. Unlike $\tilde{K}_s$ and $\tilde{K}_i$, $\tilde{K}_{LZ}$ depends on the order in which tile species are incorporated and on the rotational representative of each tile, so it can differ between trajectories that use the same species.

For a given BD rule set, each complexity measure is taken to be its minimum over the $k$ tested trajectories. For a given structure, $\tilde{K}_s$, $\tilde{K}_i$, and $\tilde{K}_{LZ}$ are then each defined as the minimum over all sampled BD rule sets that generate it. Each reported complexity thus involves two successive minimizations: first over the tested trajectories of each BD rule set, and then over all sampled BD rule sets generating the structure.

\clearpage
\section{Polycube size distributions}
\label{SM:size}

\begin{figure*}[b]
    \includegraphics[width=\textwidth]{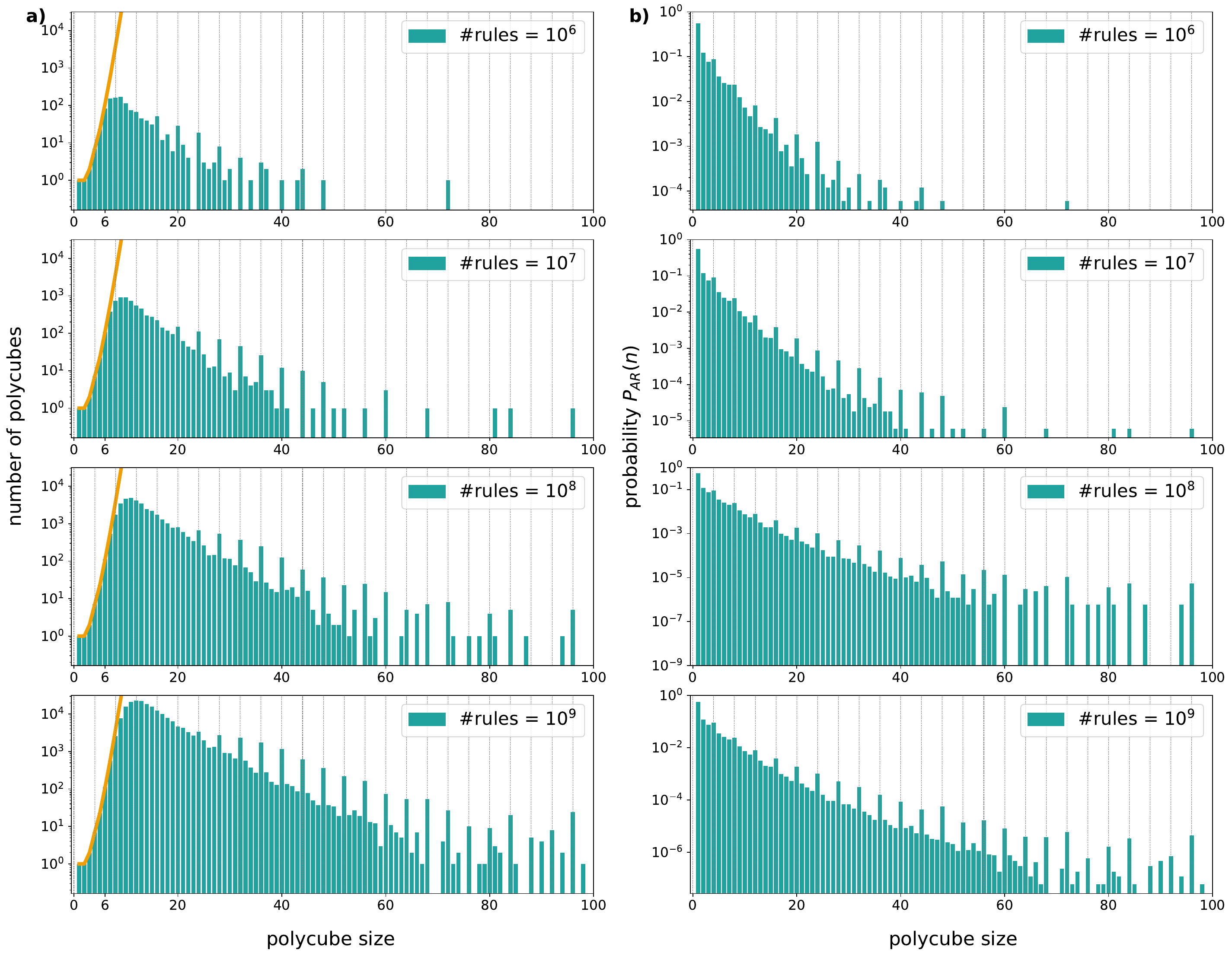}
    \caption{\textbf{Effect of sampling depth on polycube size distributions.}
    Results are obtained from the three-dimensional rule space $S^{3D}_{5s,31c}$ under seeded assembly, using $10^6$, $10^7$, $10^8$, and $10^9$ randomly sampled rule sets.
    (a) Number of distinct free polycubes generated by bounded-deterministic (BD) rule sets as a function of polycube size. The yellow line shows the number of theoretically possible free polycubes (OEIS sequence A038119)~\cite{OEIS2026}. Increasing the sampling depth recovers progressively more distinct polycubes, particularly at larger sizes.
    (b) Assembly rule probability $P_{AR}(n)$, giving the fraction of BD rule sets that produce polycubes of size $n$. Vertical gray lines indicate polycube sizes that are multiples of four.}
\label{Fig:3D_5s31c_10^6_to_10^9_nmer_hist_rule_count_fraction}
\end{figure*}

\begin{figure*}[]
    \includegraphics[width=\textwidth]{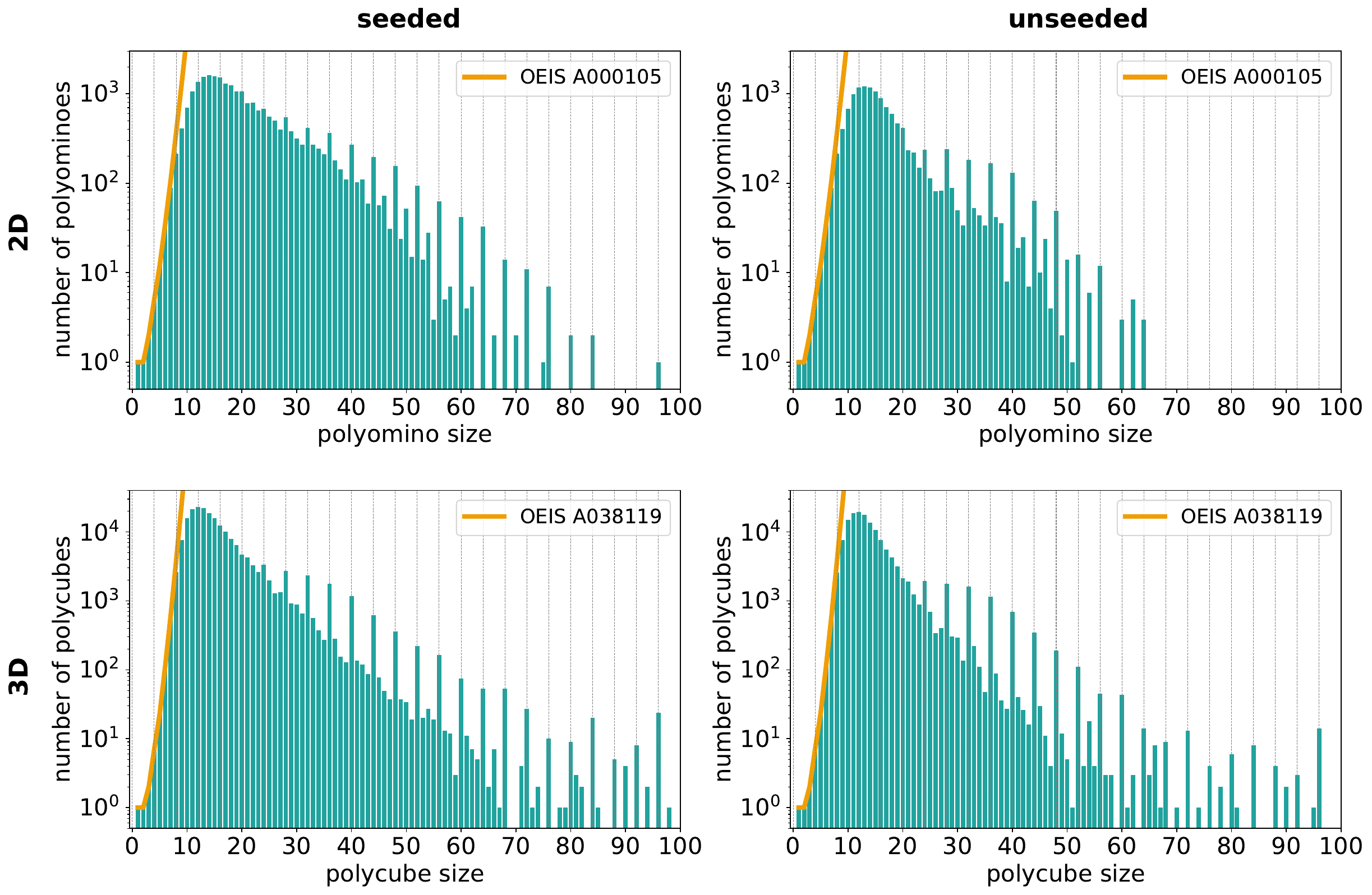}
    \caption{\textbf{Size distributions of two- and three-dimensional assemblies under seeded and unseeded growth}. Bars show the number of distinct structures generated by bounded-deterministic (BD) rule sets from $10^9$ random samples. The top row shows two-dimensional polyominoes generated from $S^{2D}_{5s,31c}$, and the bottom row shows three-dimensional polycubes generated from $S^{3D}_{5s,31c}$, under seeded (left) and unseeded (right) assembly. Yellow lines show the number of theoretically possible free polyominoes and polycubes at each size from the OEIS sequences indicated in the legends. Vertical gray lines indicate sizes that are multiples of four.}
\label{Fig:2D_3D_5s31c_10^9_nmer_hist}
\end{figure*}

\begin{figure*}[!htbp]
    \includegraphics[width=\textwidth]{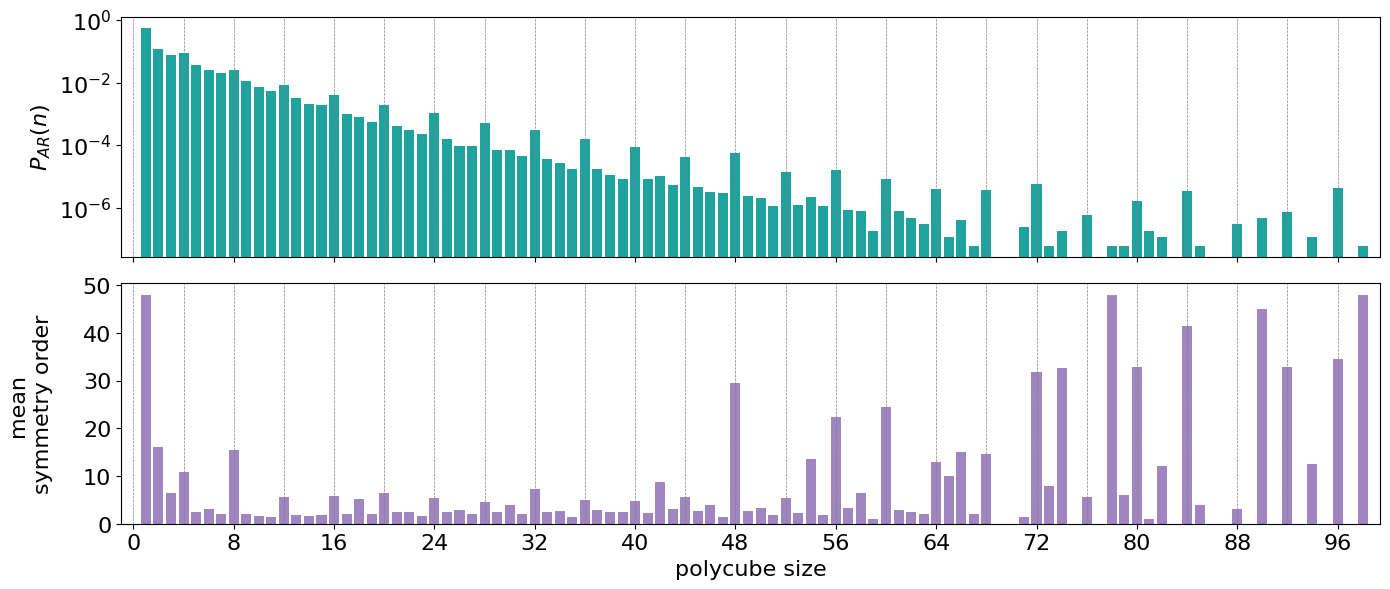}
    \caption{
    \textbf{Assembly rule probability and mean symmetry order versus polycube size.}
    Results are obtained from $10^9$ randomly sampled rule sets in the three-dimensional rule space $S^{3D}_{5s,31c}$ under seeded assembly.
    (\textit{Top}) Assembly rule probability $P_{AR}(n)$, giving the fraction of BD rule sets that generate polycubes of size $n$. $P_{AR}(n)$ generally decreases with $n$, with pronounced local peaks at particular sizes.
    (\textit{Bottom}) Mean symmetry order at each polycube size, weighted by the number of BD rule sets generating each structure.
    Several local peaks in assembly rule probability, particularly at larger sizes, coincide with elevated mean symmetry order.
    }
\label{Fig:mean_sym_order_vs_size}
\end{figure*}

\begin{figure*}[!htbp]
\includegraphics[width=\textwidth]{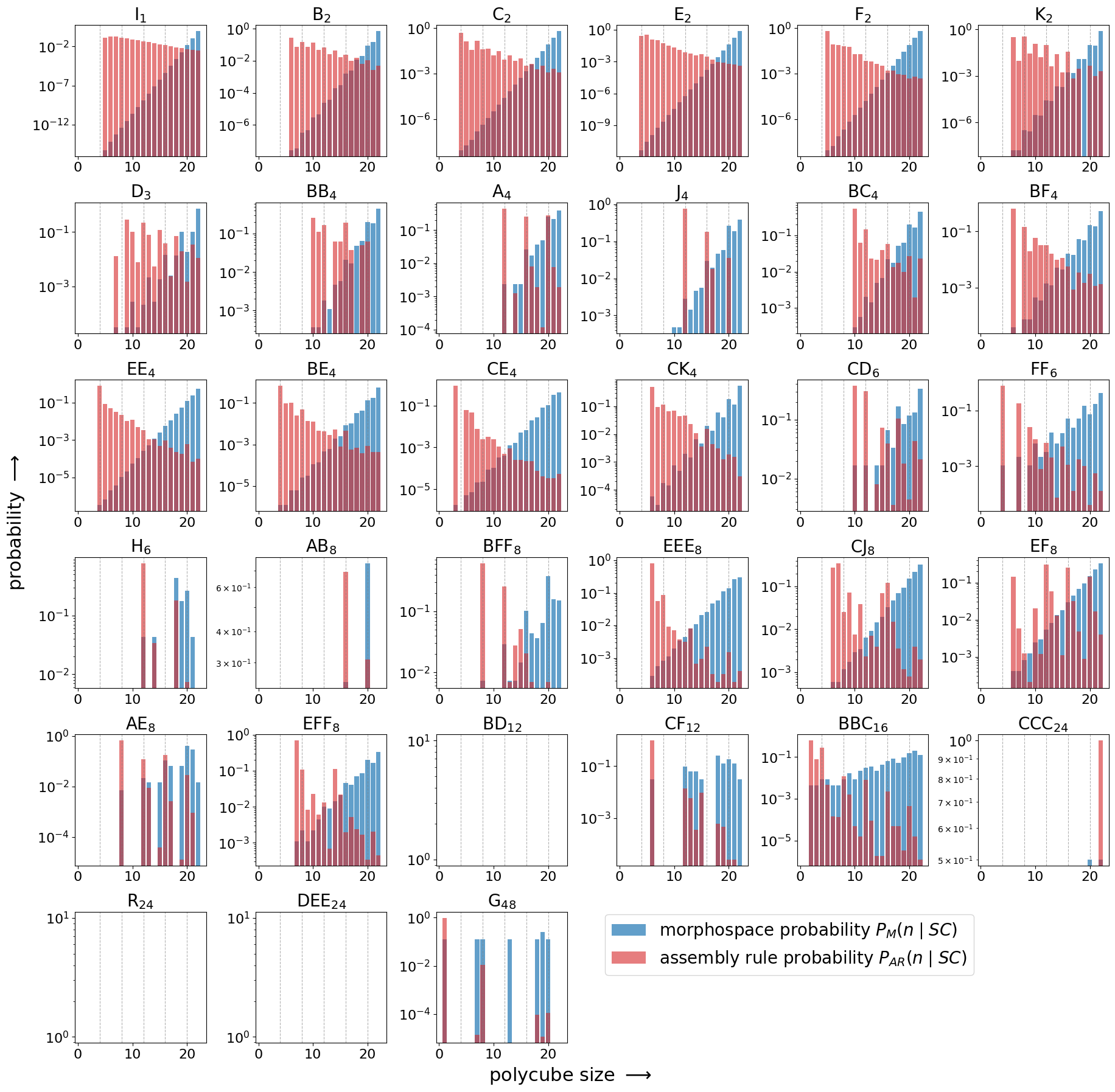}
\caption{
\textbf{Polycube size distributions within symmetry classes in morphospace and under random rule sampling.}
Probability distributions over polycube size are shown for each of the $33$ symmetry classes listed in main-text Table~\ref{Table:lunnon_sym_classes}. Blue bars show the morphospace probability $P_M(n\mid SC)$, where $SC$ denotes symmetry class, calculated from the OEIS enumerations listed in Table~\ref{Table:OEIS}. Red bars show the corresponding assembly rule probability $P_{AR}(n\mid SC)$ obtained from $10^9$ randomly sampled rule sets in the three-dimensional rule space $S^{3D}_{5s,31c}$ under seeded assembly. For each symmetry class, both distributions are independently normalized over polycube sizes up to $n=22$, the largest size for which OEIS enumerations are available for all symmetry classes. Panel titles indicate the symmetry class, with symmetry order shown as a subscript. Vertical dashed lines indicate polycube sizes that are multiples of four.}
\label{Fig:3D_OEISxsamp_size_foreachsymclass}
\end{figure*}

\begin{figure*}[!htbp]
    \includegraphics[width=\textwidth]{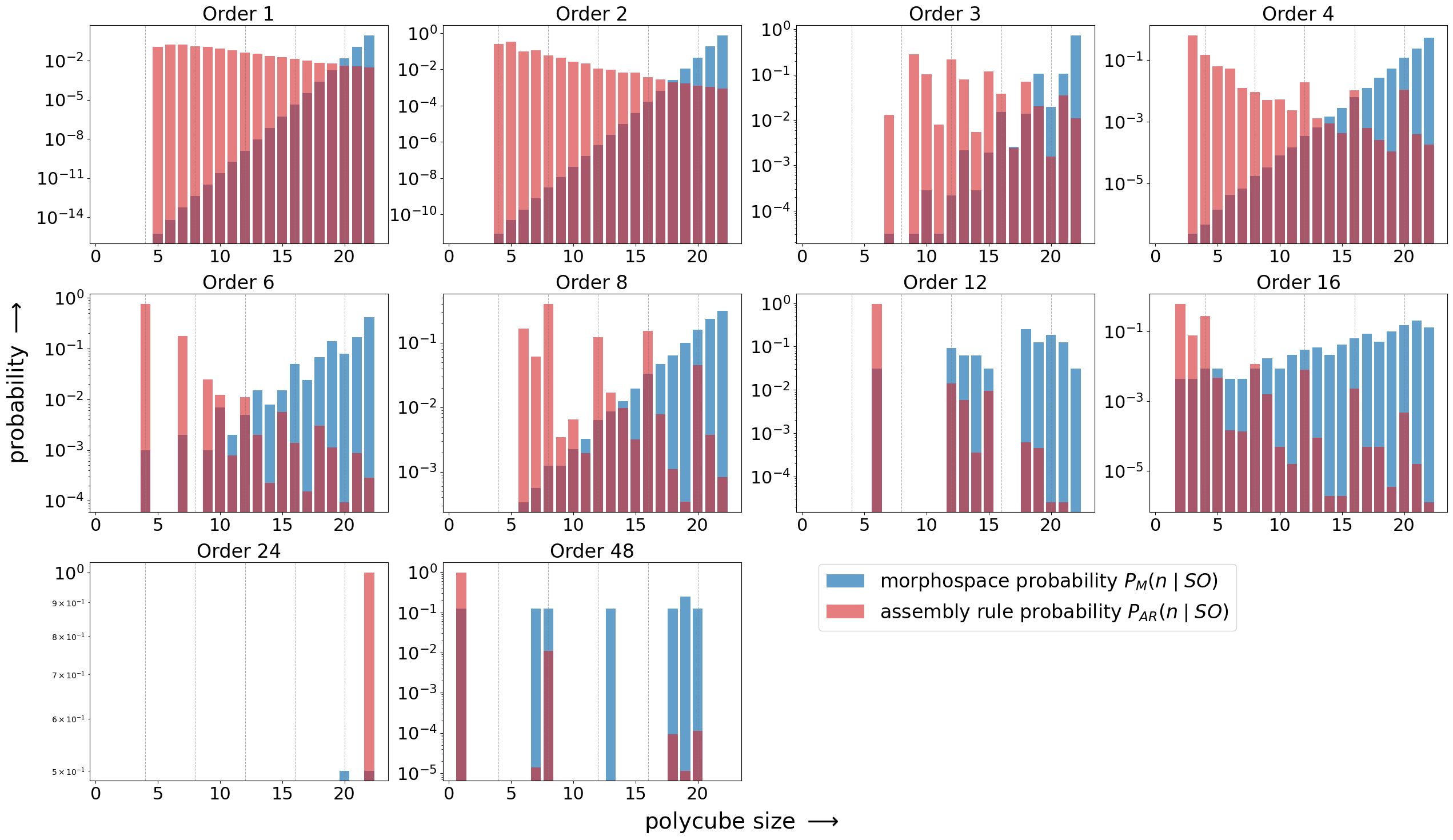}
    \caption{
   \textbf{Polycube size distributions within symmetry orders in morphospace and under random rule sampling.}
   Probability distributions over polycube size are shown for each symmetry order listed in main-text Table~\ref{Table:lunnon_sym_classes}. Blue bars show the morphospace probability $P_M(n\mid SO)$, where $SO$ denotes symmetry order, calculated from the OEIS enumerations listed in Table~\ref{Table:OEIS}. Red bars show the assembly rule probability $P_{AR}(n\mid SO)$ obtained from $10^9$ randomly sampled rule sets in the three-dimensional rule space $S^{3D}_{5s,31c}$ under seeded assembly. For each symmetry order, both distributions are independently normalized over polycube sizes up to $n=22$, the largest size for which OEIS enumerations are available for all symmetry classes. Vertical dashed lines indicate polycube sizes that are multiples of four.
    }
\label{Fig:3D_OEISxsamp_size_foreachsymorder}
\end{figure*}

\begin{table}
    \centering
    \setlength{\tabcolsep}{3pt}
    \caption{\textbf{Number of free polycubes and free polyominoes up to size $22$}.
    Values are taken from OEIS~\cite{OEIS2026} sequences A038119 (3D) and A000105 (2D).}
    \label{table:free_polycubes_polyominoes}

    \begin{tabular}{|r|r||r|r|}
        \hline
        \multicolumn{4}{|c|}{3D polycubes} \\
        \hline
        Size & Number of polycubes & Size & Number of polycubes \\
        \hline
         1 & 1         & 12 & 9,371,094 \\
         2 & 1         & 13 & 69,513,546 \\
         3 & 2         & 14 & 520,878,101 \\
         4 & 7         & 15 & 3,934,285,874 \\
         5 & 23        & 16 & 29,915,913,663 \\
         6 & 112       & 17 & 228,779,330,204 \\
         7 & 607       & 18 & 1,758,309,223,457 \\
         8 & 3,811     & 19 & 13,573,319,825,615 \\
         9 & 25,413    & 20 & 105,192,814,197,984 \\
        10 & 178,083   & 21 & 818,136,047,201,932 \\
        11 & 1,279,537 & 22 & 6,383,528,588,447,574 \\
        \hline
    \end{tabular}
    \hspace{0.4cm}
    \begin{tabular}{|r|r||r|r|}
        \hline
        \multicolumn{4}{|c|}{2D polyominoes} \\
        \hline
        Size & Number of polyominoes & Size & Number of polyominoes \\
        \hline
         1 & 1      & 12 & 63,600 \\
         2 & 1      & 13 & 238,591 \\
         3 & 2      & 14 & 901,971 \\
         4 & 5      & 15 & 3,426,576 \\
         5 & 12     & 16 & 13,079,255 \\
         6 & 35     & 17 & 50,107,909 \\
         7 & 108    & 18 & 192,622,052 \\
         8 & 369    & 19 & 742,624,232 \\
         9 & 1,285  & 20 & 2,870,671,950 \\
        10 & 4,655  & 21 & 11,123,060,678 \\
        11 & 17,073 & 22 & 43,191,857,688 \\
        \hline
    \end{tabular}
\end{table}

\begin{table}[!htbp]
    \centering
    \scriptsize
    \setlength{\tabcolsep}{3pt} 
    \caption{
    \textbf{OEIS enumeration of free polycubes by size and symmetry class\cite{OEIS2026}}.
    Free polycubes are classified into the $33$ symmetry classes listed in main-text Table~\ref{Table:lunnon_sym_classes}, with symmetry order shown as a subscript to each class label. The OEIS sequence corresponding to each symmetry class is given in parentheses:
    ${I_1}$ (A376964), ${A_4}$ (A376965), ${B_2}$ (A376966), ${C_2}$ (A376967), ${D_3}$ (A376968), ${E_2}$ (A376969), ${F_2}$ (A376970), ${H_6}$ (A376972), ${J_4}$ (A376973), ${K_2}$ (A376974), ${BB_4}$ (A376975), ${BC_4}$ (A376976), ${BE_4}$ (A376977), ${BF_4}$ (A376978), ${CE_4}$ (A376979), ${CK_4}$ (A376980), ${EE_4}$ (A376981), ${CD_6}$ (A376982), ${FF_6}$ (A376983), ${AB_8}$ (A377127), ${AE_8}$ (A376984), ${BFF_8}$ (A376985), ${CJ_8}$ (A376986), ${EEE_8}$ (A376987), ${EF_8}$ (A376988), ${EFF_8}$ (A376989), ${BD_{12}}$ (A377128), ${CF_{12}}$ (A376990), ${BBC_{16}}$ (A376991), ${CCC_{24}}$ (A377129), ${DEE_{24}}$ (A377130), ${R_{24}}$ (A377131), and ${G_{48}}$ (A376971).
    For each polycube size, the sum over all symmetry classes equals the total number of free polycubes given by OEIS sequence A038119.
    }  
    \begin{tabular}{|l|l|l|l|l|l|l|l|l|l|l|l|l|l|l|l|}
    \hline
        \textbf{size} & ${I_1}$ & ${A_4}$ & ${B_2}$ & ${C_2}$ & ${D_3}$ & ${E_2}$ & ${F_2}$ & ${H_6}$ & ${J_4}$ & ${K_2}$ & ${BB_4}$ & ${BC_4}$ & ${BE_4}$ & ${BF_4}$ & ${CE_4}$ \\ 
        \hline
        1 & 0 & 0 & 0 & 0 & 0 & 0 & 0 & 0 & 0 & 0 & 0 & 0 & 0 & 0 & 0 \\
        2 & 0 & 0 & 0 & 0 & 0 & 0 & 0 & 0 & 0 & 0 & 0 & 0 & 0 & 0 & 0 \\ 
        3 & 0 & 0 & 0 & 0 & 0 & 0 & 0 & 0 & 0 & 0 & 0 & 0 & 0 & 0 & 1 \\ 
        4 & 0 & 0 & 0 & 1 & 0 & 1 & 0 & 0 & 0 & 0 & 0 & 0 & 1 & 0 & 0 \\ 
        5 & 4 & 0 & 0 & 2 & 0 & 6 & 3 & 0 & 0 & 0 & 0 & 0 & 1 & 0 & 3 \\ 
        6 & 46 & 0 & 3 & 5 & 0 & 28 & 6 & 0 & 0 & 1 & 0 & 0 & 5 & 1 & 4 \\ 
        7 & 394 & 0 & 4 & 17 & 1 & 126 & 26 & 0 & 0 & 1 & 0 & 0 & 5 & 0 & 12 \\ 
        8 & 3025 & 0 & 37 & 49 & 0 & 520 & 72 & 0 & 0 & 20 & 0 & 0 & 21 & 2 & 13 \\ 
        9 & 22707 & 0 & 52 & 138 & 1 & 2099 & 241 & 0 & 0 & 17 & 0 & 0 & 26 & 2 & 48 \\ 
        10 & 167732 & 0 & 342 & 374 & 9 & 8429 & 623 & 0 & 1 & 192 & 1 & 1 & 93 & 12 & 61 \\ 
        11 & 1241417 & 0 & 502 & 1062 & 1 & 33676 & 2028 & 0 & 1 & 175 & 1 & 2 & 116 & 9 & 186 \\ 
        12 & 9221624 & 2 & 2836 & 2851 & 7 & 135201 & 5374 & 1 & 6 & 1632 & 5 & 7 & 392 & 37 & 231 \\ 
        13 & 68936674 & 0 & 4343 & 8010 & 68 & 543248 & 16781 & 0 & 3 & 1522 & 3 & 5 & 497 & 32 & 727 \\ 
        14 & 518574100 & 2 & 22622 & 21432 & 9 & 2195182 & 45094 & 1 & 10 & 13088 & 13 & 17 & 1616 & 134 & 941 \\ 
        15 & 3925132946 & 2 & 35405 & 60142 & 61 & 8893547 & 138971 & 0 & 12 & 12339 & 16 & 23 & 2087 & 114 & 2855 \\ 
        16 & 29878869619 & 22 & 176176 & 161386 & 473 & 36196788 & 378196 & 0 & 63 & 102846 & 58 & 78 & 6690 & 426 & 3707 \\ 
        17 & 228629549175 & 14 & 281141 & 453253 & 81 & 147739046 & 1156028 & 0 & 41 & 97374 & 47 & 62 & 8656 & 375 & 11293 \\ 
        18 & 1757697391087 & 31 & 1363112 & 1222110 & 440 & 605138811 & 3183390 & 10 & 97 & 803363 & 136 & 181 & 27493 & 1393 & 14909 \\ 
        19 & 13570818452472 & 41 & 2205171 & 3436564 & 3316 & 2485051070 & 9679303 & 4 & 124 & 760594 & 181 & 218 & 35797 & 1236 & 44955 \\ 
        20 & 105182527335313 & 213 & 10527712 & 9316409 & 614 & 10233284681 & 26929415 & 6 & 553 & 6274638 & 546 & 717 & 113728 & 4398 & 59760 \\ 
        21 & 818093680980786 & 182 & 17221126 & 26231463 & 3315 & 42234214910 & 81575661 & 1 & 401 & 5928747 & 521 & 591 & 147838 & 3934 & 180092 \\ 
        22 & 6383353461322488 & 321 & 81462884 & 71496106 & 23537 & 174693725151 & 228958790 & 0 & 821 & 49133524 & 1247 & 1606 & 469824 & 13883 & 241827 \\ \hline
    \end{tabular}

    \vspace{6pt} 

    \setlength{\tabcolsep}{3.4pt} 
    \begin{tabular}{|l|l|l|l|l|l|l|l|l|l|l|l|l|l|l|l|l|l|l|}
    \hline
        \textbf{size} & ${CK_4}$ & ${EE_4}$ & ${CD_6}$ & ${FF_6}$ & ${AB_8}$ & ${AE_8}$ & ${BFF_8}$ & ${CJ_8}$ & ${EEE_8}$ & ${EF_8}$ & ${EFF_8}$ & ${BD_{12}}$ & ${CF_{12}}$ & ${BBC_{16}}$ & ${CCC_{24}}$ & ${DEE_{24}}$ & ${R_{24}}$ & ${G_{48}}$ \\ \hline
        1 & 0 & 0 & 0 & 0 & 0 & 0 & 0 & 0 & 0 & 0 & 0 & 0 & 0 & 0 & 0 & 0 & 0 & 1 \\ 
        2 & 0 & 0 & 0 & 0 & 0 & 0 & 0 & 0 & 0 & 0 & 0 & 0 & 0 & 1 & 0 & 0 & 0 & 0 \\ 
        3 & 0 & 0 & 0 & 0 & 0 & 0 & 0 & 0 & 0 & 0 & 0 & 0 & 0 & 1 & 0 & 0 & 0 & 0 \\ 
        4 & 0 & 1 & 0 & 1 & 0 & 0 & 0 & 0 & 0 & 0 & 0 & 0 & 0 & 2 & 0 & 0 & 0 & 0 \\ 
        5 & 0 & 2 & 0 & 0 & 0 & 0 & 0 & 0 & 0 & 0 & 0 & 0 & 0 & 2 & 0 & 0 & 0 & 0 \\ 
        6 & 2 & 6 & 0 & 0 & 0 & 0 & 0 & 1 & 1 & 1 & 0 & 0 & 1 & 1 & 0 & 0 & 0 & 0 \\ 
        7 & 1 & 11 & 0 & 2 & 0 & 0 & 0 & 1 & 2 & 1 & 1 & 0 & 0 & 1 & 0 & 0 & 0 & 1 \\ 
        8 & 6 & 32 & 0 & 0 & 0 & 1 & 1 & 2 & 3 & 2 & 2 & 0 & 0 & 2 & 0 & 0 & 0 & 1 \\ 
        9 & 5 & 61 & 0 & 1 & 0 & 0 & 0 & 3 & 4 & 3 & 1 & 0 & 0 & 4 & 0 & 0 & 0 & 0 \\ 
        10 & 26 & 158 & 1 & 6 & 0 & 0 & 0 & 5 & 7 & 6 & 2 & 0 & 0 & 2 & 0 & 0 & 0 & 0 \\ 
        11 & 17 & 308 & 0 & 2 & 0 & 0 & 0 & 6 & 12 & 7 & 4 & 0 & 0 & 5 & 0 & 0 & 0 & 0 \\ 
        12 & 69 & 749 & 1 & 3 & 0 & 3 & 4 & 11 & 16 & 13 & 9 & 0 & 3 & 7 & 0 & 0 & 0 & 0 \\ 
        13 & 50 & 1481 & 0 & 15 & 0 & 2 & 1 & 16 & 29 & 20 & 8 & 0 & 2 & 8 & 0 & 0 & 0 & 1 \\ 
        14 & 235 & 3481 & 1 & 6 & 0 & 0 & 1 & 25 & 39 & 32 & 13 & 0 & 2 & 5 & 0 & 0 & 0 & 0 \\ 
        15 & 159 & 6997 & 1 & 14 & 0 & 2 & 2 & 33 & 73 & 43 & 19 & 0 & 1 & 10 & 0 & 0 & 0 & 0 \\ 
        16 & 693 & 16086 & 4 & 46 & 1 & 15 & 14 & 56 & 92 & 71 & 42 & 0 & 0 & 15 & 0 & 0 & 0 & 0 \\ 
        17 & 470 & 32692 & 2 & 22 & 0 & 9 & 6 & 81 & 170 & 109 & 37 & 0 & 0 & 20 & 0 & 0 & 0 & 0 \\ 
        18 & 2142 & 74114 & 10 & 48 & 0 & 0 & 5 & 118 & 209 & 162 & 65 & 0 & 8 & 12 & 0 & 0 & 0 & 1 \\ 
        19 & 1399 & 152114 & 5 & 133 & 0 & 9 & 9 & 162 & 405 & 226 & 78 & 0 & 4 & 23 & 0 & 0 & 0 & 2 \\ 
        20 & 6350 & 341410 & 7 & 66 & 3 & 57 & 52 & 263 & 485 & 363 & 181 & 0 & 6 & 36 & 1 & 0 & 0 & 1 \\ 
        21 & 4086 & 705978 & 8 & 161 & 0 & 40 & 22 & 380 & 932 & 553 & 152 & 0 & 4 & 48 & 0 & 0 & 0 & 0 \\ 
        22 & 19273 & 1573062 & 20 & 395 & 0 & 2 & 21 & 556 & 1095 & 811 & 298 & 0 & 1 & 30 & 1 & 0 & 0 & 0 \\ \hline
    \end{tabular}
\label{Table:OEIS}
\end{table}

\subsection{Dependence on sampling depth}

Figure~\ref{Fig:3D_5s31c_10^9_nmer_hist} in the main text shows the size distribution of BD polycubes obtained from $10^9$ rule sets sampled from $S^{3D}_{5s,31c}$. To assess how this distribution depends on sampling depth, Fig.~\ref{Fig:3D_5s31c_10^6_to_10^9_nmer_hist_rule_count_fraction} compares results obtained from $10^6$ to $10^9$ sampled rule sets. Increasing the sampling depth recovers progressively more distinct polycubes (Fig.~\ref{Fig:3D_5s31c_10^6_to_10^9_nmer_hist_rule_count_fraction}(a)). The gain is most prominent at large polycube sizes, where structures typically have low assembly rule probability and therefore require more samples to be found. Complete recovery of the theoretical morphospace is achieved only at small sizes: all $112$ free polycubes of size $6$ are recovered at sampling depths of $10^8$ and $10^9$. Figure~\ref{Fig:3D_5s31c_10^6_to_10^9_nmer_hist_rule_count_fraction}(b) shows the distribution of assembly rule probability across polycube sizes at different sampling depths. Across all panels, small polycubes account for the largest fraction of BD rule sets, followed by a rapid decrease in assembly rule probability with increasing size, with characteristic local peaks at particular sizes (typically multiples of four). Thus, although deeper sampling continues to uncover additional rare structures, the qualitative form of the assembly rule probability distribution is robust to sampling depth.

\subsection{Dependence on dimension and seeding}

Figure~\ref{Fig:2D_3D_5s31c_10^9_nmer_hist} compares the number of distinct structures obtained in two and three dimensions under seeded and unseeded assembly. In all four cases, the number of distinct assembled structures initially increases with size, reaches a maximum at intermediate sizes, and then decreases, despite the rapid growth of the theoretical morphospace with size (see Table~\ref{table:free_polycubes_polyominoes}). Consequently, an increasingly small fraction of the available morphospace is recovered at larger sizes.

Seeded assembly produces more distinct structures than unseeded assembly in both dimensions, consistent with the larger fraction of BD rule sets under seeded growth (main-text Fig.~\ref{Fig:UND_assembly}(c)). The three-dimensional system generates more distinct structures at larger sizes than the two-dimensional system, but recovers a substantially smaller fraction of the corresponding theoretical morphospace.  The distributions also exhibit local peaks at particular structure sizes, several of which occur at multiples of four. 

\subsection{Size and symmetry}

Although assembly rule probability $P_{AR}(n)$ decreases with size $n$ overall, this decrease is punctuated by pronounced local peaks (main-text Fig.~\ref{Fig:3D_5s31c_10^9_nmer_hist}).
As discussed in the main text, a natural explanation is that highly symmetric structures, which can have low assembly complexity, are available at these sizes. 
Figure~\ref{Fig:mean_sym_order_vs_size} compares $P_{AR}(n)$ with the mean symmetry order of the assembled polycubes, weighted by the number of BD rule sets generating each structure. Several local peaks in $P_{AR}(n)$, particularly at larger sizes, coincide with elevated mean symmetry order. At large sizes, however, few structures are sampled, so the mean symmetry order at a given size can be dominated by a small number of structures.

We also resolve the size distributions by symmetry classes and orders. We write $P_M(n\mid SC)$ and $P_{AR}(n\mid SC)$ for the morphospace and assembly rule probabilities of size $n$ within symmetry class $SC$, and $P_M(n\mid SO)$ and $P_{AR}(n\mid SO)$ for the corresponding probabilities within symmetry order $SO$. The morphospace probabilities are calculated from the OEIS enumerations in  Table~\ref{Table:OEIS}. Since these are available for all $33$ symmetry classes only up to size $22$, both distributions are restricted to $n\leq 22$ and normalized over this range.

Figure~\ref{Fig:3D_OEISxsamp_size_foreachsymclass} shows these distributions for each symmetry class. The morphospace distributions vary substantially between classes. Low-symmetry classes are generally increasingly dominated by larger polycubes, reflecting the rapid growth in the number of geometrically possible structures with size. In contrast, several higher-symmetry classes, such as $H_6$, $AB_8$, and $CCC_{24}$, are only possible at particular polycube sizes. Three classes, $BD_{12}$, $R_{24}$, and $DEE_{24}$, have no possible structures up to size $22$. These features reflect geometric constraints on the sizes at which particular symmetries can be realized. The distributions obtained by random sampling of assembly rules (red) differ markedly from those of the underlying morphospace (blue). BD assemblies are generally biased towards smaller polycubes, while several symmetry classes, including $A_4$, $J_4$, $FF_6$, and $BBC_{16}$, show enhanced probabilities at particular sizes. 

Figure~\ref{Fig:3D_OEISxsamp_size_foreachsymorder} shows the corresponding distributions grouped by symmetry order. At low orders, the morphospace is increasingly dominated by large sizes, while BD assemblies are biased towards small sizes. At high orders, only certain sizes support the required symmetries, so both distributions are concentrated at a few sizes. In addition, the BD distributions for orders $4$, $8$ and $16$ show peaks at sizes that are multiples of four, and those for order $3$ show peaks at multiples of three, whereas the morphospace distributions show no such peaks. At these sizes, a fourfold or threefold rotation axis can pass between cubes rather than through them, so every cube belongs to a group of four or three symmetry-related positions that can share a tile species. At other sizes, some cubes must lie on the axis and gain no such reuse, so the corresponding structures require more assembly information and are generated less often.

\subsection{Morphospace of size-$6$ polycubes and polyominoes}

\begin{figure*}
    \centering
    \includegraphics[width=1\linewidth]{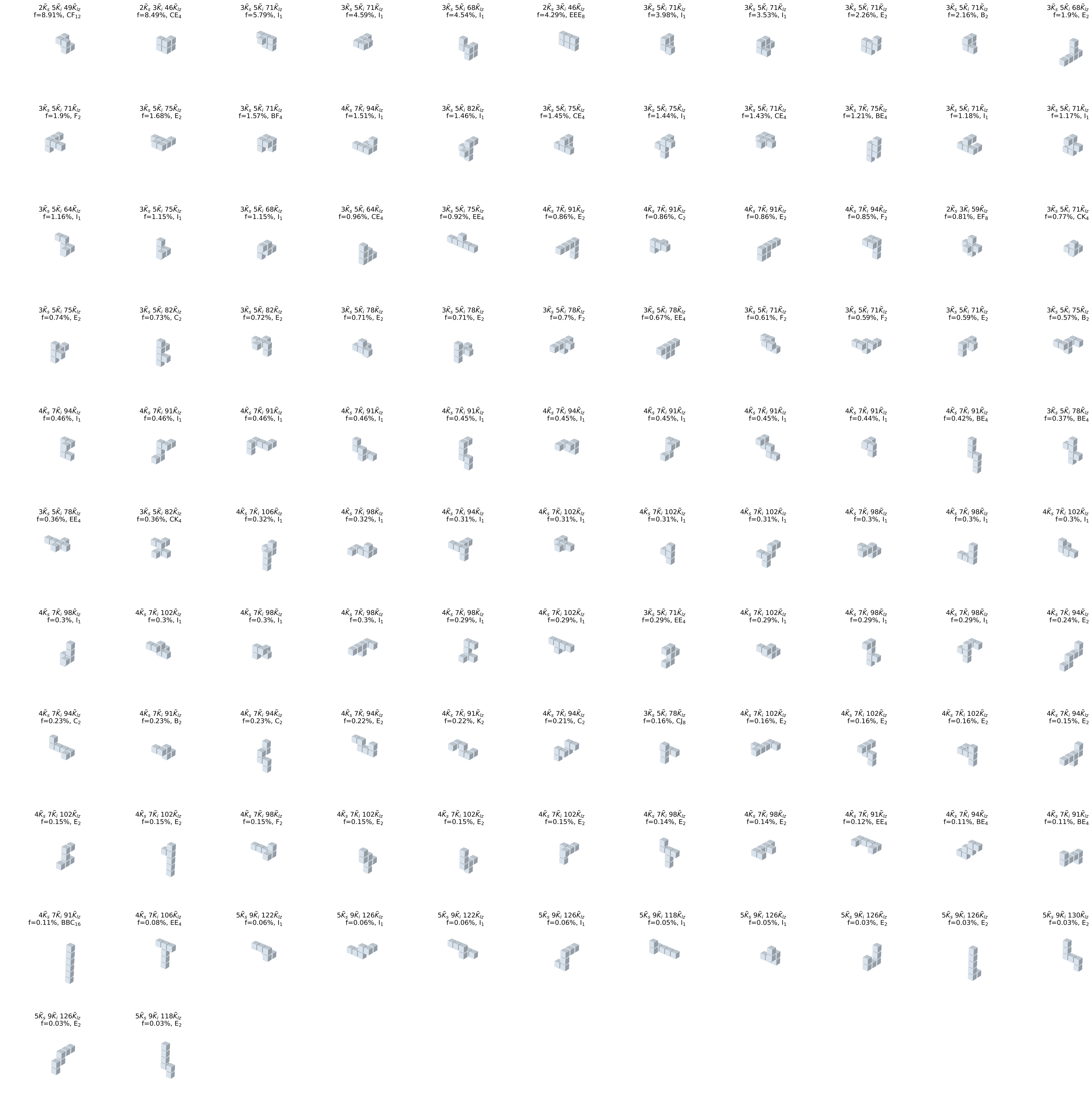}
    \caption{
    \textbf{All free polycubes of size $6$ generated by bounded-deterministic (BD) rule sets.}
    All $112$ possible free polycubes of size $6$ are recovered from $10^9$ randomly sampled rule sets in $S^{3D}_{5s,31c}$ under seeded assembly. Each polycube is labeled with its species complexity $\tilde{K}_s$, interface complexity $\tilde{K}_i$, Lempel--Ziv complexity $\tilde{K}_{LZ}$, assembly frequency $f$, and symmetry class, with symmetry order shown as a subscript. Here, $f$ is the size-normalized assembly rule probability $P_{AR}(p\mid n=6)$ expressed as a percentage.}
    \label{fig:6mer_zoo_3D}
\end{figure*}

\begin{figure*}
    \centering
    \includegraphics[width=0.8\linewidth]{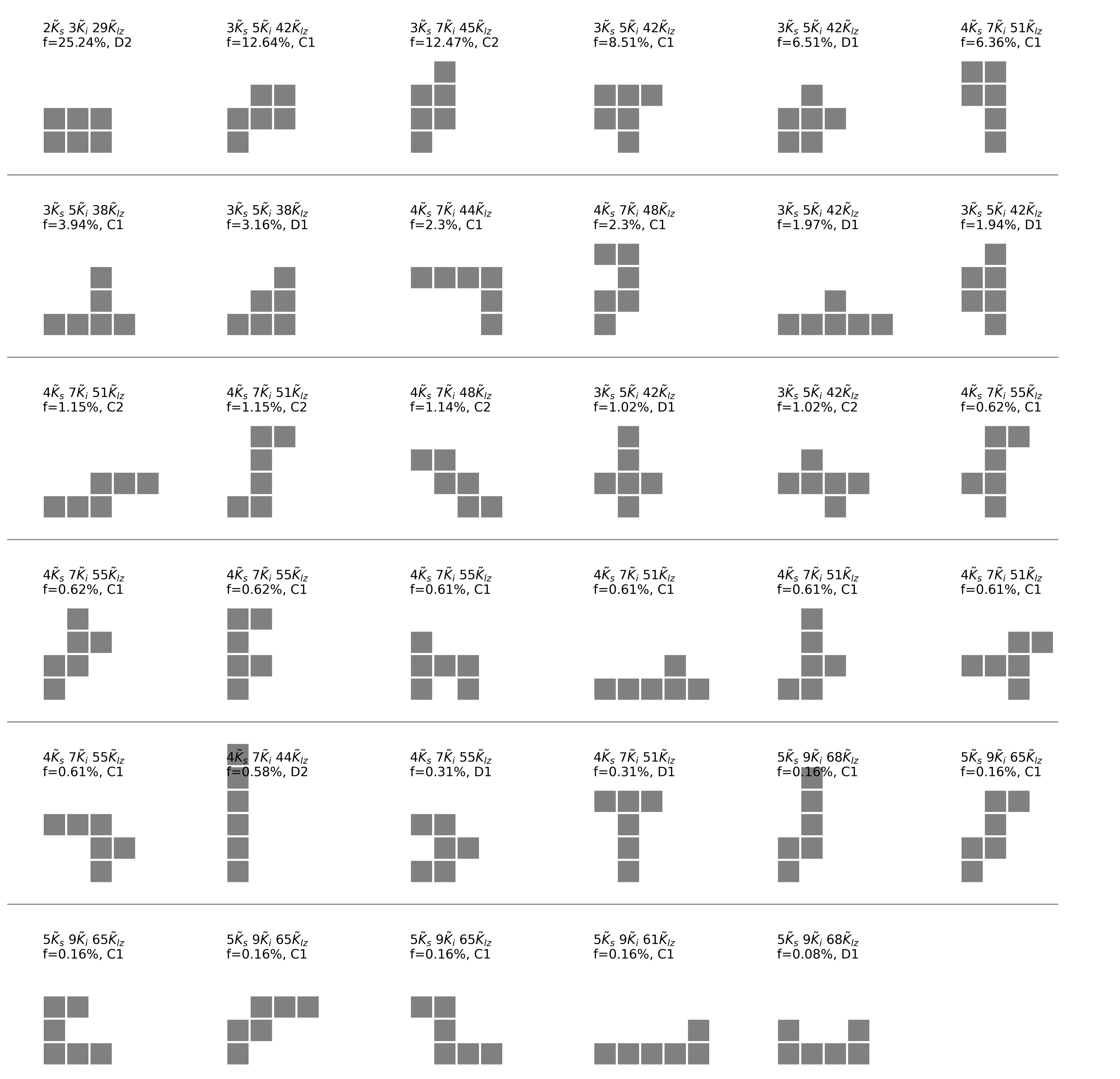}
    \caption{
    \textbf{All free polyominoes of size $6$ generated by bounded-deterministic (BD) rule sets.}
    All $35$ possible free polyominoes of size $6$ are recovered from $10^9$ randomly sampled rule sets in $S^{2D}_{5s,31c}$ under seeded assembly. Each polyomino is labeled with its species complexity $\tilde{K}_s$, interface complexity $\tilde{K}_i$, Lempel--Ziv complexity $\tilde{K}_{LZ}$, assembly frequency $f$, and symmetry class. Here, $f$ is the size-normalized assembly rule probability $P_{AR}(p\mid n=6)$ expressed as a percentage.
    }
    \label{fig:6mer_zoo_2D}
\end{figure*}

Size $6$ is the largest size at which the full morphospace is recovered by sampling $10^9$ rule sets in both two- and three-dimensional rule spaces, $S^{2D}_{5s,31c}$ and $S^{3D}_{5s,31c}$, respectively. Figures~\ref{fig:6mer_zoo_3D} and~\ref{fig:6mer_zoo_2D} show all $112$ free polycubes and all $35$ free polyominoes of this size, respectively, ordered by decreasing assembly rule
probability and labeled by their complexities and symmetries.

\clearpage
\section{Assembly rule probability versus assembly complexity}
\label{SM:prob_vs_comp}

Dingle \emph{et al.}~\cite{Dingle2018} showed that, for a broad class of computable input--output maps, the probability $P(p)$ of obtaining an output $p$ upon random sampling of inputs is bounded from above by a function that decreases exponentially with the complexity of $p$, $P(p)\lesssim 2^{-a\tilde{K}(p)-b}$, where $\tilde{K}(p)$ is a computable approximation to the Kolmogorov complexity and $a>0$ and $b$ are map-dependent constants. Moreover, most of the probability weight is expected to lie relatively close to this bound~\cite{Dingle2020}. Here we examine both properties for the polycube self-assembly map.

\subsection{Empirical upper bounds}
Figure~\ref{Fig:freq_vs_comp_with_fit} shows $\log P_{AR}(p)$ against species complexity $\tilde{K}_s$, interface complexity $\tilde{K}_i$, and Lempel--Ziv complexity $\tilde{K}_{LZ}$ for three-dimensional (top two rows) and two-dimensional (bottom two rows) assembly, each under seeded and unseeded growth. In all panels, the highest probability at each complexity decreases approximately linearly with complexity on the logarithmic scale, consistent with an exponential upper bound on $P_{AR}$. The dashed lines are fitted upper bounds that lie on or above all plotted points and minimize the mean vertical distance to them. Structures generated by fewer than three BD rule sets are omitted since their probabilities cannot be estimated reliably from so few occurrences.

\subsection{Distribution of probability weight below the bound}
To see how probability is distributed below the bound, we sort the structures in Fig.~\ref{Fig:freq_vs_comp_with_fit} at each complexity value by decreasing $P_{AR}(p)$ and color in blue the most probable structures that together account for at least $50\%$ of the probability weight at that value; the remaining structures are shown in
yellow. At most complexities, the blue points lie close to the bound (on a log scale), while the more numerous yellow points extend over several orders of magnitude below it. Half of the probability weight is therefore carried by a few structures near the bound.

\subsection{Dependence on sampling depth}
\begin{figure*}[!htbp]
    \includegraphics[width=\textwidth]{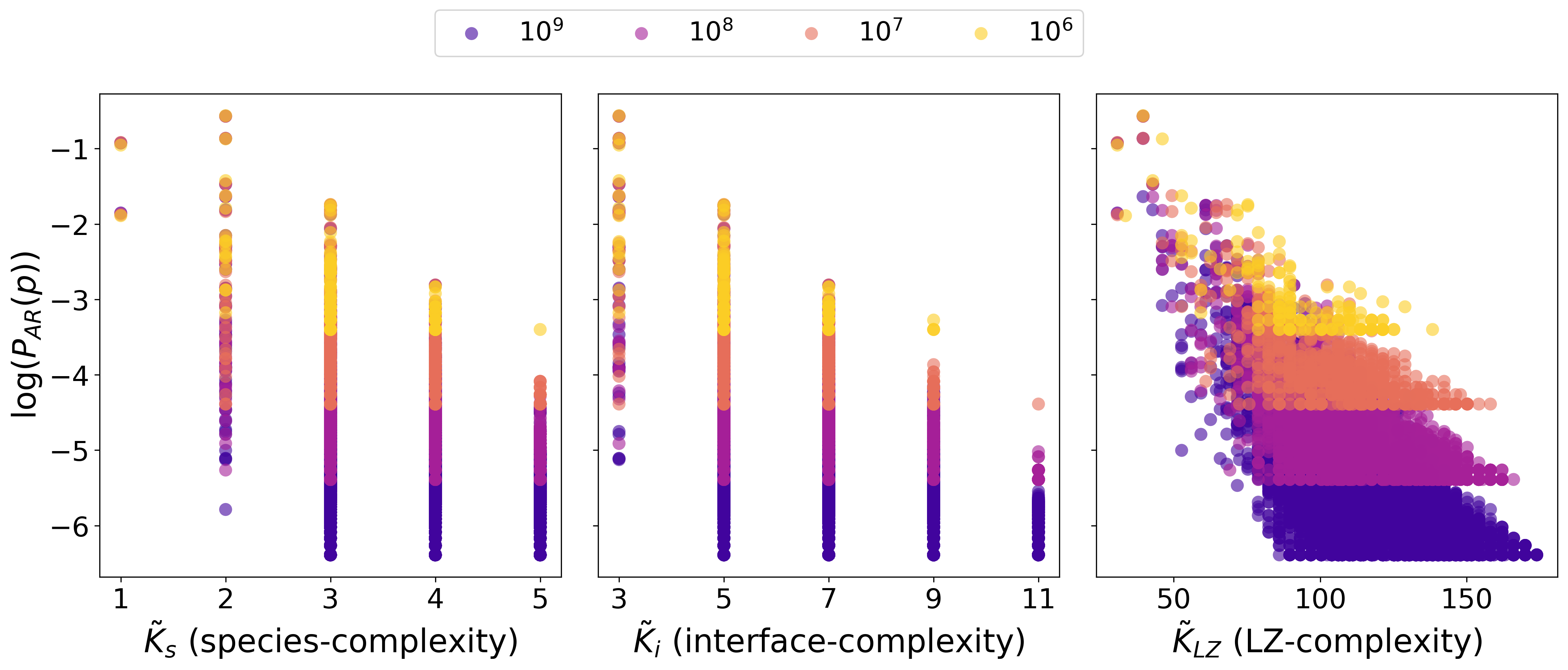}
    \caption{
    \textbf{Assembly rule probability versus assembly complexity at different sampling depths.}
    The logarithm of assembly rule probability $P_{AR}(p)$ is shown against species complexity $\tilde{K}_s$ (left), interface complexity $\tilde{K}_i$ (middle), and Lempel--Ziv complexity $\tilde{K}_{LZ}$ (right) for polycubes generated by bounded-deterministic (BD) rule sets in the three-dimensional rule space $S^{3D}_{5s,31c}$ under seeded assembly. Colors indicate the number of randomly sampled rule sets (legend), each depth being a separate sample. Single cubes are excluded, and $P_{AR}(p)$ is normalized over the remaining BD rule sets at each depth. Structures generated by fewer than three BD rule sets at each sampling depth are also excluded.}
    \label{Fig:freq_vs_comp_nrules}
\end{figure*}

Figure~\ref{Fig:freq_vs_comp_nrules} shows $\log P_{AR}(p)$ against complexity for $10^6$, $10^7$, $10^8$, and $10^9$ sampled rule sets in $S^{3D}_{5s,31c}$ under seeded assembly.
The main effect of shallower sampling is a shortening of the low-probability, high-complexity tail of the distribution, for two reasons. First, $\tilde K_s$, $\tilde K_i$, and $\tilde K_{LZ}$ are each estimated as the minimum complexity among the sampled BD rule sets generating a given structure, so fewer samples make it less likely that the true minimal rule set for a given structure has been found, inflating its apparent complexity. Second, and more importantly, a structure observed only a few times at low sampling depth is intrinsically hard to assign a reliable $P_{AR}(p)$, and such structures are systematically under-represented until sampling is deep enough to resolve them (structures generated by fewer than three BD rule sets are excluded from each panel for this reason). 

\subsection{Simplicity bias at fixed polycube size}

\begin{figure*}[!htbp]
    \includegraphics[width=\textwidth]{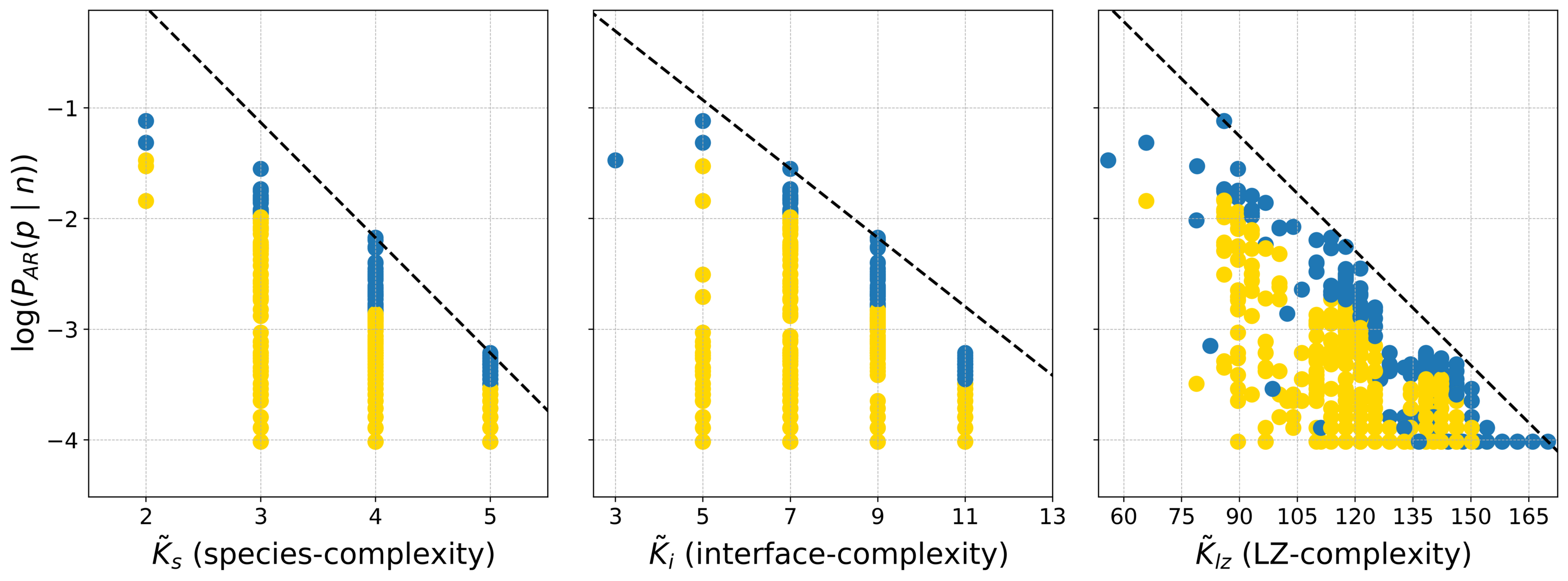}
    \caption{
    \textbf{Upper bounds on assembly rule probability as a function of assembly complexity, for size-$20$ polycubes.}
    The logarithm of the size-normalized assembly rule probability, $\log P_{AR}(p\mid n)$, is shown against species complexity $\tilde{K}_s$ (left), interface complexity $\tilde{K}_i$ (middle), and Lempel--Ziv complexity $\tilde{K}_{LZ}$ (right) for size-$20$ polycubes generated by bounded-deterministic (BD) rule sets in $S^{3D}_{5s,31c}$ under seeded assembly. $P_{AR}(p\mid n)$ is normalized over all size-$20$ structures. Structures generated by fewer than three BD rule sets are omitted. Dashed lines show the fitted upper bound on each probability--complexity distribution. At each complexity value, blue points collectively account for the upper $50\%$ of the probability weight, while yellow points account for the remaining $50\%$.
    }
    \label{Fig:freq_vs_comp_20mer_with_fit}
\end{figure*}

Larger structures generally have higher assembly complexity and lower assembly rule probability (main-text Fig.~\ref{Fig:2D_3D_freq_vs_comp_for_all_complexities}), so the observed relationship between complexity and probability may partly reflect differences in structure size. To separate these effects, Fig.~\ref{Fig:freq_vs_comp_20mer_with_fit} plots the logarithm of the size-normalized assembly rule probability, $\log P_{AR}(p\mid n)$, against assembly complexity for polycubes of fixed size $n=20$ in $S^{3D}_{5s,31c}$ under seeded assembly. The upper envelope decreases approximately linearly with complexity on the logarithmic scale, while the blue points, which together account for half of the assembly rule probability at each complexity, generally lie close to this envelope. Thus, the approximately exponential upper bound characteristic of simplicity bias persists even when structure size is held fixed.

\begin{figure*}[!htbp]
    \includegraphics[width=\textwidth]{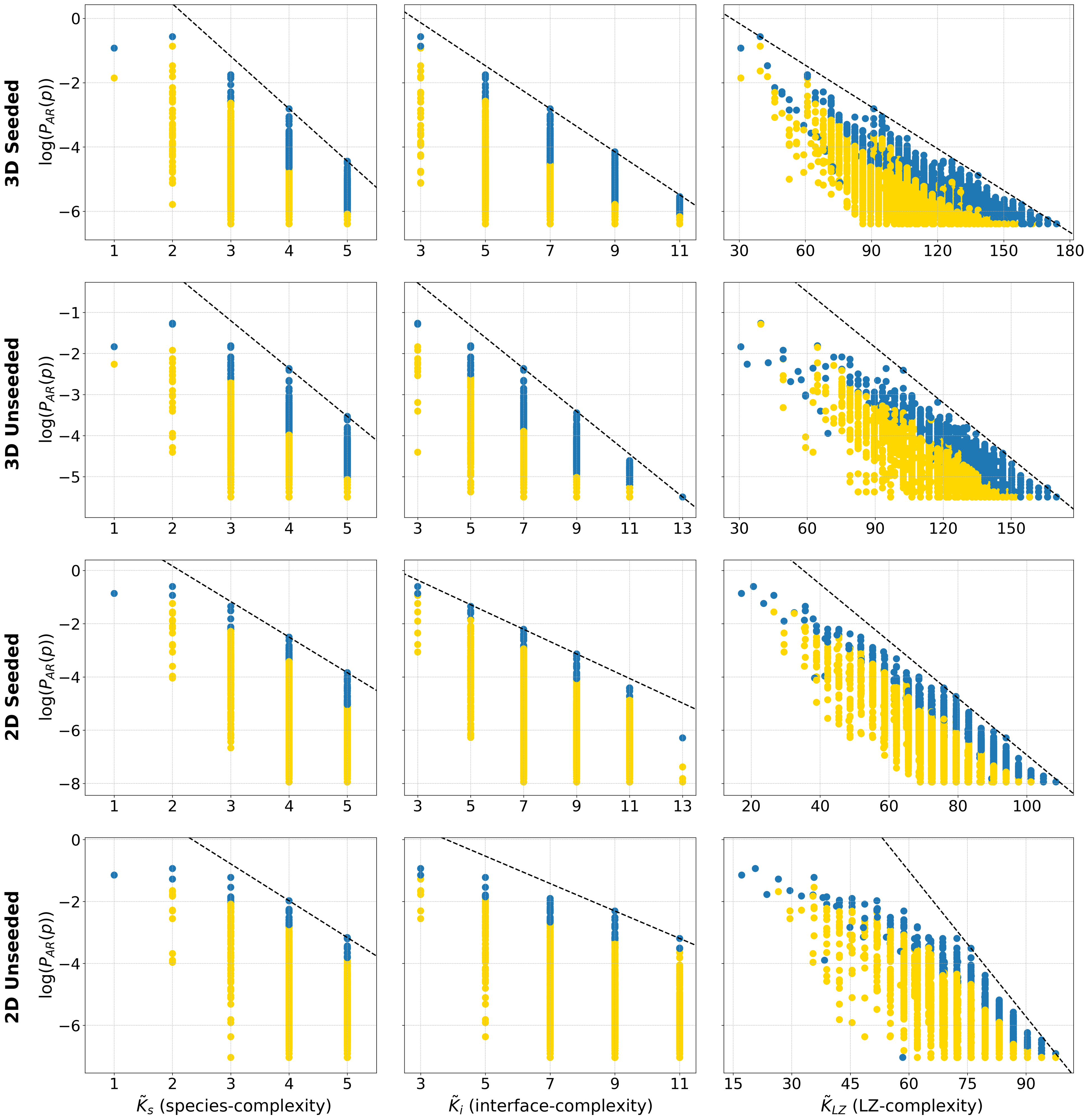}
    \caption{
    \textbf{Upper bound on assembly rule probability as a function of assembly complexity.}
    The logarithm of assembly rule probability, $\log P_{AR}(p)$, is shown against species complexity $\tilde{K}_s$ (left), interface complexity $\tilde{K}_i$ (middle), and Lempel--Ziv complexity $\tilde{K}_{LZ}$ (right). From top to bottom, rows show three-dimensional seeded, three-dimensional unseeded, two-dimensional seeded, and two-dimensional unseeded assembly. Single cubes are excluded, and $P_{AR}(p)$ is normalized over the remaining BD rule sets. Dashed lines show the fitted upper bound on each probability--complexity distribution. At each complexity value, blue points collectively account for the upper $50\%$ of the probability weight, while yellow points account for the remaining $50\%$. Structures generated by fewer than three BD rule sets are excluded from the plot.}
    \label{Fig:freq_vs_comp_with_fit}
\end{figure*}

\clearpage
\section{Rank, complexity and symmetry distributions for assemblies of fixed size}

\subsection{Rank distributions}

\begin{figure*}[!htbp]
    \includegraphics[width=\textwidth]{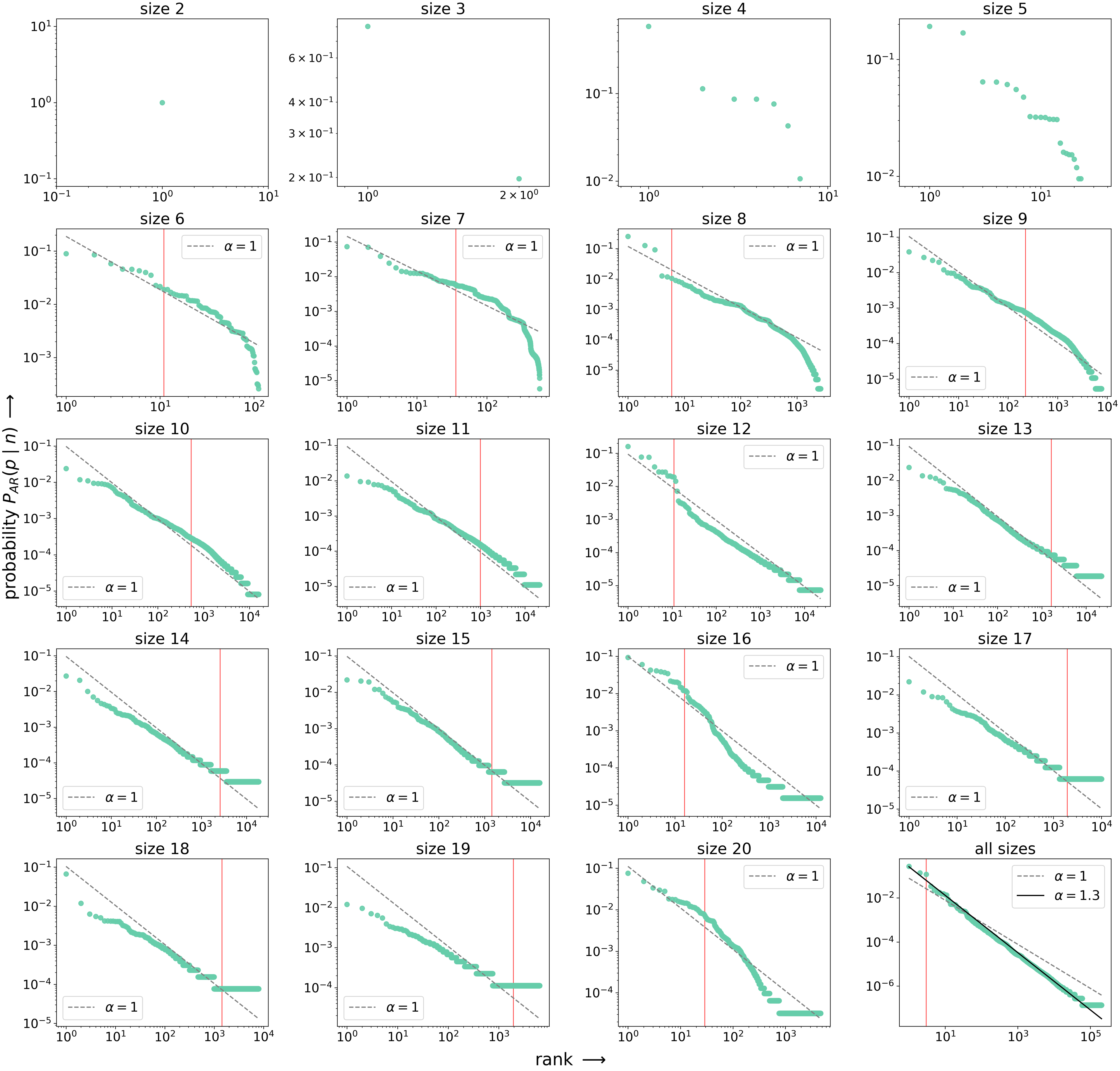}
    \caption{
    \textbf{Assembly rule probability versus rank for polycubes of fixed size.}
    For each polycube size $n$ from \(2\) to \(20\), distinct polycubes generated by bounded-deterministic (BD) rule sets are ordered from highest to lowest size-normalized assembly rule probability \(P_{AR}(p\mid n)\). In the last panel, all sizes are combined and $P_{AR}(p)$ is normalized over all BD rule sets except those producing a single-tile assembly. Results are obtained from \(10^9\) randomly sampled rule sets in the three-dimensional rule space \(S^{3D}_{5s,31c}\) under seeded assembly. The gray dashed line is the power law $P(r)=r^{-\alpha}/H_{N,\alpha}$ with exponent $\alpha=1$ (Zipf's law), where $r$ is the rank, $H_{N,\alpha}=\sum_{r=1}^{N}r^{-\alpha}$, and $N$ is the number of distinct polycubes in the panel. In the last panel, the black solid line is the same power law with $\alpha=1.3$. The red vertical line marks the rank at which the cumulative assembly rule probability reaches 50\%: the polycubes to its left together account for half of the BD rule sets in that panel (shown for \(n \geq 6\) and for all sizes combined).
    }
    \label{Fig:rank_superplot_3D}
\end{figure*}

\begin{figure*}[!htbp]
    \includegraphics[width=\textwidth]{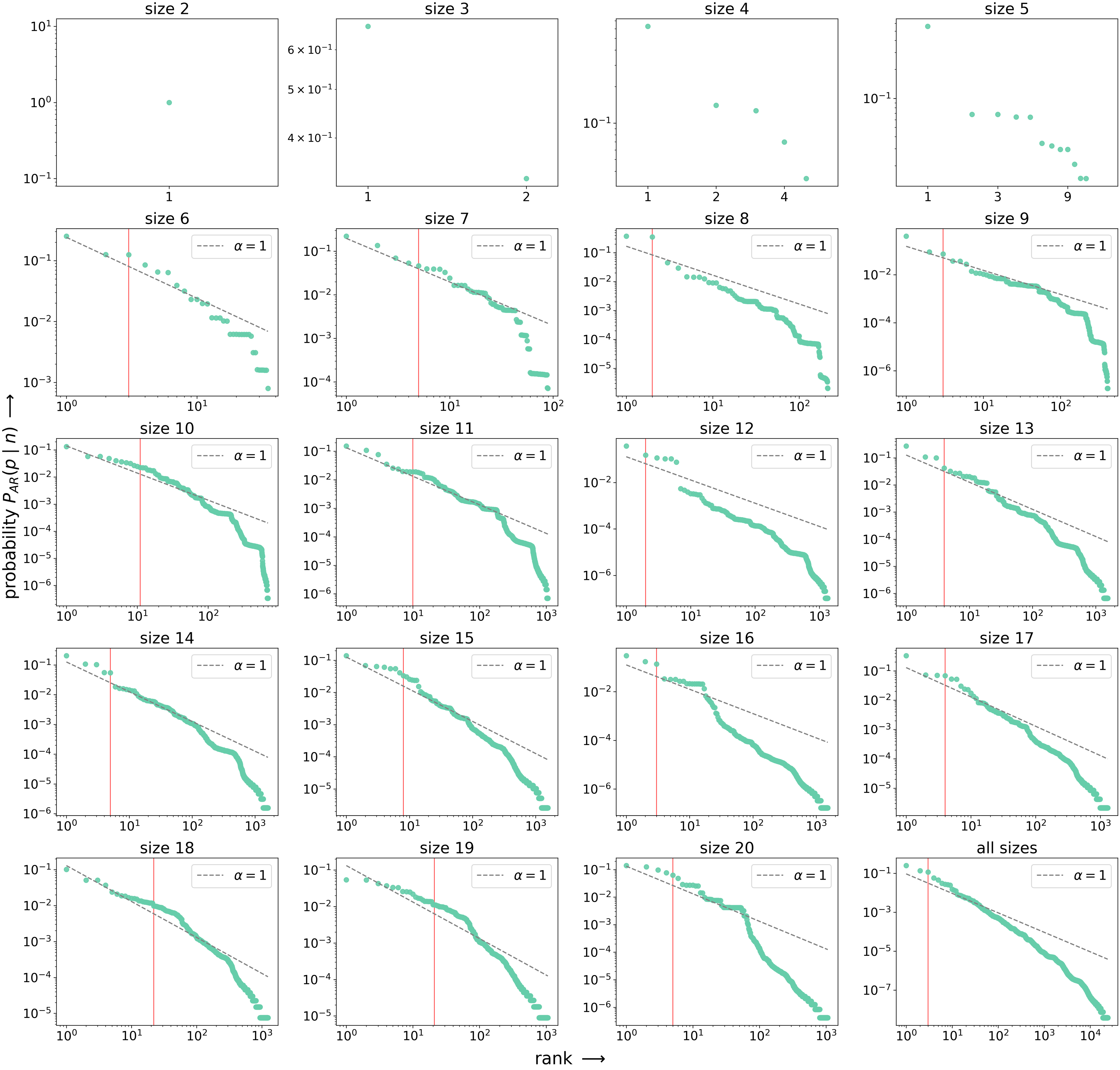}
    \caption{
    \textbf{Assembly rule probability versus rank for polyominoes of fixed size.} For each polyomino size $n$ from $2$ to $20$, distinct polyominoes generated by bounded-deterministic (BD) rule sets are ordered from highest to lowest size-normalized assembly rule probability $P_{AR}(p\mid n)$. In the last panel, all
    sizes are combined and $P_{AR}(p)$ is normalized over all BD rule sets
    except those producing a single-tile assembly. Results are obtained from \(10^9\) randomly sampled rule sets in the two-dimensional rule space \(S^{2D}_{5s,31c}\) under seeded assembly.  The gray dashed line is a reference for Zipf's law with exponent $1$, $P(r)=1/(rH_N)$, where $r$ is the rank, $H_N=\sum_{r=1}^{N}1/r$, and $N$ is the number of distinct polyominoes in the panel. The red vertical line marks the rank at which the cumulative assembly rule probability reaches 50\%: the polyominoes to its left together account for half of the BD rule sets in that panel (shown for \(n \geq 6\) and for all sizes combined).
    }
    \label{Fig:rank_superplot_2D}
\end{figure*}

To examine how assembly rule probability varies among structures of the same size $n$, we rank the distinct structures by their size-normalized assembly rule probability $P_{AR}(p\mid n)$. Figure~\ref{Fig:rank_superplot_3D} shows the resulting rank distributions for polycubes of sizes $2$--$20$ in $S^{3D}_{5s,31c}$, and Fig.~\ref{Fig:rank_superplot_2D} shows the corresponding distributions for polyominoes in $S^{2D}_{5s,31c}$. The last panel of each figure combines all sizes, with $P_{AR}(p)$ normalized over all BD rule sets except those producing a single-tile assembly. 

At sizes $n\leq6$ the distributions contain every possible structure from the theoretical morphospace; at larger sizes only a decreasing fraction of the morphospace is recovered. For $n\geq6$, $P_{AR}(p\mid n)$ decreases rapidly with rank and spans several orders of magnitude, so some structures are generated by many more BD rule sets than others and the assembly map does not sample the morphospace  uniformly. 

To characterize the decay, we compare it with power laws $P(r)=r^{-\alpha}/H_{N,\alpha}$, where $r$ is the rank, $N$ is the number of distinct structures in the panel, and $H_{N,\alpha}=\sum_{r=1}^{N}r^{-\alpha}$ is a normalization factor ensuring that the reference probabilities sum to one. For $\alpha=1$, this gives Zipf's law, $P(r)=1/(rH_N)$, shown as the gray dashed line. For the last panel in Fig.~\ref{Fig:rank_superplot_3D}, we also show a steeper power law with $\alpha=1.3$ as a black solid line.

At fixed size, the distributions are not described by a single power law. The flat plateaus at the lowest probabilities correspond to structures generated by only one or a few BD rule sets and reflect the resolution of finite sampling rather than the true form of the distribution. When all polycube sizes are combined, the rank distribution follows an approximate power law with $\alpha\approx1.3$ over about five decades, decaying faster than Zipf's law.

The red vertical lines mark the rank at which the cumulative assembly rule probability reaches 50\%, that is, the number of most probable structures that together account for half of the BD rule sets producing structures of that size. This number is strikingly small at sizes that are multiples of four. For polycubes, half of the BD rule sets at $n=8$, $12$, $16$ and $20$ are carried by fewer than ~$30$ structures, compared with hundreds to thousands at sizes $9$--$19$ that are not multiples of four. As discussed in main-text Sec.~\ref{subsec:why_symmetries}, this reflects the efficient reuse of tile species in structures with fourfold rotational symmetry, which is possible when the rotation axis passes between cubes, and hence when the size is a multiple of four. The same pattern is present but weaker for polyominoes, where the number of structures needed is small at most sizes. 

\subsection{Assembly complexity distributions}

\begin{figure*}
    \centering
    \includegraphics[width=\textwidth]{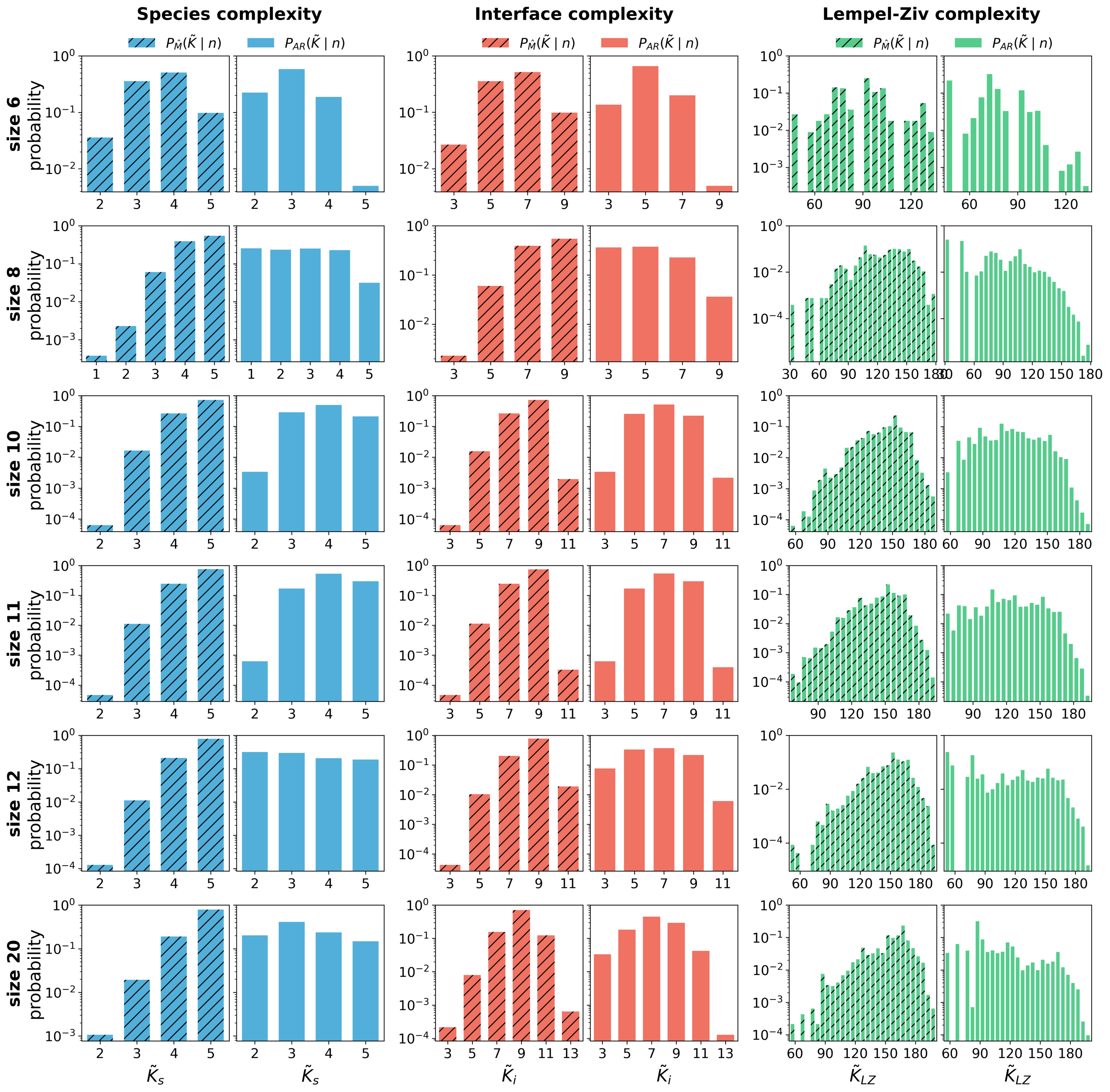}
    \caption{
    \textbf{Assembly complexity distributions across polycube sizes.}
    Distributions are shown for polycubes generated by bounded-deterministic (BD) rule sets from $10^9$ randomly sampled rule sets in the three-dimensional rule space $S^{3D}_{5s,31c}$ under seeded assembly. Rows correspond to polycube sizes $6$, $8$, $10$, $11$, $12$, and $20$. The three column groups show species complexity $\tilde{K}_s$, interface complexity $\tilde{K}_i$, and Lempel--Ziv complexity $\tilde{K}_{LZ}$. Within each group, the left panel shows the sampled morphospace probability $P_{\hat{M}}(\tilde{K}\mid n)$, with each distinct polycube counted once, while the right panel shows the assembly rule probability $P_{AR}(\tilde{K}\mid n)$, obtained by counting the BD rule sets generating polycubes in each complexity bin. All probabilities are shown on a logarithmic scale.
    }
    \label{fig:P(K)_superplot}
\end{figure*}

Figure~\ref{fig:P(K)_superplot} shows assembly complexity distributions for polycubes of sizes $6$, $8$, $10$, $11$, $12$, and $20$; sizes $6$ and $20$ are also shown in the top and middle rows of main-text Fig.~\ref{Fig:P(K)dist}. Results are obtained from $10^9$ randomly sampled rule sets in $S^{3D}_{5s,31c}$ under seeded assembly. For each size, the sampled morphospace probability $P_{\hat{M}}(\tilde{K}\mid n)$ is the fraction of distinct recovered polycubes in each complexity bin, so every structure counts once, whereas the assembly rule probability $P_{AR}(\tilde{K}\mid n)$ is the fraction of BD rule sets producing polycubes in each bin, so every structure counts as often as it is generated. The three column groups show species complexity $\tilde{K}_s$, interface complexity $\tilde{K}_i$, and Lempel--Ziv complexity $\tilde{K}_{LZ}$, the last in bins of width $5$. At size $6$ all $112$ polycubes are recovered, so $P_{\hat{M}}$ equals the theoretical morphospace distribution; at larger sizes it describes only the structures found.

For every size and every complexity measure, $P_{AR}(\tilde{K}\mid n)$ is shifted towards lower complexity than $P_{\hat{M}}(\tilde{K}\mid n)$. At size $6$, for example, $P_{\hat{M}}(\tilde K_s\mid n)$ peaks at $\tilde{K}_s=4$ whereas $P_{AR}(\tilde{K}_s\mid n)$ peaks at $\tilde K_s=3$; moreover, $\tilde K_s=5$ accounts for about $10\%$ of the structures but less than $1\%$ of the BD rule sets. At size $20$, most assembled structures have the maximum species complexity, $\tilde{K}_s=5$. Complex structures therefore greatly outnumber simple ones, but simple structures are generated by many more BD rule sets than complex ones. Since sampling also misses rare high-complexity structures, the true morphospace is expected to be at least as skewed towards high complexity as the sampled one, so the shift of $P_{AR}$ relative to the full morphospace is expected to be at least as large as shown.

\clearpage
\subsection{Symmetry distributions}

\begin{figure*}[t!]
    \includegraphics[width=\textwidth]{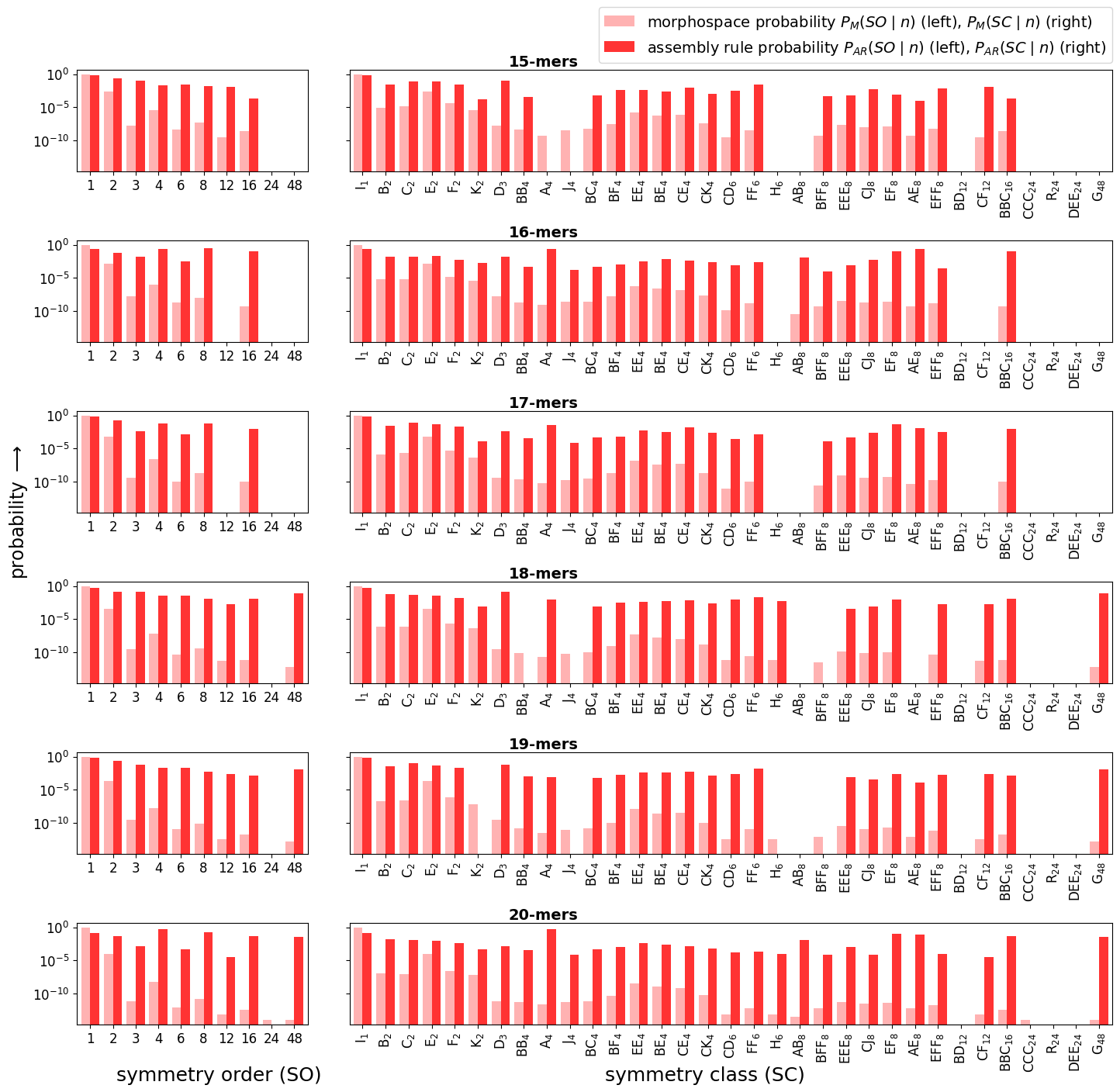}
    \caption{
    \textbf{Symmetry distributions of polycubes at fixed size.}
    Each row corresponds to a fixed polycube size from $15$ to $20$, with assembly rule probabilities obtained from $10^9$ randomly sampled rule sets in the three-dimensional rule space $S^{3D}_{5s,31c}$ under seeded assembly.
    (\textit{Left}) Morphospace probability $P_M(SO\mid n)$ (light red) and assembly rule probability $P_{AR}(SO\mid n)$ (dark red) across symmetry orders ($SO$).
    (\textit{Right}) Morphospace probability $P_M(SC\mid n)$ (light red) and assembly rule probability $P_{AR}(SC\mid n)$ (dark red) across the $33$ symmetry classes ($SC$) listed in main-text Table~\ref{Table:lunnon_sym_classes}.
    }
\label{Fig:3D_prob_order_and_class_5s31c_10^9_15to20mer}
\end{figure*}

Main-text Fig.~\ref{Fig:3D_prob_order_and_class_5s31c_10^9_20mer}(a) compares morphospace and assembly rule probability distributions across symmetry orders and symmetry classes for polycubes of size $20$. Figure~\ref{Fig:3D_prob_order_and_class_5s31c_10^9_15to20mer} extends this comparison to polycube sizes $15$--$20$. Here $P_M$ (light bars) is the theoretical morphospace probability, calculated from the OEIS enumerations (Table~\ref{Table:OEIS}), and $P_{AR}$ (dark bars) is obtained from $10^9$ randomly sampled rule sets under seeded assembly.

Across these sizes, $P_M(SO\mid n)$ falls by many orders of magnitude from order $1$ to the highest orders, reflecting the scarcity of highly symmetric structures in the morphospace. In contrast, higher symmetry orders account for a disproportionately large fraction of the assembly rule probability $P_{AR}(SO\mid n)$. This enrichment is particularly pronounced for symmetry order $48$: at sizes where such structures occur (i.e., $18$-$20$ here), they constitute only a small fraction of the morphospace but account for a substantially larger fraction of the assembly rule probability. Highly symmetric structures are therefore strongly enriched under random sampling of assembly rules.

Resolving the distributions by symmetry class reveals variation that symmetry order alone does not capture. Classes of the same order can differ in $P_{AR}$ by several orders of magnitude: at size $20$, for example, $A_4$ has by far the highest probability among the order-$4$ classes, whereas $J_4$ lies several orders of magnitude lower. The relative probabilities of classes also change with size. Class $A_4$ receives a large share of the BD rule sets at sizes $16$ and $20$ but only a very small share at size $19$, and none of its possible structures of size $15$ was found in our sampling. At sizes that are not multiples of four, a fourfold rotation axis must pass through some cubes, which gain no reuse from the symmetry. How much a symmetry reduces assembly complexity therefore depends not only on the symmetry itself, but also on whether the size and topology of the structure allows every cube to share its tile species with symmetry-related copies.

\begin{figure*}[t!]
    \centering
    \includegraphics[width=0.8\textwidth]{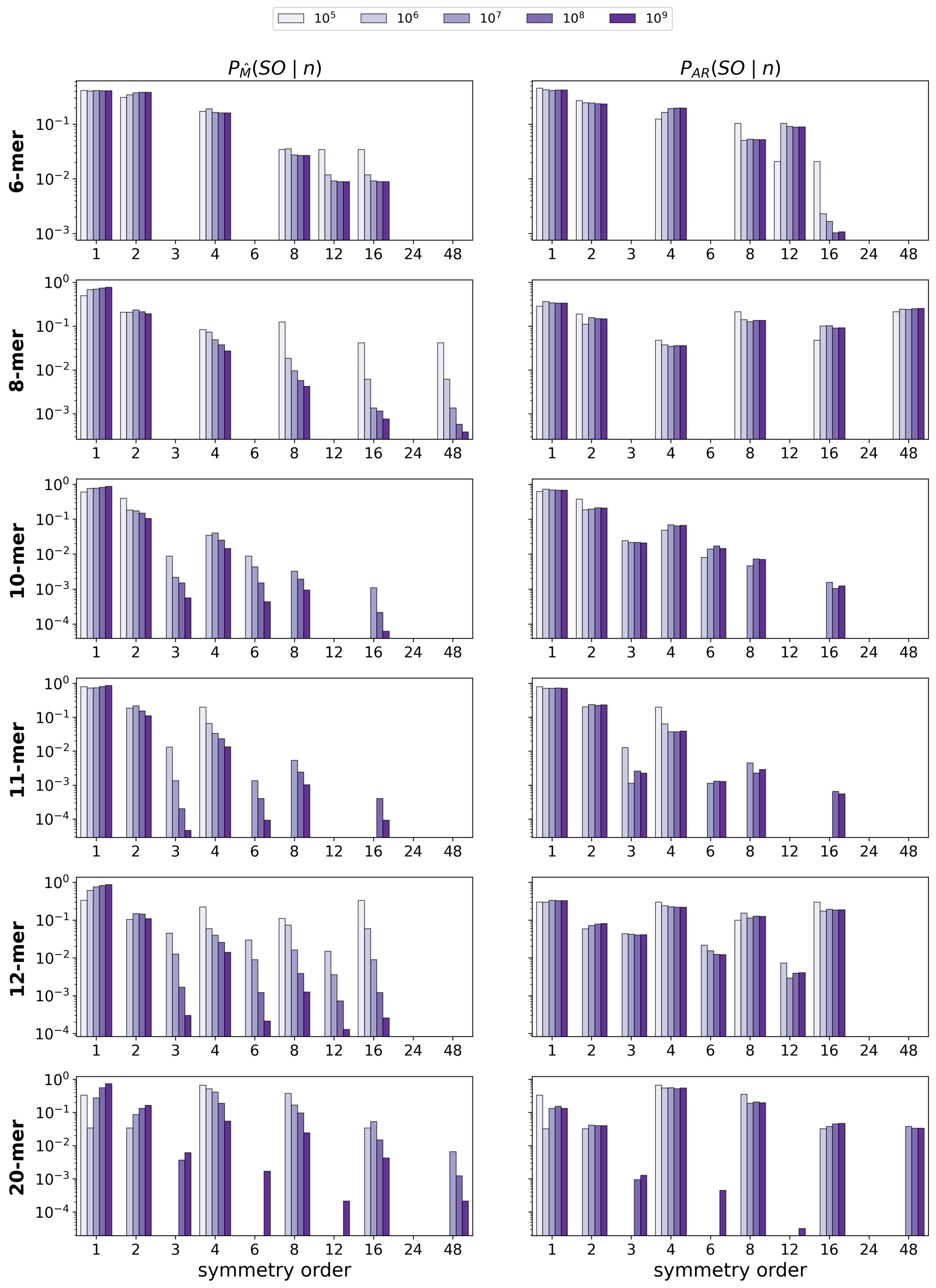}
    \caption{
    \textbf{Symmetry order distributions across polycube sizes and sampling depths.}
    Distributions are shown for polycubes generated by bounded-deterministic (BD) rule sets at sampling depths from $10^5$ to $10^9$ randomly sampled rule sets in the three-dimensional rule space $S^{3D}_{5s,31c}$ under seeded assembly. Each row corresponds to a fixed polycube size, and colors indicate sampling depth. The left column shows the sampled morphospace probability $P_{\hat{M}}(SO\mid n)$ and the right column shows the assembly rule probability $P_{AR}(SO\mid n)$ across symmetry orders ($SO$). Increasing the sampling depth continues to change the sampled morphospace distributions, particularly at larger polycube sizes, while the assembly rule probability distributions show comparatively smaller changes at the largest sampling depths. For size $6$, all $112$ possible free polycubes are recovered at sampling depths of $10^8$ and $10^9$.
    }
    \label{Fig:P(symorder)_nrules_diffsizes}
\end{figure*}

Figure~\ref{Fig:P(symorder)_nrules_diffsizes} shows the dependence of symmetry order distribution on sampling depth for sizes $6$, $8$, $10$, $11$, $12$, and $20$, from $10^5$ to $10^9$ sampled rule sets. Symmetry order is fixed by the geometry of a polycube once it is identified, so, unlike minimal assembly complexity, it does not change when further rule sets generating the same structure are found. The sampled
morphospace distribution $P_{\hat M}(SO\mid n)$ (left) nevertheless changes strongly with depth at larger sizes, because deeper sampling keeps recovering additional, mostly low-symmetry, structures. By contrast, $P_{AR}(SO\mid n)$ (right) is essentially unchanged at the deepest depths, so $10^9$ sampled rule sets are sufficient to determine it.

\begin{figure*}
\centering
\includegraphics[width=\textwidth]{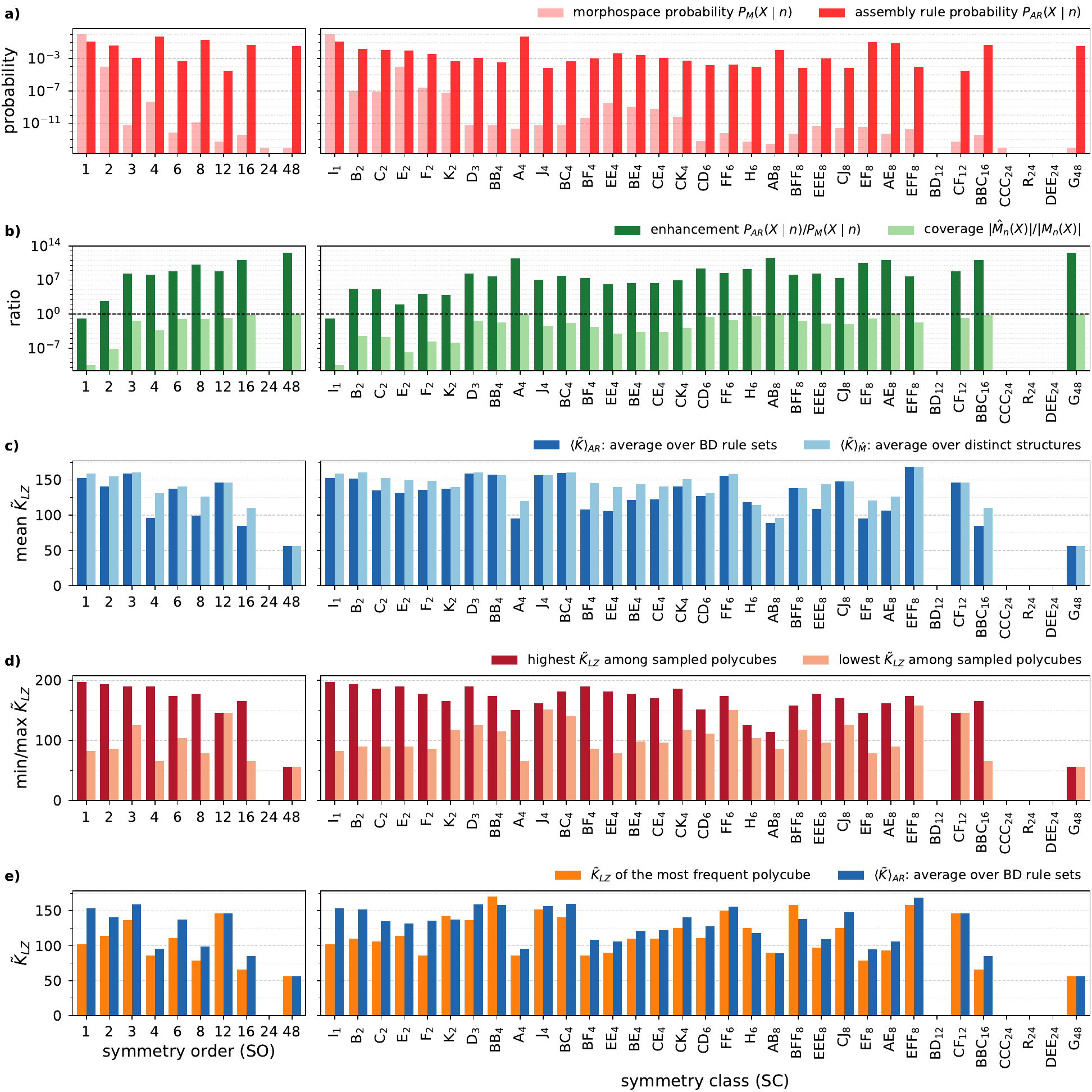}
\caption{
\textbf{Extended version of main-text Fig.~\ref{Fig:3D_prob_order_and_class_5s31c_10^9_20mer} for polycubes of size $n = 20$.}
Panels (a)--(c) are as in main-text Fig.~\ref{Fig:3D_prob_order_and_class_5s31c_10^9_20mer}: (a) morphospace and assembly rule probabilities, (b) enhancement factor $P_{AR}(X \mid n)/P_M(X \mid n)$ and morphospace coverage $|\hat{M}_n(X)|/|M_n(X)|$, and (c) mean Lempel--Ziv complexities $\langle \tilde{K}_{LZ} \rangle_{AR}$ and $\langle \tilde{K}_{LZ} \rangle_{\hat{M}}$, for each symmetry order (left) and class (right).
(d) Highest and lowest $\tilde{K}_{LZ}$ among the sampled polycubes in each symmetry order or class.
(e) $\tilde{K}_{LZ}$ of the most frequently assembled polycube in each symmetry order or class, compared with $\langle \tilde{K}_{LZ} \rangle_{AR}$.
}
\label{Fig:3D_prob_enhance_complexity_minmax_top_20mer}
\end{figure*}

\begin{figure*}
\centering
\includegraphics[width=\textwidth]{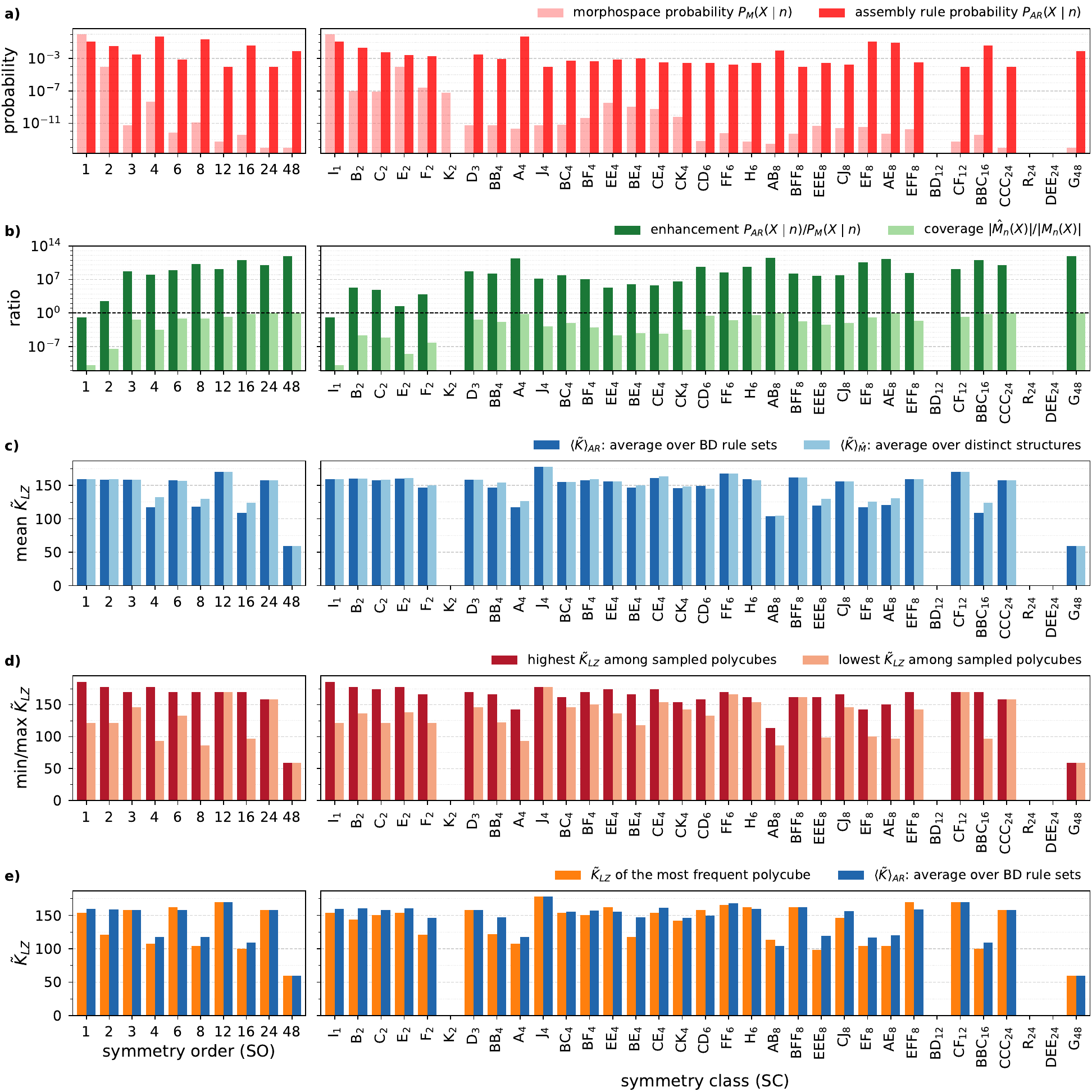}
\caption{
\textbf{Unseeded version of Fig.~\ref{Fig:3D_prob_enhance_complexity_minmax_top_20mer}.}
Panels (a)--(e) are as in Fig.~\ref{Fig:3D_prob_enhance_complexity_minmax_top_20mer}, but for polycubes of size $n=20$ obtained under unseeded assembly in $S^{3D}_{5s,31c}$.
}
\label{Fig:3D_prob_enhance_complexity_minmax_top_20mer_unseeded}
\end{figure*}

Figure~\ref{Fig:3D_prob_enhance_complexity_minmax_top_20mer} extends the symmetry analysis of main-text Fig.~\ref{Fig:3D_prob_order_and_class_5s31c_10^9_20mer} by showing the minimum and maximum Lempel--Ziv complexities and the complexity of the most frequently assembled structure within each symmetry order and class. The corresponding results under unseeded assembly are shown in Fig.~\ref{Fig:3D_prob_enhance_complexity_minmax_top_20mer_unseeded}.

\clearpage
\section{Calculating assembly complexity of simple structures}
\label{SM:tile_comp}

In this section we derive the minimal species complexity $\tilde{K}_s(N)$ and interface complexity $\tilde{K}_i(N)$ (including neutral interface type) for deterministic assembly of two simple structures of size $N$: a straight rod, and an L-shape assembly obtained by moving one end cube of the rod perpendicularly. In the main text these quantities are estimated as minima over sampled rule sets, so they are upper bounds on the true minima; here we obtain the exact values.

\paragraph*{Setting.}
As in the main text, each non-neutral (active) interface type binds only to its complementary partner, which is distinct from itself. Tile species are identified up to the $24$ proper rotations of the cube (classes $\boldsymbol{I}$–$\boldsymbol{D}$ in main-text Table~\ref{Table:cubic_operations}). Orientations play no role here: orientation matching only selects the rotation of an attaching cubic tile about the bond axis and never prevents two complementary faces from binding. Both structures considered here are planar, so the results hold in both the two- and three-dimensional models, and the arguments apply to both seeded and unseeded assembly.

Both structures are paths of $N$ tiles, in which each tile is bonded to its neighbors along the path and to nothing else. For growth to stop at the target structure, every face that does not form a bond must be neutral. Each end tile therefore has one active face and each interior tile has two, so an end tile can never be the same species as an interior tile.

\subsection{Linear assembly}
\label{SM_subsec:linear}

\begin{proposition}
\label{prop:rod}
For a linear rod of $N$ tiles,
\begin{equation}
\label{eqn_linear_assembly}
\tilde{K}_s(N)=\lfloor N/2\rfloor+1,
\qquad
\tilde{K}_i(N)=2\lfloor N/2\rfloor+1,
\end{equation}
that is, $\tilde{K}_s=(N+1)/2$ and $\tilde{K}_i=N$ for odd $N$, and
$\tilde{K}_s=N/2+1$ and $\tilde{K}_i=N+1$ for even $N$.
\end{proposition}

\begin{figure*}[tbp]
\centering
\begin{tikzpicture}[x=1.25cm,y=1.25cm]
 
\pcpanel{-0.3}{1.55}{a}
\pcnote{0.4}{1.55}{Rod, $N=4$: 3 species, 5 interface types}
\pctile{0}{0}{spA}{T_1}{}{a}{}{}
\pctile{1}{0}{spB}{T_2}{}{c}{}{b}
\pctile{2}{0}{spC}{T_3}{}{b}{}{d}
\pctile{3}{0}{spA}{T_4}{}{}{}{a}
\draw[dashed,gray] (2,-0.15) -- (2,1.15);
\draw[-{Stealth[length=4pt]},gray] (2.2,-0.35) arc[start angle=0,end angle=300,radius=0.2];
\pcnote{4.2}{0.5}{\parbox{2.6cm}{\scriptsize central bond maps to itself:\\ $T_2\neq T_3$}}
 
\pcpanel{7.2}{1.55}{b}
\pcnote{7.9}{1.55}{Rod, $N=5$: 3 species, 5 interface types}
\pctile{7.5}{0}{spA}{T_1}{}{a}{}{}
\pctile{8.5}{0}{spB}{T_2}{}{c}{}{b}
\pctile{9.5}{0}{spD}{T_3}{}{d}{}{d}
\pctile{10.5}{0}{spB}{T_4}{}{b}{}{c}
\pctile{11.5}{0}{spA}{T_5}{}{}{}{a}
\draw[dashed,gray] (10,1.0) -- (10,1.2); \draw[dashed,gray] (10,-0.15) -- (10,0);
\draw[-{Stealth[length=4pt]},gray] (10.2,-0.35) arc[start angle=0,end angle=300,radius=0.2];
 
\pcpanel{-0.3}{-1.15}{c}
\pcnote{0.4}{-1.15}{Rod, further reuse in $N=4$ destroys determinism}
\pcnote{0}{-1.75}{\scriptsize $c=a$: repeat (unbounded)}
\pctile{0}{-3}{spA}{T_1}{}{a}{}{}
\pctile{1}{-3}{spB}{T_2}{}{a}{}{b}
\pctile{2}{-3}{spB}{T_2}{}{a}{}{b}
\pctile{3}{-3}{spB}{T_2}{}{a}{}{b}
\node[font=\large] at (4.45,-2.5) {$\cdots$};
\pcnote{7.5}{-1.75}{\scriptsize $c=b$: skip (terminal trimer)}
\pctile{7.5}{-3}{spA}{T_1}{}{a}{}{}
\pctile{8.5}{-3}{spB}{T_2}{}{b}{}{b}
\pctile{9.5}{-3}{spA}{T_1}{}{}{}{a}
\pcnote{10.7}{-2.5}{\scriptsize 3 tiles, not 4}
 
\pcpanel{-0.3}{-3.75}{d}
\pcnote{0.4}{-3.75}{Rod, opposite-order reuse at bonds 1 and 3 ($N=5$)}
\pcnote{0}{-4.35}{\scriptsize intended}
\pctile{0}{-5.6}{spA}{T_1}{}{a}{}{}
\pctile{1}{-5.6}{spB}{T_2}{}{c}{}{b}
\pctile{2}{-5.6}{spC}{T_3}{}{b}{}{d}
\pctile{3}{-5.6}{spD}{T_4}{}{e}{}{a}
\pctile{4}{-5.6}{spE}{T_5}{}{}{}{f}
\pcnote{0}{-6.2}{\scriptsize turn-around: $T_3,T_2,T_1$ follow in reverse order}
\pctile{0}{-7.45}{spA}{T_1}{}{a}{}{}
\pctile{1}{-7.45}{spC}{T_3}{}{d}{}{b}
\pctile{2}{-7.45}{spB}{T_2}{}{b}{}{c}
\pctile{3}{-7.45}{spA}{T_1}{}{}{}{a}
\pcnote{4.2}{-6.95}{\scriptsize 4 tiles, not 5}
 
\pcpanel{7.2}{-3.75}{e}
\pcnote{7.9}{-3.75}{L-shape, $N=4$: 3 species, 5 interface types}
\pctile{7.5}{-6.15}{spA}{T_1}{}{a}{}{}
\pctile{8.5}{-6.15}{spB}{T_2}{}{c}{}{b}
\pctile{9.5}{-6.15}{spF}{T_3}{b}{}{}{d}
\pctile{9.5}{-5.15}{spA}{T_4}{}{}{a}{}
\pcnote{10.8}{-5.1}{\parbox{2.6cm}{\scriptsize $T_4$ is $T_1$ rotated:\\ the two ends share\\ a species}}
 
\end{tikzpicture}
\caption{
\textbf{Minimal assembly complexity for the rod and L-shape.}
Each square is a tile (in 2D or 3D, the latter viewed from above), labeled by its species; only active faces are labeled, all other faces being neutral, and, within each panel, tiles of the same color belong to the same species, possibly rotated. Complementary pairs are $a\leftrightarrow b$, $c\leftrightarrow d$, and $e\leftrightarrow f$.
(a) Rod of $N=4$ tiles (Proposition~\ref{prop:rod}). The half-turn symmetry (gray) relates $T_1$ and $T_4$, which are therefore the same species, but maps the central bond onto itself, so $T_2$ and $T_3$ must be distinct.
(b) Rod of $N=5$ tiles. The central tile is mapped onto itself and carries $d$ on both faces.
(c) Further reuse in the $N=4$ rule set destroys determinism. Setting $c=a$ lets $T_2$ bind to copies of itself, so growth is unbounded, whereas setting $c=b$ allows a terminal trimer.
(d) If the pair $a\leftrightarrow b$ is used in opposite order at bonds $1$ and $3$ of a rod of $N=5$ tiles, which are not related by the symmetry, a turned-around copy of $T_3$ can attach to $T_1$, after which $T_2$ and $T_1$ follow in reverse order, which can terminate after $1+3=4$ tiles.
(e) L-shape of $N=4$ tiles (Proposition~\ref{prop:L}). The only reuse is between the two end tiles: the vertical bond reuses the pair of the first bond, so $T_4$ is a rotation of $T_1$.}
\label{Fig:simple_structures}
\end{figure*}
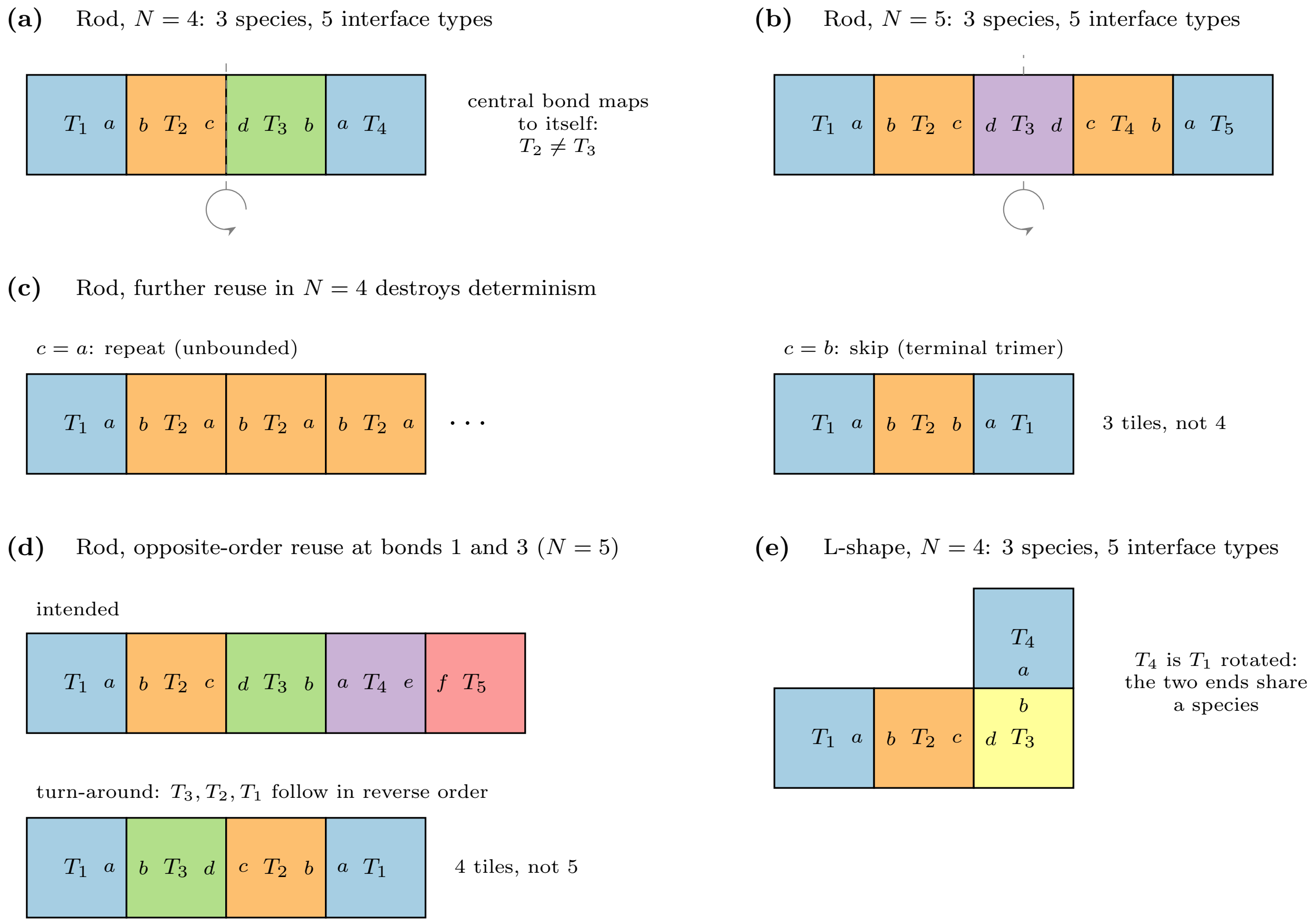

\begin{proof}
\textbf{Fully addressable starting point.}
Giving each of the $N-1$ bonds its own complementary pair, and setting all other faces to neutral interface type $0$, makes every tile a distinct species. This construction is deterministic, but it uses $2(N-1)+1$ interface types and $N$ species, and ignores the symmetry of the rod.

\textbf{Using the symmetry.}
A rod is unchanged by a half-turn about an axis through its midpoint, perpendicular to the rod. This half-turn swaps the tile at position $p$ with the tile at position $N+1-p$, and swaps the corresponding bonds. Since tiles can rotate, the two tiles of each symmetry-related pair can be the same species, and the two corresponding bonds can use the same complementary pair.

\emph{Odd $N$.} The central tile is mapped onto itself, and the $N-1$ bonds  form $(N-1)/2$ symmetry-related pairs. The rod can therefore be assembled  using $(N-1)/2$ complementary pairs, that is, $N$ interface types including  the neutral type, and $(N+1)/2$ species. For example, for $N=5$ (Fig.~\ref{Fig:simple_structures}(b)), with  $a\leftrightarrow b$ and $c\leftrightarrow d$, and listing only the west and  east faces,
\[
T_1:(0,a), \quad T_2:(b,c), \quad T_3:(d,d), \quad T_4:(c,b), \quad
T_5:(a,0).
\]
Here $T_5$ is a rotation of $T_1$ and $T_4$ a rotation of $T_2$. The central  tile carries $d$ on both faces, which works because both its neighbors bind  to it through $c$.  

\emph{Even $N$.} Now the central bond, between the two middle tiles, is  mapped onto itself. It needs its own complementary pair, with one member on  each of the two middle tiles, so these two tiles are different species. The  rod can therefore be assembled using $N/2$ complementary pairs, that is,  $N+1$ interface types, and $N/2+1$ species. For example, for $N=4$ (Fig.~\ref{Fig:simple_structures}(a)),  
\[
T_1:(0,a), \qquad T_2:(b,c), \qquad T_3:(d,b), \qquad T_4:(a,0),
\]
where $T_4$ is a rotation of $T_1$, while $T_2$ and $T_3$ are distinct.

In both cases, growth is deterministic. The interface types are placed so  that whenever an exposed interface can be bound by more than one tile or  face, the alternatives are related by the half-turn symmetry. In the $N=4$  example, for instance, an exposed $a$ can bind either $T_2$ or $T_3$, and  growth then proceeds as $T_1T_2T_3T_1$ or as $T_1T_3T_2T_1$, which is the  same rod viewed from the other end. Every growth pathway therefore produces the rod, from any seed.

\textbf{Why no further reuse is possible.}
Consider what happens if we try to reuse more in the $N=4$ example (Fig.~\ref{Fig:simple_structures}(c)). Setting $c=a$ (and hence $d=b$) turns $T_2$ into $(b,a)$, which binds to copies of itself, so the rod can grow forever. Setting $c=b$ (and hence $d=a$) turns $T_2$ into $(b,b)$, so the trimer $T_1T_2T_1$ can also form and stop.

These two failures illustrate the general problem. If the same complementary pair is used at two bonds that are not related by the
symmetry, tiles intended for one bond can also attach at the other, creating alternative growth pathways. One of three things can then happen:
\begin{itemize}
\item the part of the rod between the two bonds is \emph{skipped}, giving a rod that is too short (as for $c=b$);
\item that part is \emph{repeated}, giving a rod that is too long or grows without bound (as for $c=a$);
\item if the two bonds carry the pair in opposite order, growth can turn around and run back along the rod, giving a rod whose length is correct only if the two bonds are symmetry related (Fig.~\ref{Fig:simple_structures}(d)).
\end{itemize}
The seed may lie to the left of both bonds, between them, or to the right of both, and in each case at least one of these alternatives is available. The argument therefore holds for any seed position, and so applies to both seeded and unseeded assembly. Hence a complementary pair can be shared
only by symmetry-related bonds, and the $N-1$ bonds need at least $\lfloor N/2\rfloor$ distinct pairs. With the neutral type, $\tilde{K}_i\geq 2\lfloor N/2\rfloor+1$.

The same reasoning applies to tiles: two interior positions can be the same species only if they are symmetry related, and, as noted above, an end tile can never share a species with an interior tile. For even $N$ there is one further restriction. The two middle tiles are symmetry related, but they cannot be the same species. If they were, the central bond would join two copies of the same face, and hence two faces of the same interface type, which cannot bind. Hence $\tilde{K}_s\geq\lfloor N/2\rfloor+1$, and the constructions above are minimal.

This central constraint is the origin of the parity dependence in Eq.~\eqref{eqn_linear_assembly}: extending an odd rod by one tile requires an additional species and complementary pair, whereas extending an even rod by one tile requires neither. 
\end{proof}

\subsection{L-shaped assembly}
\label{SM_subsec:L-shape}

\begin{proposition}
\label{prop:L}
For an L-shape of $N\geq3$ tiles, consisting of a straight row of $N-1$ tiles with one further tile attached perpendicularly to one end of the row,
\begin{equation}
\label{eqn_L_assembly}
\tilde{K}_s(N)=N-1,
\qquad
\tilde{K}_i(N)=2N-3.
\end{equation}
\end{proposition}

\begin{proof}
Place the row along the $x$-axis and number its tiles from left to right as $1,\ldots,N-1$, and let tile $N$ sit directly north of tile $N-1$, in the $+y$ direction. Bonds $1,\ldots,N-2$ are then horizontal, and the last bond is vertical.

\textbf{Construction.}
Unlike the rod, the L-shape (with $N>3$) has no symmetry that maps the path onto itself, so the symmetry-based reuse used for the rod is not available. The only reuse we exploit is between the two end tiles, which both bind through a single face. Listing faces in (N, E, S, W) order, with out-of-plane faces neutral, we take
\[
\begin{aligned}
T_1 &: (0,\;a_1,\;0,\;0),\\
T_p &: (0,\;a_p,\;0,\;b_{p-1}), \qquad p=2,\ldots,N-2,\\
T_{N-1} &: (b_1,\;0,\;0,\;b_{N-2}),\\
T_N &: (0,\;0,\;a_1,\;0),
\end{aligned}
\]
where $a_p\leftrightarrow b_p$. Each horizontal bond has its own pair $(a_p,b_p)$, and the vertical bond reuses the pair $(a_1,b_1)$ of the first bond, so that $T_N$ is simply $T_1$ rotated. The construction therefore uses $N-1$ species and $N-2$ complementary pairs, that is, $2N-3$ interface types including the neutral type.

The only choice during growth arises when a copy of $T_1$ exposes $a_1$, which can bind either $T_2$ or the corner tile $T_{N-1}$. In the first case it becomes the left end of the arm, and in the second the tile on top of the corner. Either way, growth then proceeds uniquely to the same L-shape.

For example, for $N=4$ (Fig.~\ref{Fig:simple_structures}(e)), with $a\leftrightarrow b$ and $c\leftrightarrow d$,
\[
T_1=(0,a,0,0), \quad T_2=(0,c,0,b), \quad T_3=(b,0,0,d), \quad
T_4=(0,0,a,0)
\]
assemble the L-shape using five interface types and three species. As for the rod, further reuse fails. Setting $c=a$ (and hence $d=b$) makes $T_2$ bind to copies of itself, so growth is unbounded. Setting $c=b$ (and hence $d=a$) lets $T_4$ attach directly to $T_2$, so the straight trimer $T_1T_2T_4$ can also form.

\textbf{Why no further reuse is possible.}
The skip, repeat, and turn-around argument used for the rod applies here as well. If two horizontal bonds share a complementary pair, skipping or repeating the part between them changes the number of tiles. Turning around gives either a straight rod or a path that passes through the corner twice and so has two bends. None of these is the L-shape, so every horizontal bond needs its own pair.

The vertical bond can share a pair only with the first bond. Sharing it with any other horizontal bond $p\geq2$ would allow the top tile or the corner tile to attach directly to the arm at bond $p$, giving a path of $N-p+1$, $p+1$, or $N+p-1$ tiles, or allow part of the arm to be repeated. None of these outcomes is the target L-shape. Sharing with the first bond is exactly the reuse made in the construction. The $N-1$ bonds therefore need at least $N-2$ distinct pairs, and with the neutral type $\tilde{K}_i\geq 2N-3$.

For species, the same argument shows that no two interior tiles along the straight part of the arm can be the same species. The corner tile must also be distinct from these tiles, because its two active faces are at right angles, whereas theirs are opposite, and a rotation cannot change the angle between two faces. The $N-2$ interior tiles therefore require distinct species. An end tile, with its single active face, cannot share a species with an interior tile, so the two ends require at least one additional species. Hence $\tilde{K}_s\geq N-1$. Together with the construction above, these lower bounds establish Eq.~\eqref{eqn_L_assembly}.
\end{proof}

\paragraph*{Comparison.}
The rod and the L-shape differ only in the position of a single cube, yet their minimal complexities differ substantially. For large $N$, the rod needs about $N/2$ species and $N$ interface types, whereas the L-shape needs about $N$ species and $2N$ interface types. The difference comes from symmetry: the half-turn symmetry of the rod lets tiles and interfaces be reused in symmetry-related pairs, whereas the L-shape has no such symmetry and lies close to the fully addressable limit of $N$ species and $2N-1$ interface types.

\subsection{Rule-set diversity of the square and rod tetrominoes}
\label{SM_subsec:square_rod}

\begin{figure*}
\includegraphics[width=\textwidth]{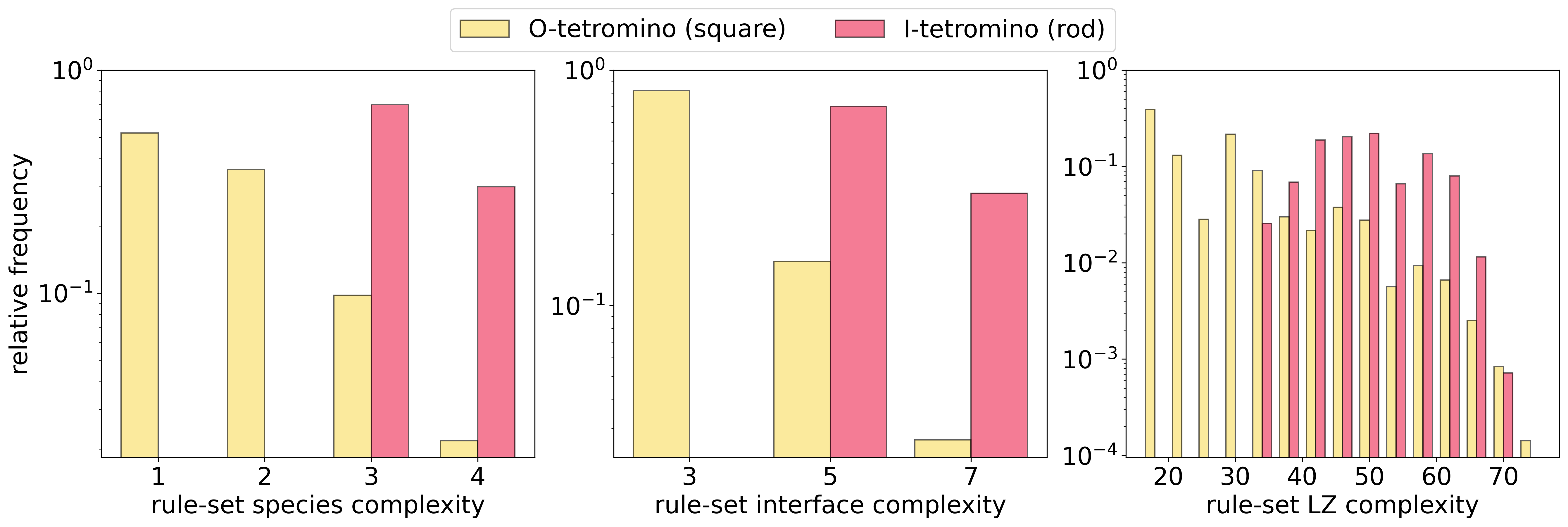}
\caption{
\textbf{Complexity distributions of rule sets generating the square and rod tetrominoes.}
Distributions of individual rule-set complexities for the O-tetromino (square) and I-tetromino (rod) in the two-dimensional rule space $S^{2D}_{4s,17c}$, obtained from $10^8$ randomly sampled rule sets under seeded assembly. Panels show species complexity, interface complexity, and Lempel--Ziv complexity. Instead of assigning each structure the minimum complexity of its generating rule sets, here we retain the complexity of every sampled rule set and show the resulting distribution. The square is generated by a broader range of rule-set complexities than the rod. The minimum assembly complexities are one tile species and three interface types for the square, and three tile species and five interface types for the rod.}
\label{Fig:input_comp_dist}
\end{figure*}

Although each structure is assigned a single assembly complexity, based on the simplest BD rule set that generates it, many BD rule sets of different complexities may produce the same structure. Structures that admit compact assembly descriptions are expected to be compatible with a larger and more diverse set of rule sets than structures that require more specific assembly instructions, and therefore to have higher assembly rule probability. To examine this, Fig.~\ref{Fig:input_comp_dist} shows the distributions of complexities of all individual BD rule sets generating the two-dimensional square and rod of size $4$, sampled from $S^{2D}_{4s,17c}$ under seeded assembly. The square is generated by substantially more rule sets, spanning a wider range of complexities, than the rod, consistent with its much higher assembly rule probability. The minimal complexities found by sampling, three species and five interface types for the rod, agree with Proposition~\ref{prop:rod}.

\clearpage
\section{Robustness of our results to simulation and polycube model parameters}
\label{sec:robustness}

\subsection{Assembly size and spatial cutoffs}

To identify unbounded growth, we impose a maximum linear extent of $D_{\max}=16$ tiles and a maximum assembly size of $N_{\max}=100$ tiles. This spatial cutoff was also used in previous polyomino assembly studies~\cite{Johnston2022}. We expect the linear cutoff to be sufficient for identifying most unbounded assemblies.
We found empirically that, for the $10^9$ samples we use, it is extremely rare to find structures close to these cutoffs in size.  We therefore expect the effect of this cutoff to be small for the properties we are studying here.  However, for larger numbers of allowed species and interfaces, or for longer runs, these cutoffs may need to be adjusted.  

\subsection{Sensitivity to determinism testing}
\label{SM_subsec:kmax}

The determinism test uses up to $k=25$ stochastic assembly trajectories per rule set. This value was chosen as a compromise between sensitivity to non-deterministic growth and computational cost, given that up to $10^9$ rule sets are sampled per rule space. A comparable number was used by Johnston \emph{et al.}~\cite{Johnston2022}, who tested determinism using $20$ repeated assembly trajectories.

To assess the sensitivity of our classification to the number of trajectories, we re-tested all rule sets initially classified as BD in a sample of $10^6$ rule sets from $S^{3D}_{5s,31c}$, using an independent set of up to $k=100$ trajectories. Only $0.25\%$ of these rule sets were reclassified as non-deterministic. These reclassifications occurred predominantly among rule sets generating rare structures, while the assembly rule probabilities of frequently observed structures remained essentially unchanged. Thus, increasing $k$ has little effect on the results, although non-determinism along unsampled trajectories cannot be ruled out.

\subsection{Dependence on rule-space parameters}

\begin{figure*}[!htbp]
    \centering
    {\raggedright\textbf{(a)}\par}
    \includegraphics[width=\textwidth]{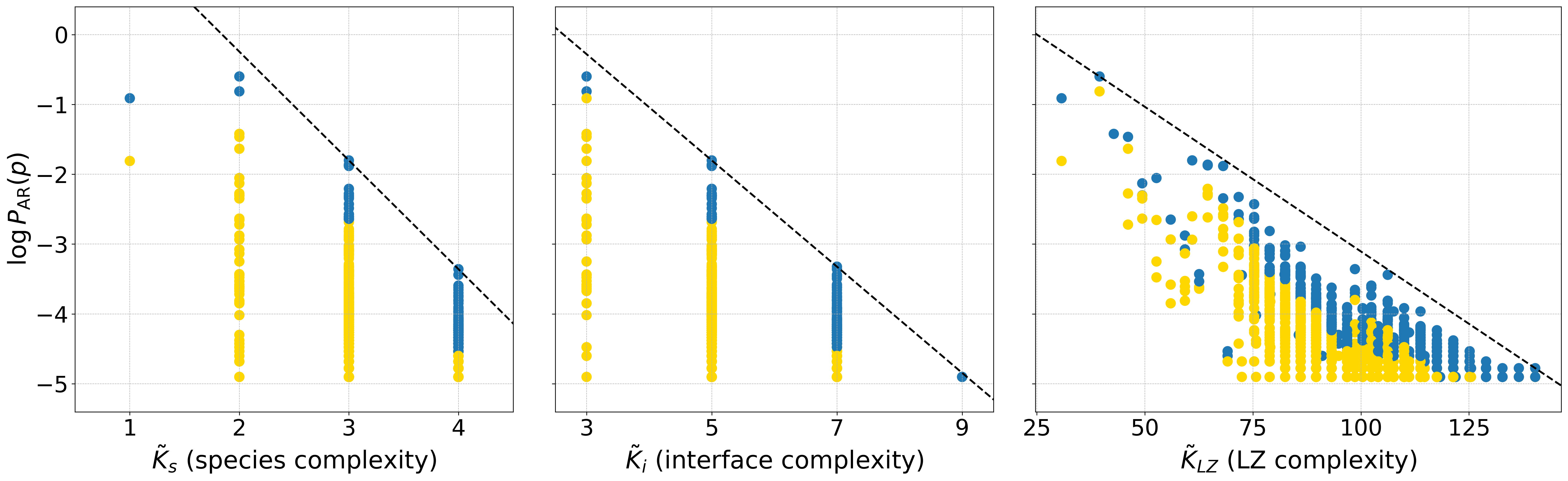}
    \vspace{8pt}
    {\raggedright\textbf{(b)}\par}
    \includegraphics[width=\textwidth]{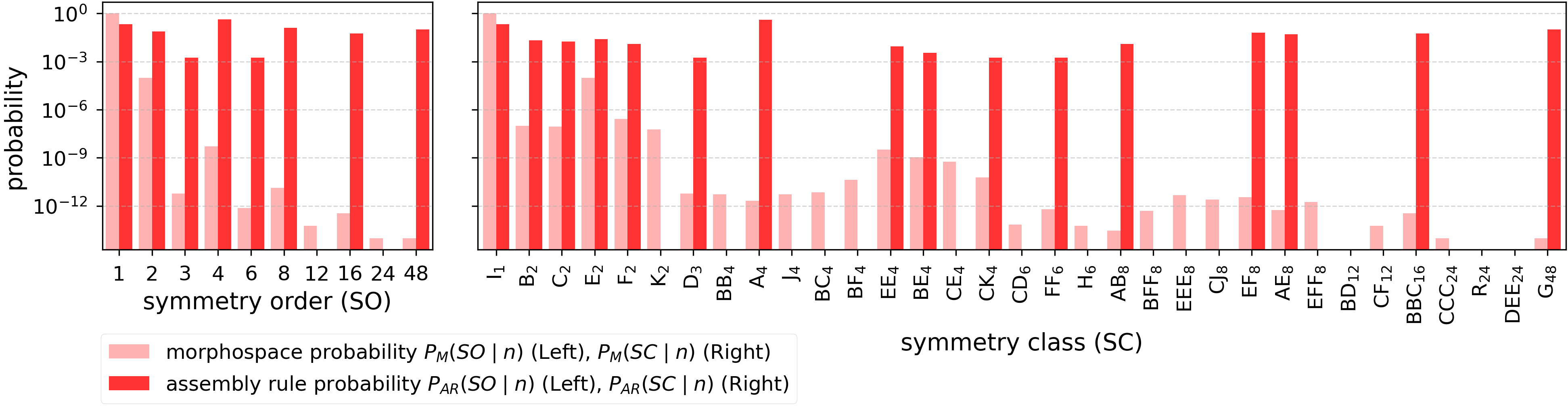}
    \caption{
    \textbf{Robustness of the main results to the rule-space parameters.}
    Results are obtained from $10^8$ randomly sampled rule sets in the three-dimensional rule space $S^{3D}_{4s,17c}$ under seeded assembly, for comparison with the rule space $S^{3D}_{5s,31c}$ used throughout the main text.
    (a) Logarithm of the assembly rule probability, $\log P_{AR}(p)$, against species complexity $\tilde{K}_s$ (left), interface complexity $\tilde{K}_i$ (middle), and Lempel--Ziv complexity $\tilde{K}_{LZ}$ (right), as in Fig.~\ref{Fig:freq_vs_comp_with_fit}. Single cubes are excluded, $P_{{AR}}(p)$ is normalized over the remaining BD rule sets, and structures generated by fewer than three BD rule sets are omitted. Dashed lines show the fitted upper bounds. At each complexity value, blue points collectively account for the upper $50\%$ of the probability weight, while yellow points account for the remaining $50\%$.
    (b) Morphospace probability (light bars) and assembly rule probability (dark bars) for polycubes of size $n=20$, across symmetry orders, $P_M(SO\mid n)$ and $P_{{AR}}(SO\mid n)$ (left), and across symmetry classes, $P_M(SC\mid n)$ and $P_{{AR}}(SC\mid n)$ (right), as in main-text Fig.~\ref{Fig:3D_prob_order_and_class_5s31c_10^9_20mer}(a). The morphospace probabilities are calculated from the OEIS enumerations in Table~\ref{Table:OEIS} and are therefore identical to those in Fig.~\ref{Fig:3D_prob_order_and_class_5s31c_10^9_20mer}(a); only the assembly rule probabilities depend on the rule space.
    }
    \label{Fig:robustness_4s17c}
\end{figure*}

The main results are obtained from the rule space $S^{3D}_{5s,31c}$, which is a compromise between expressivity (how many kinds of polycubes can be made) and computational efficiency; larger rule-spaces are more expensive to sample.  To test the effect of varying the number of tile species and interface types, we also sampled rule sets from $S^{3D}_{4s,17c}$. As shown in Fig.~\ref{Fig:robustness_4s17c}, the relationships between assembly rule probability and both assembly complexity and symmetry are quite similar in the two rule spaces.
The qualitative trends should not be that sensitive  to our choice of $s=5$ and $c=31$.   It would be interesting to do these analyses  with larger numbers of species and interface types, but that is also computationally more expensive.   

\subsection{Dependence on sampling depth}

The full morphospace of polycube structures is incredibly large.  Sampling inevitably only captures a small fraction of the full morphospace of possibilities.    To  assess the effect of sampling depth, we compare results from $10^6$ to $10^9$ randomly sampled rule sets in $S^{3D}_{5s,31c}$ under seeded assembly. Increasing the number of samples recovers additional distinct polycubes, particularly larger and less frequently assembled structures, while the overall assembly rule probability distribution across sizes changes little (Fig.~\ref{Fig:3D_5s31c_10^6_to_10^9_nmer_hist_rule_count_fraction}). Similarly, the assembly rule probability distributions across symmetry orders at fixed size become comparatively stable at the largest sampling depths (Fig.~\ref{Fig:P(symorder)_nrules_diffsizes}).

\subsection{Minimal rule-set calculations}


We estimate assembly complexity by taking the minimum complexity among sampled BD rule sets that produce a particular structure. This approach implies that longer sampling can improve estimates of the minimum assembly complexity, as additional rule sets may provide simpler descriptions of structures already observed (see, e.g., Fig.~\ref{Fig:freq_vs_comp_nrules}). Empirically, we find that increasing the sampling depth has little effect on the estimated minimum complexities of the most frequently assembled structures, partly because simpler rule sets tend to be sampled more frequently than more complex ones. We expect sampling errors to be more significant for less frequently assembled structures. Formal methods, such as the SAT-based inverse-design solver of Ref.~\onlinecite{Bohlin2023}, could be used to determine the true minimum complexities, but are computationally expensive. Since the most frequently assembled structures account for much of the probability weight in our analysis, we do not expect more exhaustive minimization to substantially affect our main findings.

\subsection{Dependence on seeded and unseeded assembly}

The main results are obtained under seeded assembly, where the first tile species initiates growth.  We also perform unseeded assembly, where the initial tile species is chosen randomly from the assembly kit. Unseeded assembly produces a significantly smaller fraction of BD rule sets (main-text Fig.~\ref{Fig:UND_assembly}(c)), but the overall relationships between assembly rule probability, complexity, and symmetry remain qualitatively similar. The probability--complexity distributions retain their approximately exponential upper bounds (Fig.~\ref{Fig:freq_vs_comp_with_fit}), while the size distributions show similar patterns under both growth conditions (Fig.~\ref{Fig:2D_3D_5s31c_10^9_nmer_hist}). The corresponding symmetry distributions for size-$20$ polycubes are compared in Figs.~\ref{Fig:3D_prob_enhance_complexity_minmax_top_20mer} and~\ref{Fig:3D_prob_enhance_complexity_minmax_top_20mer_unseeded}.

\subsection{Effects of cubic lattice geometry}

The cubic lattice restricts the possible symmetries of assembled structures to those compatible with its geometry. The relative abundance of different symmetry classes in morphospace and their assembly rule probabilities will depend on the  underlying lattice.
Investigating  the relationships between symmetry, assembly complexity, and assembly rule probability  for other lattices or off-lattice systems remains an open direction for future work.

\subsection{Limitations of the assembly dynamics}

Our model considers sequential, irreversible attachment of tiles, whereas more general models could allow reversible attachment and detachment, variable bond strengths, and kinetic effects, as in the kinetic Tile Assembly Model (kTAM)~\cite{winfree1998thesis}. Similarly, our model allows binding only between distinct complementary interface types, whereas more general interaction rules could permit self-binding interfaces through the interaction matrix $A_{ij}$, as well as cooperative attachment, as in the abstract Tile Assembly Model (aTAM)~\cite{winfree1998thesis}.
It would be interesting to investigate  the  relationships between assembly rule probability and complexity under these more general interaction rules and assembly dynamics.

\putbib
\end{bibunit}

\end{document}